\documentclass[11pt]{article}

\usepackage[english]{babel}

\usepackage[numbers,square,sort&compress]{natbib}
\usepackage{graphicx}
\usepackage[colorlinks=true, allcolors=blue]{hyperref}

\usepackage{amsfonts}
\usepackage{amsthm} 
\usepackage[singlespacing]{setspace}
\usepackage[headheight=1in,margin=1in]{geometry}
\usepackage[inline]{enumitem}
\usepackage{makecell}
\usepackage{tabularx}
\usepackage{mathtools}
\usepackage{amsmath}

\usepackage{bbm}
\usepackage{multirow}
\usepackage{xcolor}
\usepackage{float}
\usepackage [autostyle, english = american]{csquotes}
\usepackage{epigraph}
\usepackage{algorithm}
\usepackage{algorithmic}
\usepackage{booktabs}
\usepackage{amssymb}
\usepackage{authblk}

\usepackage{xr}  
\usepackage{setspace}
\setcitestyle{aysep={}} 

\title{\bf Emergent Directedness in Social Contagion}
\author[1]{Fabian Tschofenig}
\author[1\thanks{Corresponding author: dguilb@stanford.edu}]{Douglas Guilbeault}
\affil[1]{Stanford Graduate School of Business, 655 Knight Way, Stanford, CA 94305, USA}

\date{}

\begin{document}
\maketitle

\begin{abstract}
An enduring challenge in contagion theory is that the pathways contagions follow through social networks exhibit emergent complexities that are difficult to predict using network structure. Here, we address this challenge by developing a causal modeling framework that (\textit{i}) simulates the possible network pathways that emerge as contagions spread and (\textit{ii}) identifies which edges and nodes are most impactful on diffusion across these possible pathways. This yields a surprising discovery. If people require exposure to multiple peers to adopt a contagion (a.k.a., `complex contagions’), the pathways that emerge often only work in one direction. In fact, the more complex a contagion is, the more asymmetric its paths become. This emergent directedness problematizes canonical theories of how networks mediate contagion. Weak ties spanning network regions – widely thought to facilitate mutual influence and integration – prove to privilege the spread contagions from one community to the other. Emergent directedness also disproportionately channels complex contagions from the network periphery to the core, inverting standard centrality models. We demonstrate two practical applications. We show that emergent directedness accounts for unexplained nonlinearity in the effects of tie strength in a recent study of job diffusion over LinkedIn. Lastly, we show that network evolution is biased toward growing directed paths, but that cultural factors (e.g., triadic closure) can curtail this bias, with strategic implications for network building and behavioral interventions.

\end{abstract}

\onehalfspacing

\newpage
\section{Introduction}
Despite decades of research and investment into the study and application of contagion theory, a persistent and foundational challenge remains: how contagions spread through social networks is often unpredictable. This is especially true for the spread of ``complex contagions", which are behaviors that are risky, costly, and novel, such that people require influence from multiple social contacts to incentivize adopting the behavior \cite{centola2007complex, guilbeault2018complex}. Complex contagions are central to cultural dynamics, including the spread of social norms \cite{centola2005emperor, centola2007complex, monsted2017evidence, puglisi2008cultural, guilbeault2021experimental}, innovative technologies \cite{karsai2014complex}, fake news \cite{tornberg2018echo, monsted2017evidence}, ethical choices \cite{centola2010spread, traag2016complex}, emotional states \cite{pinus2025emotion}, and political movements \cite{romero2011differences, state2015diffusion, saetre2025protests}. Yet, the interaction between complex contagions and social network structure is fundamentally difficult to predict. For example, recent work shows that it is NP-hard to predict the global spreading patterns of complex contagions using only the local neighborhood connectivity of the agents initiating the contagion (i.e., the `seeds') \cite{chaitanya2022hardness}. A given agent's local network connectivity has limited predictive power over the distribution of network connections several steps beyond this agent's neighborhood. In large enough networks, the set of possible pathways for a complex contagion beyond a node's neighborhood is often vast and unknown. By consequence, studying how complex contagions spread requires an approach common to studying complex systems \cite{wolfram2003new, epstein2012generative}; given the inability to predict diffusion pathways from initial conditions, one must instead observe the spreading patterns that dynamically emerge as the contagion diffuses in real-time. The necessity of this observational approach, to date, has imposed significant limitations on the generalizability and applicability of complex contagion theory to empirical, policy-relevant interventions. 

A core reason why the diffusion of complex contagions is difficult to predict is because they traverse networks along pathways that are qualitatively different than the pathways traversed by simple contagions -- i.e., contagions that only require exposure to a single `infected' contact to trigger spreading, such as biological pathogens or easy information \cite{guilbeault2018complex}. As recent work explains \cite{guilbeault2021topological}, most standard metrics of network structure -- such as shortest path length, clustering coefficients, and betweenness centrality -- assume simple contagion dynamics by treating single-tie connections as an indication of the ability for nodes to influence each other. Consider the pervasive use of path length to characterize the degrees of social separation between actors \cite{milgram1967small, newman2018networks}, for which chains of single ties are deemed sufficient for facilitating spreading and thus for capturing social distance via the number of diffusion steps separating agents. However, for complex contagions, single ties are insufficient to enable spreading \cite{centola2007complex, guilbeault2021topological}. Instead, to adopt a complex contagion, people require exposure to influence from multiple distinct actors in their network, which is typically modeled by a node-level adoption threshold (e.g., the total number of adopters or the fraction of one's neighbors to whom one needs to be exposed to trigger behavior change). Nodes which are connected through a chain of single-ties are frequently unable to spread complex contagions to each other due to the lack of sufficient clustering and local reinforcement for sustaining the activation of nodes' thresholds along the pathway. As a result, standard measures of network distance and centrality developed for tracking simple contagions tend to be poor approximations of complex contagion dynamics. 

To address this, recent work by Guilbeault \& Centola (2021) \cite{guilbeault2021topological} shows how complex contagions spread through chains of sufficiently wide bridges (i.e.,\textit{ complex paths}) that sustain local reinforcement at each step in the chain, supporting new measures of social distance and connectivity that better capture complex contagion spreading dynamics. Integral to the measure of complex paths is the insight that the distribution of agents' thresholds directly impacts the pathways a complex contagion can traverse on a graph, revealing that spreading dynamics are not reducible to network structure alone \cite{guilbeault2021topological}. On the one hand, this helps to explain why measures of static network structure fall short of modeling complex contagions, since complex paths are determined by the interaction of network structure with the distribution of agents' resistance and social influence requirements. Yet, on the other hand, this suggests that the spreading dynamics of complex contagions are highly sensitive to the threshold distribution, exposing a world of unexplained heterogeneity in the distribution of possible complex paths that emerge during a spreading process. This raises the question of whether progress can be made not by attempting to define static network structures that can predict the spread of complex contagions \textit{a priori}, but rather by identifying and mathematically characterizing the space of possible complex paths that emerge as a result of the interaction between threshold distributions and underlying network structure. 

At present, no such method exists for modeling and mathematically characterizing the counterfactual space of complex paths that emerge across possible seed sets and threshold distributions for a given network. To address this gap, we develop a novel causal modeling framework that identifies which edges and nodes are most causally impactful for inducing adoption across the suite of possible complex paths that can emerge for a given graph across a comprehensive range of counterfactual spreading scenarios. Unlike empirical studies of contagion which typically observe only a single spreading event originating from a single set of seeds (providing a sparse and unrepresentative sample of possible complex paths), our method discovers properties of diffusion dynamics that generalize across counterfactual spreading events and thereby informs theories of how spreading dynamics unfold \textit{in expectation}. Characterizing the space of possible complex paths is necessary for representing network structure in terms of the connectivity and differential spreadability among nodes for complex contagion. Without a counterfactual approach, it is exceedingly difficult to make inferences about the differential spreading advantage of nodes who happen to \textit{not} be selected as initial seeds in a single empirical spreading process. Our counterfactual approach reveals that there are fundamental properties of both network structure and diffusion dynamics that are only learnable when aggregating across a comprehensive set of possible spreading pathways for complex contagions, unconstrained from limited empirical observations. While numerically iterating over the full space of complex paths for large networks is combinatorically infeasible, we develop and validate a scalable sampling procedure that results in stable and convergent rankings of expected causal impact of nodes and edges on complex diffusion. This allows us to extract a representative sample of the dominant complex paths for large networks that drive their diffusion dynamics. 

To precisely isolate and measure properties of possible complex paths, we apply our method to the simplest canonical model of contagion dynamics, which assumes that agents observe and are influenced by each of their social contacts' behavior state via undirected, symmetric ties; that is, we assume that all agents have equal influence capability over each other (e.g., no actor is more charismatic or persuasive than any other), and that each agent observes and influences each other symmetrically (see formal model specification in the following section). These simplifying assumptions are common across canonical models of social contagion \cite{newill1964generalization, granovetter1983threshold, centola2013homophily, flache2014small, dellaposta2015liberals, centola2005emperor, centola2007complex, centola2015social, centola2018behavior, guilbeault2021topological} \footnote{This assumption originates in models of simple contagion, such as the spread of biological pathogens or easy information, which typically hold if agent \textit{x} has the potential to infect or inform \textit{y}, then \textit{y} can do the same for \textit{x}. Yet, assumptions of influence symmetry also extend to complex contagions. Standard models of complex contagion assume that if \textit{x} can spread a complex contagion to \textit{y} with the help of their shared friend \textit{z}, then in theory \textit{y} could also do the same to \textit{x} with \textit{z}'s support \cite{centola2007complex, guilbeault2021topological}.}. It is also common in efforts to map empirical social networks to represent dyadic ties as symmetric and undirected in nature (e.g., by connecting two people if they reciprocally report each other as friends) \cite{wasserman1994social, easley2010networks,  burt2013social, burt1994measuring, banerjee2013diffusion, burt2024guanxi, paluck2016changing, airoldi2024induction}. This assumption is also built directly into popular social media platforms, which often represent two people as connected in a network (e.g., as `friends' on Facebook or `connections' on LinkedIn) if and only if both agree to this connection \footnote{Even in cases where observed social networked involve explicitly directed ties by design (such as follower-following relations on the social media platform `X'), it is nevertheless still common for researchers to compress these networks into an undirected graph prior to applying standard network analytic techniques.}. Applying our method under the parsimonious assumption of undirected dyadic ties reveals a striking property of network structure and diffusion dynamics that significantly explains the failure of prior network measures to accurately predict complex contagions, namely: most possible complex paths can only be traversed in one direction and are thereby functionally directed despite each individual tie being defined in an undirected manner and all node-level characteristics being identical. We demonstrate both through simulations and mathematical proofs that the more complex a contagion is (in terms of the amount of social reinforcement required), the more functionally directed its possible complex paths become. We refer to this form of directedness as \textit{emergent}, since it is neither a property of the network structure nor the threshold distribution alone, but rather of their interaction. Observing emergent directedness in this setting is especially striking given that our modeling assumptions are as strict as possible for eliminating directedness. This suggests it is a fundamental property of complex contagions that they induce directionality in diffusion in an otherwise undirected system. 

Importantly, the capacity for combinations of undirected, single ties to exhibit emergent, functionally directed properties is not unique to the standard complex contagion model. We show that this finding is highly replicable across a range of sensitivity and generality tests, including (i) the examination of alternative reinforcement-based contagion models such as the linear threshold model \cite{kempe2003maximizing} and the independent cascade model \cite{shakarian2015independent} (Figure~\ref{fig:boxplot_symmetry_LTM}, \ref{fig:boxplot_symmetry_LTM_Gauss} and \ref{fig:boxplot_symmetry_ICM}), and (ii) the integration of noisy stochastic subthreshold adoptions that intermix simple and complex contagion dynamics \cite{eckles2024long} (Figure~\ref{fig:combined_boxplots_noisy} and \ref{fig:kdeplot_symmetry_RC_RCS_GI_NOISY_SINGLE}) . As our simulations and mathematical analyses indicate, it is remarkably hard to observe complex contagion dynamics in any model that do \textit{not} lead to emergent directedness among ties and spreading paths -- especially as the level of local reinforcement required increases -- suggesting a level of generality that characterizes diffusion dynamics across models and empirical contexts. As such, these findings may have theoretical and practical implications for network modeling across fields. We briefly summarize our findings and methods below before presenting our results in detail. 

\subsection{Overview of Findings}
We begin by using our measures to evaluate the canonical construct of weak ties that connect distant parts of the network -- those ties that underlie the famous ``strength of weak ties" argument and the formation of small-world networks \cite{granovetter1973strength,watts1998collective}. Our analyses focus on a structural definition of weak ties as ties that span distant network regions (also referred to as `long' ties) \cite{centola2007complex, park2018strength}) \footnote{This is distinct from additional, more qualitative definitions of weak ties in terms of interpersonal connection characteristics, such as interaction frequency, familiarity, or affective closeness (where weakness implies lower levels of these relationship qualities). Granovetter's classic introduction of the concept of `weak' ties includes both structural and interpersonal, affective characteristics, and the broader literature continues to ambiguously evoke both aspects with this theoretical construct. In following with research on complex contagions \cite{centola2007complex}, we focus our analysis on testing theories pertaining exclusively to the structural definition of tie weakness.}. Classic network theory argues not only that weak ties accelerate the spread of contagions, but that they do so in a reciprocal fashion, increasing mutual influence and the convergence of information and behavior throughout the network \cite{watts1998collective, sporns2006small}. As expressed by one popular account of why small-world networks characterized by long ties are unique: ``The conjunction of local clustering and global interaction provides a structural substrate for the coexistence of functional segregration and integration" (pg. 19219) \cite{sporns2006small}. While this may be true for simple contagions such as information diffusion, this integrative balance can easily break down once contagions require local reinforcement from multiple peers. Through both mathematical proofs and computational simulations (involving artificial and real empirical social networks), we show that when weak ties are able to spread complex contagions, they do so in a highly directed fashion, preferentially enabling the spread of contagions from one network region to the other. Surprisingly, of the weak ties capable of spreading complex contagions, flow patterns become increasingly asymmetric the longer these weak ties become. In this way, our findings suggest that the strength of weak ties may be inseparable from their role in perpetuating global influence inequalities rather than mutual integration in the spread of culture across communities. We then build on this finding to provide a novel account of unexplained u-shaped heterogeneity in the impact of weak ties in a recent large-scale study of job diffusion over LinkedIn \cite{rajkumar2022causal}. Our methods reveal that ties with moderate strength are the most causally impactful on spreading dynamics across a wide range of networks and hypothetical threshold distributions. 

Our methods further uncover inversions to mainstream network theory regarding the role of centralities in influence dynamics. We show that as the local reinforcement required for a contagion increases, contagions are more likely to flow into rather than out of the network's core (i.e., into the nodes with the highest degree). This provides a mechanistic interpretation of a puzzling result reported across a number of empirical studies which observe outsized influence in the network periphery for the rise and spread of social innovations, including political movements and new ideas \cite{steinert2017spontaneous, hassanpour2016leading, vicinanza2023deep}. To the extent that the concept of a network `core' is intended to indicate influence centrality in a graph (where, for example, degree is treated as an indicator of influence according to simple contagion frameworks), then through the lens of complex contagions, we find that -- in terms of influence centrality -- the periphery becomes the core. 

A full exploration of the practical applications of our theory is beyond the scope of this study, which focuses on theoretically identifying and characterizing emergent directedness. For now, we conclude by using our measures to gain insight into endogenous and strategic bridge formation in social networks, a central topic in organizational design \cite{clement2018searching} and contagion policy \cite{centola2021change}. As we examine the distribution of bridges in simulated and empirical networks, we find that it is rare to observe symmetric bridges that equalize the spread of complex contagions between network regions. In our supplementary appendix, we provide mathematical proofs establishing that there are many more possible ways to form asymmetric than symmetric bridges between two communities, especially when induced by random tie rewirings as is characteristic of small-world networks. It follows that absent a social mechanism that selects for symmetry in bridge formation, the formation of asymmetric directed bridges in network evolution is strictly more likely. Our analyses also examine whether strategic initiatives can resist the pull of asymmetric bridges by promoting tie formation via triadic closure, which refers to an increased likelihood for new ties to form between mutual friends rather than between outsiders \cite{mosleh2025tendencies}. However, our simulations show that pressures toward triadic closure have to be strikingly strong to reliably mold evolving bridges into symmetrical channels. Implications for strategy in policy and organizational design are discussed. 

In sum, we present our methods and findings as follows: (1) We begin by defining the general influence model for complex contagion. (2) Next, we briefly describe our method for modeling the causal significance of ties and nodes in counterfactual spreading processes for any network. (3) Third, we demonstrate across a wide range of simulated and empirical networks that increasing the complexity of a contagion increases the rate of asymmetry in the possible causal paths that emerge. (4) We then use our measures to examine the ``strength of weak ties" and present evidence that weak ties are most often asymmetric and functionally directed for complex contagions. (5) We further show how this insight helps explain a central puzzle in contagion theory -- namely, the recent empirical finding in a large-scale study of job diffusion over Linkedin that the causal impact of weak ties follows an inverted u-shape, with moderately weak ties being most impactful, deviating from prior theories of both simple and complex contagions \cite{rajkumar2022causal}. (6) We then generalize beyond single weak ties and examine the directedness of wide bridges -- i.e., chains of locally clustered ties. We show that the specific directionality of bridges shifts as a function of the complexity of a contagion: for simpler contagions, bridges direct contagions from the core to the periphery, but for more complex contagions, this pattern inverts. (7) Finally, we examine the implications of these findings for understanding random and strategic bridge formation in network evolution. 

\begin{figure}[H]
    \centering
    \includegraphics[width=1\linewidth]{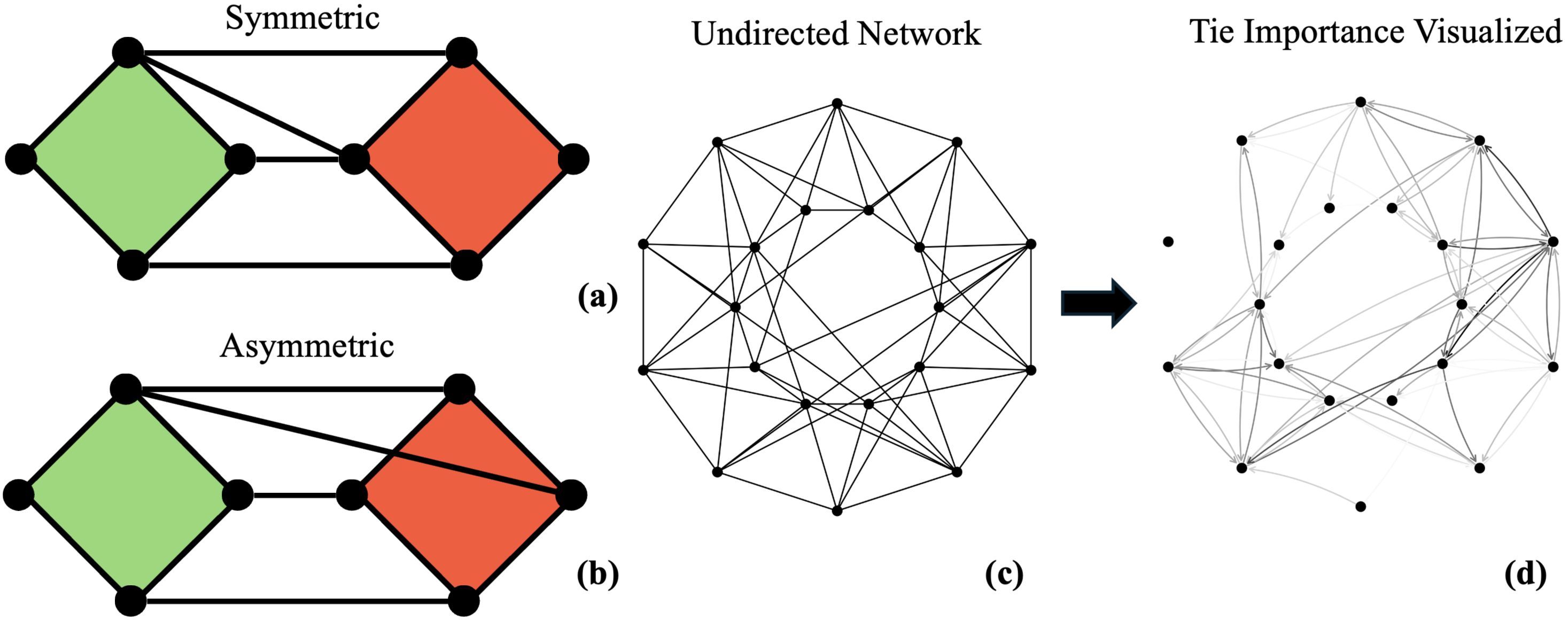}
    \caption{\textbf{An Example of Emergent Directedness in Complex Contagion.} 
    All panels depict complex contagions with a threshold $T = 2$. 
    (a) A symmetric bridge between the red and green community where contagion can spread in both directions. 
    (b) An asymmetric bridge where it is possible to spread from the red to the green neighborhood but not in the reverse direction. 
    (c) An exemplary undirected network before analysis. 
    (d) The same network visualized with edge opacity scaled by our measure of Causal Tie Importance, revealing emergent asymmetries and directed pathways along which complex contagions are most likely to flow, even in this structurally undirected system. 
    This visualization demonstrates how the interaction between network topology and complex contagion dynamics induces emergent directional biases.}
    \label{fig:Panel_Asymmetry}
\end{figure}

\section{Overview of Methods}
We study threshold-based diffusion on a network $G=(V,E)$, following the General Influence Model for complex contagions~\cite{guilbeault2021topological}. Each node $i \in V$ adopts a binary state $\sigma_i^k(t) \in \{0,1\}$, where $\sigma_i^k(t) = 1$ indicates that $i$ is active at time $t$ in the scenario defined by seed set $R_k$.

\paragraph{Seed Sets.} We define two algorithms for identifying seed sets:
$$
R^{RS} \quad \text{and} \quad R^{CS}.
$$
The first, $R^{RS}$, contains random node subsets of $V$ of a fixed size; the second, $R^{CS}$, contains clustered subsets in which every seed node has at least one seed neighbor. A particular $R_k$ is drawn from either $R^{RS}$ or $R^{CS}$ to initialize the diffusion at $t = t_0$.

\paragraph{Thresholds.} Each node $i$ has a threshold $T_i$, the minimum number of active neighbors required for $i$ to become active. For relative thresholds,
$$
T_i = \lceil \theta_i \, |N[i]| \rceil,
$$
where $\theta_i$ is the fraction of neighbors that must be active, $N[i]$ is the (closed) neighborhood of $i$, and $\lceil \cdot \rceil$ denotes the ceiling function.

\paragraph{Update Rule.} The process evolves in synchronous steps:
$$
\sigma_i^k(t+1) = 
\begin{cases}
1, & \text{if } \sum_{j \in N[i]} \sigma_j^k(t) \ge T_i,\\
0, & \text{otherwise}.
\end{cases}
$$
Hence, node $i$ becomes active at $t+1$ if at least $T_i$ of its neighbors were active at $t$. We iterate until \textit{convergence}, which is reached when no node changes state.

Under this model, one might naturally expect that every edge in an undirected network would aggregate into bidirectional influence -- i.e., that if A and B are connected enough with $T$ or more shared neighbors, then both A and B are capable of spreading a complex contagion requiring $T$ reinforcements to each other. This intuition is captured by panel Figure~\ref{fig:Panel_Asymmetry}(a), which displays a symmetric bridge. In this case, it is true that on average, without any systematic differences in where a contagion is seeded in the network, it is expected that the green and red neighborhoods are equally capable of spreading a complex contagion of $T = 2$ to each other. Yet, Figure~\ref{fig:Panel_Asymmetry}(b) (green region vs.\ red region) visualizes a simple counterexample by presenting an asymmetric bridge. In this case, rewiring only a single tie (while keeping the overall bridge size constant) can entirely disrupt the ability for the green neighborhood to spread the contagion into the red neighborhood, while the red neighborhood continues to enjoy its ability to spread contagions into the green region. The asymmetry of this bridge is an emergent property of the complex contagion; both bridges in (a) and (b) are equally symmetric in terms of simple contagion. Despite all ties in all cases being defined as undirected at the dyadic level with no node-level differences in influence capability, the threshold $T = 2$ imposes a \textit{de facto} one-way path of influence at the structural level in panel (b). Although this asymmetry is obvious in this toy example, hidden directional paths are far more difficult to detect in larger, more intricate networks, even though they will prove to be considerably more common than their symmetric counterparts. 

The asymmetric flow dynamics introduced by complex contagions introduces a measurement challenge. If social contagions flow preferentially in one direction across an edge, continuing to treat the network as undirected introduces errors in predicting, analyzing and manipulating diffusion outcomes. To correctly capture the constraints imposed by the contagion process, we transition to a directed representation, where edges reflect the observed, causal vectors of influence rather than the raw, static connectivity of the graph. By switching to a directed representation -- from Figure~\ref{fig:Panel_Asymmetry}(c) to (d) -- we transform the network from a static skeleton of potential simple pairwise interactions into a causally grounded model of emergent influence dynamics that arise as a function of contagion complexity. This redefinition enables us to correctly infer -- across a comprehensive range of possible initial seeding configurations -- which ties consistently contribute to the spread of complex contagions, which ties remain inactive, and where bottlenecks or reinforcement clusters among ties occur within the broader directed flow landscape, all as a function of the complexity of the contagion spreading. 

To quantitatively uncover these asymmetries, we introduce our causal modeling approach, which provides two novel measures: Causal Tie Importance and Causal Node Importance. Unlike traditional centrality measures, which assess influence based on static network properties, our approach reconstructs influence dynamics and trajectories directly from the diffusion process itself. For each activated node in a diffusion process, we trace back along the sequence of prior activations recursively to identify the causal contributors to its activation. We adopt this approach because influence dynamics in complex contagions are emergent and often computationally irreducible \cite{guilbeault2021topological, chaitanya2022hardness}, meaning the only way to determine causal importance is to follow the actual activation sequence, which often cannot be determined \textit{a priori} based on local network features. Our method, by design, ensures that no external assumptions about influence directionality are imposed. 

To account for the many possible positions of seed nodes, we perform a large range of simulation runs with a representative sample of different random seed configurations for each graph (in section \ref{sec:extensive_symmetry_figs} and \ref{sec:convergence} in the supplementary appendix, we demonstrate the statistical representativeness and estimation stability of our sampling approach, while also replicating our results across different seeding approaches, clustered or dispersed, and across a wide range of both synthetic and real-world networks; the number of seeds and simulation runs for each network analysis are specified in the results section). Then, for each graph, we aggregate the frequency with which each node and tie appears in these causal activation chains across all simulated seeding configurations (see Figure~\ref{fig:Panel_Asymmetry} c-d for an example). By remaining agnostic to the initial seeding configuration and aggregating across a representative sample of possible seeds -- seeking, ultimately, to capture the general flow dynamics across seeding configurations -- our results not only aim to identify what is expected, on average, from any seeding initialization, but also more generally to capture the endogeneous diffusion dynamics of cultural emergence and change, whereby new behaviors, ideas, and attitudes arise and spread from within a network (e.g., via innovation or mutation) without outside strategic intervention. In the next section, we provide a condensed version of our mathematical measures. The extended ``Materials and Methods" section at the end of this study provides full derivations. 

\subsection*{Mathematical Formalisms (Condensed)}
\subsubsection*{Causal Subgraphs}
Once convergence occurs after spreading from seed set $R_k$ -- such that spreading in the General Influence Model is converged --
we extract the \textit{induced subgraph} $I_k\subseteq G$ comprising all active nodes and their edges. For each active node $m\in V(I_k)$, its \textit{causal subgraph} $C_{k,m}$ is the set of nodes in $I_k$ that causally contributed to $m$'s activation at some earlier time $\tau^k_j < \tau^k_m$.  
We then define:
\begin{itemize}
\item \textbf{Causal Node Importance} $\mathrm{NI}(i)$: The count of how many causal subgraphs $C_{k,m}$ include node $i$.
\item \textbf{Causal Tie Importance} $\mathrm{TI}(i,j)$: The count of how often $(i,j)$ appears in these causal subgraphs, with $i$ activating prior to $j$.
\end{itemize}
For comparability across networks with different sizes and structures, both $\mathrm{NI}(i)$ and $\mathrm{TI}(i,j)$ are normalized by the global maximum over all node and edge values, respectively, within each graph. We avoid min-max normalization because in networks with highly homogeneous importance distributions, the minimum and maximum values may be close or identical—leading to unstable or meaningless rescaling. This max-only normalization preserves relative rankings while ensuring scores lie on a shared $[0,1]$ scale.
\subsubsection*{Causal Flow Symmetry}
To measure the macroscopic symmetry (or asymmetry) of causal flows, we define the \textit{Causal Flow Symmetry Measure} $\Xi_s$. For every edge $(i,j)$, we compare $\mathrm{TI}(i,j)$ vs. $\mathrm{TI}(j,i)$, then calculate their Pearson correlation across the entire network \footnote{Our measures reproduce similar results when using alternative statistical measures (e.g., cosine distance); however, the sensitivity of the Pearson correlation measure to outliers is advantageous for identifying meaningful heterogeneity in the effects of contagion complexity on spreading dynamics as a function of network structure.}. Thus,
$$
\Xi_s = \mathrm{Corr}\bigl(\{\mathrm{TI}(i,j)\}, \{\mathrm{TI}(j,i)\}\bigr),
$$
where $\Xi_s=1$ indicates perfectly balanced (symmetric) tie importance in both directions, and lower or negative values indicate greater asymmetry.

\subsubsection*{Uncovering Flow Dynamics Between Core and Periphery}
We can now apply our causal modeling framework to introduce a measure that is able to quantify core-periphery flow dynamics.
\begin{enumerate}
\item $\Delta S_{ij} = \mathrm{TI}(i,j) - \mathrm{TI}(j,i)$, the difference in tie importance between the two directions of $(i,j)$.  
\item $\Delta k_{ij} = k(j) - k(i)$, the difference in degrees of $i$ and $j$.
\end{enumerate}

We then compute $\rho(\Delta S, \Delta k)$, the Pearson correlation between $\Delta S$ and $\Delta k$. A negative $\rho$ suggests that most flow proceeds from high-degree (core) to lower-degree (peripheral) nodes, whereas positive values suggest peripheral-to-core influence.

\subsubsection*{Correlation Between Node Importance and Degree}
To assess how a node's degree influences its overall \textit{Causal Node Importance}, we calculate  
$$
\rho(\mathrm{NI(i)}, k(i)),
$$
i.e., the correlation between $\mathrm{NI}(i)$ (the causal importance of node $i$ in spreading) and $i$'s degree $k(i)$. Positive correlations imply that higher-degree nodes tend to have higher importance in driving contagion, while near-zero or negative values suggest that degree alone does not strongly determine a node's causal influence---or that lower-degree nodes may sometimes be more causally impactful, if negative.

\subsubsection*{Degree-Normalized Tie Importance Correlation}
Finally, we examine \textit{degree-normalized} node importance by considering $\frac{\mathrm{NI(i)}}{k(i)}$. This ratio indicates how influential a node is relative to its connectivity. We then compute:
$$
\rho\bigl(\frac{\mathrm{NI(i)}}{k(i)}, k(i)\bigr),
$$
the correlation between the degree-normalized importance and the degree itself. This reveals whether highly connected nodes exhibit disproportionately high (or low) causal importance beyond what one would predict from their degree alone. A strong positive correlation suggests that large-degree nodes enjoy \textit{more than linear} increases in their causal importance, while weak or negative correlations indicate that degree alone does not guarantee greater causal relevance in the diffusion process.

Taken together, these metrics---Causal Node/Tie Importance, Causal Flow Symmetry, correlations with degree, and degree-normalized correlations---allow for a comprehensive view of both microscopic (node and edge-level) and macroscopic (network-wide) contagion patterns, enabling us to pinpoint pivotal nodes, key channels of influence, and asymmetries in spreading dynamics.

\section{Results}
Through our counterfactual simulation method, we have introduced a novel approach to quantitatively measure the causal importance of ties and nodes in diffusion processes, revealing emergent directedness in flow dynamics as a function of contagion complexity. With these foundations in place, we turn to our empirical findings, which demonstrate how complex contagions reshape traditional understandings of influence and network structure through a combination of computational simulations on synthetic and empirical networks.

\subsection{More Complex, More Asymmetric}
\label{sec:complex_asymmetries}
When contagions are simple ($T$ = 1), any node can spread a contagion to any other node to which it is connected. While nodes may differ in how many other nodes they can quickly reach (for example, as a function of their degree centrality), under simple contagion, all nodes can eventually spread the contagion to all other nodes within a continuous connected component. Influence potential is thus trivially symmetric between all nodes for simple contagions. However, our theory identifies how complex contagions readily break this symmetry, causing stark differences in nodes' ability to spread complex contagions, not just in general, but also to each other (such that neighborhood A can spread a complex contagion to B but not the reverse; see Figure~\ref{fig:Panel_Asymmetry} for an example). Here, we show that the frequency of asymmetric bridges in a network is a direct function of how complex a contagion is (i.e., how much local reinforcement is required to trigger adoption). In supplementary analyses, we demonstrate this intuition through both computational experiments (Appendix~\ref{app:computational_bridges}) and mathematical proofs (Appendix~\ref{app:math_proof_bridges}), showing how higher adoption thresholds impose stricter conditions on bridge connectivity: not only do they make it harder for any given neighborhood to trigger complex contagion, but they also reduce the likelihood that two separate neighborhoods can reciprocally satisfy these conditions, thereby decreasing the prevalence of symmetric bridges.

\begin{figure}
    \centering
    \includegraphics[width=1\linewidth]{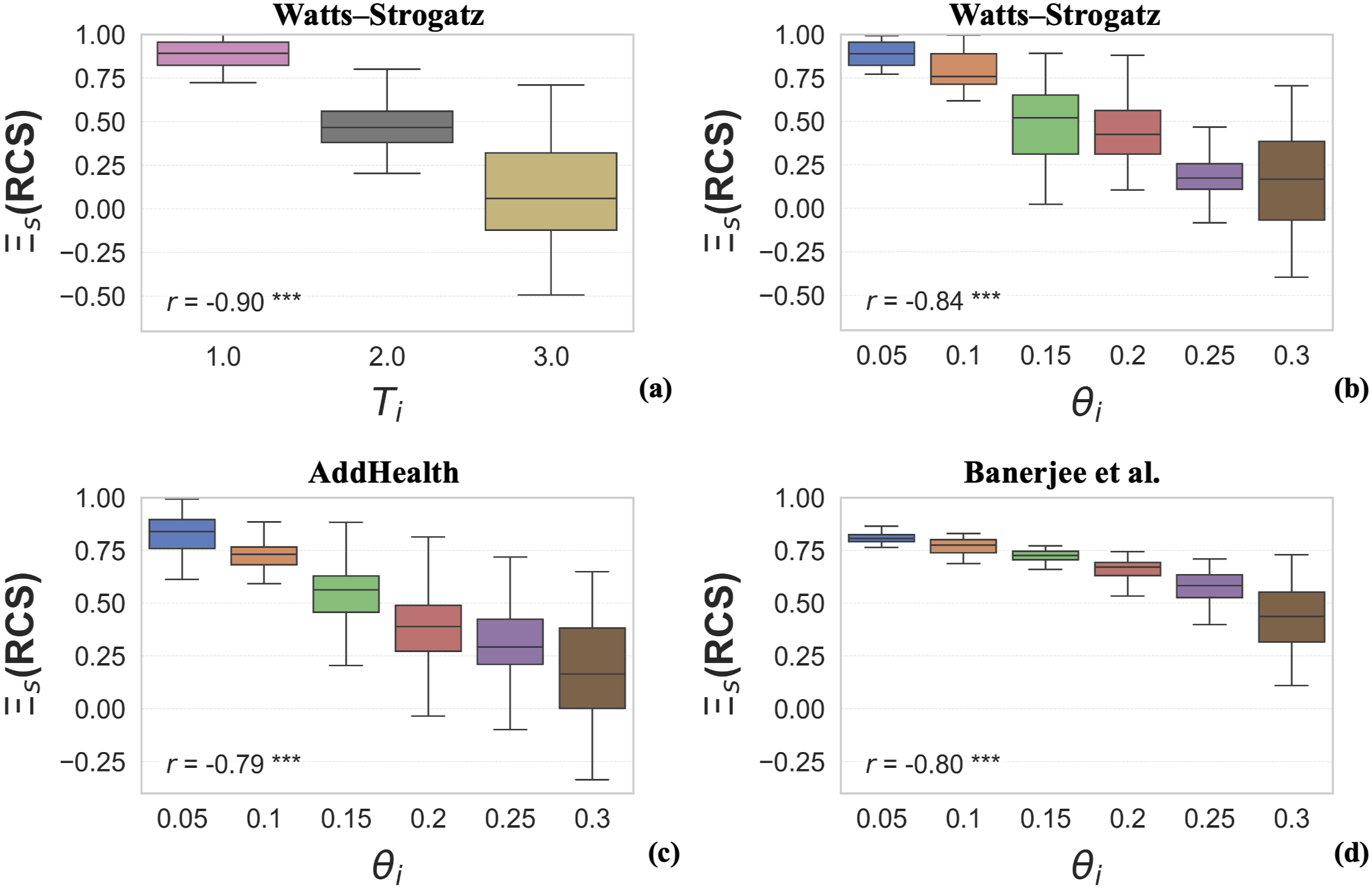}
    \caption{\textbf{Effect of threshold values on symmetry.} The boxplots display the distribution of the symmetry measure $\Xi_s(\text{RCS})$, where higher values indicate more symmetric spreading dynamics. Panel (a) shows results for absolute thresholds $T_i$ on Watts--Strogatz graphs ($n = 200$, $k = 8$), while panels (b)–(d) report results for relative thresholds $\theta_i$ on Watts--Strogatz (b), AddHealth (c), and Banerjee et al. (d) networks. For (a) and (b), 1000 Watts–Strogatz graphs were generated using 100 evenly spaced $\beta$ values in $[0, 1]$, with 10 graphs per $\beta$. For each giant component on each network, we define a simulation scenario based on its specific graph structure and a given threshold value. For each scenario, $100N$ independent and randomly chosen clustered seed sets comprised of $5\%$ of all nodes are generated to compute the symmetry measure, where $N$ denotes the number of nodes in the graph. Across all settings, the boxplots reveal a consistent pattern: increasing the threshold systematically reduces symmetry, indicating a shift toward more directional and asymmetric contagion pathways.}
    \label{fig:panel_declining_symmetry_increasing_threshold}
\end{figure}

In this section, we demonstrate these results statistically by conducting a large range of simulation experiments across both synthetic and empirical social networks. Specifically, we examine the effect of threshold complexity on the rate of asymmetric bridges in three canonical network datasets in contagion research: (i) a large number of Watts-Strogatz graphs generated across a continuum of random rewiring rates ([0,1]), yielding a full range of graphs from regular lattices and small-world networks to fully random graphs; (ii) the Addhealth dataset, which contains the full friendship networks among high school students in 86 distinct schools across the U.S. \cite{harris2011national}; and (iii) the Banerjee et al. (2013) \cite{banerjee2013diffusion} village dataset, which maps social and economic relationships among households in 43 unique rural Indian villages (see section \ref{sec:network_datasets} in the appendix for more details on each network dataset). 

We begin by presenting results across a suite of graphs generated using the Watts-Strogatz (WS) model. This is the approach used to demonstrate the canonical small-world phenomena \cite{watts1998collective}, and has since been adopted as a foundational approach to demonstrating the effects of topology on complex contagions as well \cite{centola2007complex, guilbeault2021topological}. The Watts-Strogatz (WS) model provides a controlled framework for studying small-world network properties because it begins with a highly structured k-regular lattice (i.e., a network in which all nodes have the same number of connections defined by \textit{k}) and gradually introduces randomly rewired weak ties that create shortcuts across the networks while holding the degree distribution constant. Under high rewiring rates, the lattice is converted into a random graph structure. Aggregating across a large range of small-world networks (\textit{k} = 8), we find that contagions with higher thresholds are more likely to flow in an asymmetric, directed fashion, whereby a given neighborhood A is able to spread the contagion to neighborhood B, but not the reverse (see Figure~\ref{fig:panel_declining_symmetry_increasing_threshold}). These results replicate regardless of whether we use absolute (\textit{T}) or fractional ($\theta$) thresholds. Figure~\ref{fig:panel_declining_symmetry_increasing_threshold}(c) and ~\ref{fig:panel_declining_symmetry_increasing_threshold}(d) further replicate these findings on the Addhealth and Banerjee datasets, respectively, each of which exhibit considerable variation in their degree distribution and level of clustering across networks, thereby demonstrating the stability and generalizability of our findings across diverse topologies. All panels in Figure ~\ref{fig:panel_declining_symmetry_increasing_threshold} display the Pearson correlation between threshold and symmetry across all simulations; and in all cases, the correlation is strongly negative and highly significant at the \textit{p} $<$ .0000001 level. These results are highly robust across diffusion models involving complex contagion — namely, the linear threshold model \cite{kempe2003maximizing} (Appendix~\ref{app:LTM}) and the independent cascade model \cite{shakarian2015independent} (Appendix~\ref{app:ICM}) — while also being robust to the inclusion of noisy thresholds that intermix simple and complex contagion dynamics \cite{eckles2024long} (Appendix~\ref{app:noisy}). We note, however, that even for simple contagions (T = 1), it is not possible for our symmetry measure to perfectly reach 1 in empirical networks. This residual asymmetry arises from degree heterogeneity: nodes with higher degree occupy positions that are more frequently traversed in the network from any given seed set, which can bias the overall likelihood of influence flowing more strongly in one direction along a tie. Supporting this, we observe that symmetry levels increase in networks with more homogeneous degree distributions, indicating that structural variance — not the contagion mechanism — is the source of asymmetry under simple contagion (comparing panel A and B to C and D of Figure~\ref{fig:panel_declining_symmetry_increasing_threshold}). The minimal level of asymmetry under simple contagions constitutes the ceiling of how symmetric global spreading dynamics can be for a graph, against which the systematic reduction of symmetry induced by complex contagions can be compared. 

Next, we decompose these networks to examine which ties are most likely to be asymmetric, and thus which ties are most likely to be driving the asymmetric flow patterns that arise as a function of contagion complexity. 

\subsection{Small Worlds, Big Asymmetries}
\label{sec:small_worlds_asymmetries}
\begin{figure}
    \centering
    \includegraphics[width=1\linewidth]{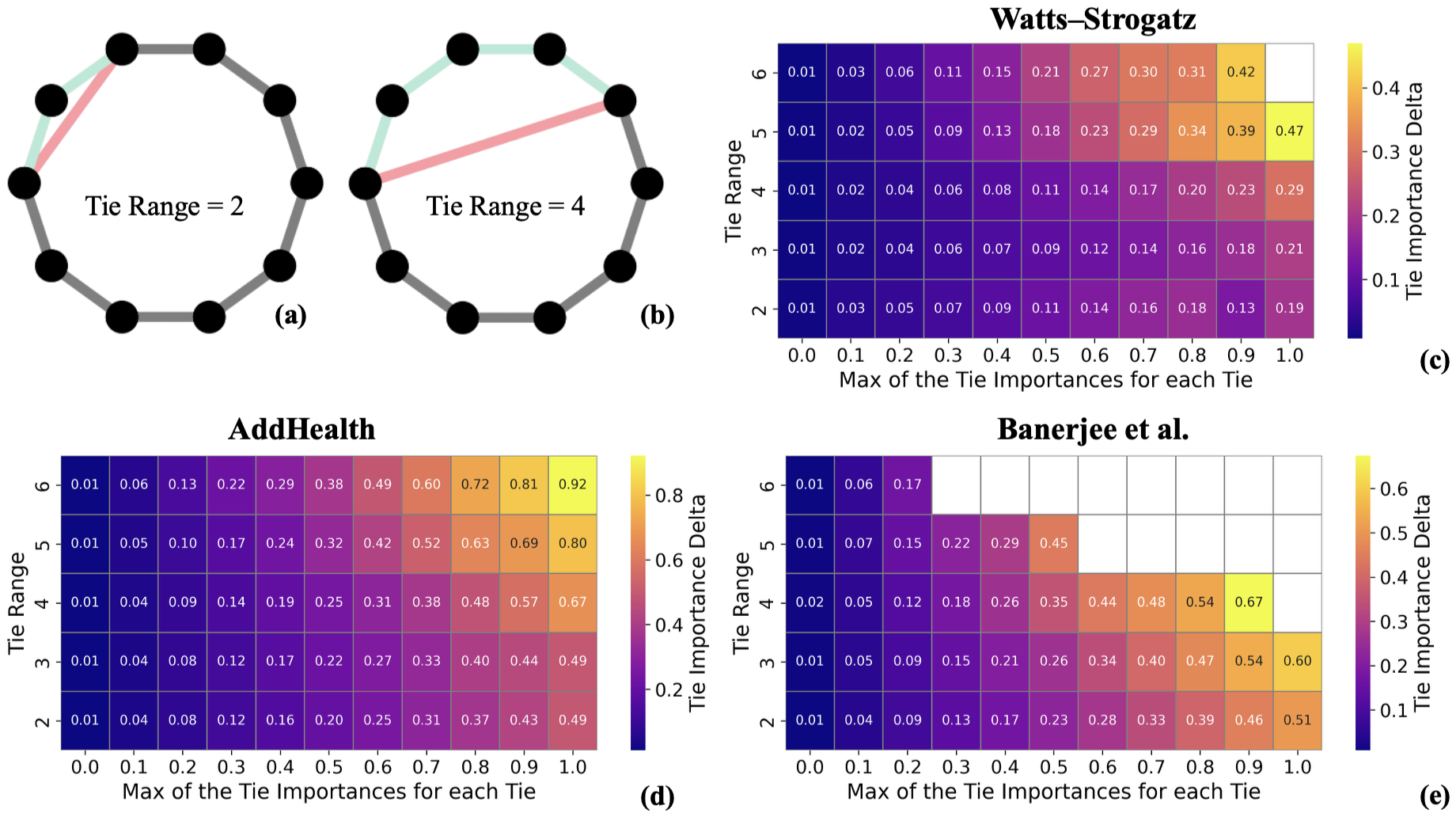}
    \caption{\textbf{Structural Dependencies Between Tie Range and Causal Tie Importance Asymmetry Across Graph Types.} (a--b) Schematic representations of Tie Range, defined as the second-shortest path between two adjacent nodes. In (a), the red tie spans a range of 2, while in (b), it spans a range of 4. (c--e) Heatmaps of Tie Importance asymmetry ($\Delta$) as a function of Tie Range and the maximum importance of each tie, across (c) 1000 generated Watts--Strogatz graphs ($n = 400$, $k = 8$) using 100 evenly spaced $\beta$ values in $[0, 1]$, with 10 graphs per $\beta$, (d) AddHealth, and (e) Banerjee et al.\ networks with threshold $T = 2$. For each giant component on each network, we define a simulation scenario based on its specific graph structure and a given threshold value $\theta_i$. Each scenario was repeated 10 times, $10N$ independent and randomly chosen clustered seed sets comprised of $5\%$ of all nodes are generated to compute the maximum importance of each tie, where $N$ denotes the number of nodes in each graph. The x-axis denotes the maximum CPC value across both directions of each tie. The y-axis shows the Tie Range. The color spectrum indicates the strength of Tie Importance Delta, which captures the level of emergent directedness, with higher deltas (warmer colors) indicating more directedness (i.e., more asymmetric flow favoring one direction along a tie).}
    \label{fig:tie_range_heatmaps}
\end{figure}

To better understand the role played by different ties in complex contagion dynamics, we use the concept of Tie Range—a structural property that captures how embedded a tie is within the local network topology \cite{park2018strength, lyu2022investigating}. The `Range' of a referent tie is defined as the second shortest path between the nodes that this referent tie connects, meaning it measures the length of the shortest alternative route connecting the two nodes without using the referent tie (Figures~\ref{fig:tie_range_heatmaps}(a) and (b)). For example, in Figure~\ref{fig:tie_range_heatmaps}(a), the red tie has a Tie Range of 2, whereas in Figure~\ref{fig:tie_range_heatmaps}(b), it has a Tie Range of 4. In what follows, we examine whether ties with different ranges contribute symmetrically to contagion flow, or whether certain types of weak ties disproportionately favor influence in one direction. This is a direct way to investigate the implications of emergent directedness for the ``strength of weak ties" hypothesis, since Tie Range is an established way to measure weak ties as those which are structurally long ties \cite{centola2007complex, park2018strength}.

For this analysis, we use our counterfactual simulation method to track how specific ties contribute to the spread of a complex contagion. First, we introduce the Causal Flow Symmetry measure ($\Xi_s$), which provides a macroscopic assessment of symmetry in a network’s diffusion process. This measure is computed as the Pearson correlation between the causal importance of edges in both directions across all edges in the network. A high value of $\Xi_s$ close to $1$ indicates that contagion spreads symmetrically through ties, while a low value close to $0$ reveals strong asymmetries in diffusion. Second, we define spreading density, which quantifies the overall reach of the contagion by measuring the fraction of the network that ultimately adopts it.  

As above, we begin by applying our measure to the continuum of graphs generated via the Watts-Strogatz procedure. This setting is ideal for isolating the effects of structurally weak ties and small-world network architecture on emergent directedness. The Watts-Strogatz procedure interpolates between two structural extremes: a regular lattice, where each node is connected to its nearest neighbors, and a random graph, where ties are rewired with some probability $\beta$, introducing weak ties which span otherwise far apart nodes. This transition is crucial for understanding social contagion processes, as real-world networks often exhibit a balance between local cohesion and global reach \cite{holme2002growing}. Weak ties are those that span structural holes in networks and are traditionally thought to facilitate integration by supporting the mutual exchange of cultural content (i.e., social contagions) between distant communities \cite{granovetter1973strength, park2018strength}. However, their role in social contagion remains unclear, as past research has produced contradicting findings on whether weak ties are crucial or irrelevant in the spread of complex contagions \cite{centola2007complex, eckles2024long}. 

Figure~\ref{fig:tie_range_heatmaps}(c) presents a heatmap of Tie Importance Delta across all ties from all WS graphs examined, where the x-axis represents the maximum importance of each tie, and the y-axis denotes Tie Range. The color intensity indicates the degree of asymmetry in contagion flow, revealing that as Tie Range increases, weak ties become increasingly unidirectional in their influence. In other words, ties that span greater structural distances are not neutral conduits but instead channel influence predominantly in one direction. Highly important ties, in particular, exhibit stark asymmetries. Statistical analyses confirm that the longer weak ties are, the more they are asymmetric. Thus, far from fostering cultural integration, the strength of weak ties in complex contagions appears to lay in their ability to give some network regions disproportionate influence in diffusion dynamics over others. Figure~\ref{fig:tie_range_heatmaps}(d) and ~\ref{fig:tie_range_heatmaps}(e) again show that this exact same pattern replicates strongly across the Addhealth and Banerjee et al. village datasets, respectively. 

Supplementary analyses using OLS regression show that these results are robust to a wide range of statistical controls -- including fixed effects for network and dataset -- yielding highly significant correlations between the level of emergent directedness (i.e., the tie importance delta) and (i) tie range (\textit{p} $<$ .0001 in all datasets), (ii) max tie importance (\textit{p} $<$ .0001 in all datasets), and (iii) their interaction (\textit{p} $<$ .0001 in all datasets) (see Table ~\ref{app:regression_no_interaction}~$\&$~\ref{app:regression_with_interaction}). These models account for a meaningful amount of the variance in the extent of emergent directedness at the individual tie level: R\textsuperscript{2} is approximately 0.61, 0.69, and 0.25 when applying these models to the AddHealth, Banerjee et al. village data, and the Watts-Strogatz graphs, respectively (see Table~\ref{app:regression_with_interaction}). 

In additional supplementary analyses, we delve deeper into the mechanics of when emergent asymmetric flow dynamics begin to characterize diffusion in small-world networks generated via the Watts-Strogatz procedure. Decomposing these findings by the level of randomness (the rate of tie rewiring) in WS graphs reveals a striking pattern: as $\beta$ (the rewiring probability) increases, spreading density rises, but symmetry exhibits a distinct decline before recovering at high $\beta$ values (Figure~\ref{fig:rewiring_dip_plot}). This sudden dip in symmetry reveals that the introduction of weak ties does not simply accelerate diffusion; it also induces global asymmetry in the diffusion dynamics. At low $\beta$, contagion spreads symmetrically within clustered neighborhoods, constrained by local reinforcement that is uniform throughout the network. As rewiring increases, a few weak ties favorably fall into place to dramatically enhance spreading, yet these ties overwhelmingly favor one direction over the other, amplifying asymmetries in influence. At high $\beta$, as the network approaches total randomness, symmetry again rebounds, but spreading density collapses due to the erosion of local clustering and wide bridges, which is essential for complex contagions. A goldilocks zone of structural forces is gleaned: when networks are just random enough, spreading is maximized but influence is unequally distributed, i.e., directed; but when networks become too random, influence becomes increasingly equalized (symmetric) across nodes, but the overall spreading of the contagion declines. Consistent with the claim that small-world networks -- those with intermediate randomness -- capture consistent features of empirical social networks, our analyses of complex diffusion in the Addhealth and Banerjee et al. suggest that their balance of local clustering and global connectivity is well-tuned to promote the spread of complex contagions, but in a highly directed fashion. 

\subsection{The Nonlinear Impact of Tie Strength}
Here we show that emergent directedness can shed light on a pressing puzzle concerning the effects of weak ties on job diffusion over LinkedIn, as reported by a recent empirical study examining 20 million people and over 2 billion network connections from the platform ~\cite{rajkumar2022causal}. While this study claims to provide large-scale causal evidence in support of the ``strength of weak" ties hypothesis, its main results highlight a striking and unexpected deviation from the canonical strength of weak ties theory. Rather than observing a negative linear relationship between tie strength and job diffusion (whereby the weakest ties are most impactful), they observe an inverted u-shape relationship between tie strength and job diffusion impact; that is, they find that ``moderately weak" ties (or, alternatively, moderately strong ties) are most impactful for triggering the diffusion of novel job opportunities. As the authors explain, this heterogeneity is not accounted for by previous theories, neither by the strength of weak ties nor by complex contagion theory (which implies that stronger, more locally embedded ties should most consistently facilitate job transmission). Rajkumar et al. (2022) document this empirical deviation from prior theory, but leave the problem of explaining this deviation to future work. In what follows, we show how capturing emergent directedness in the causal impact of ties on spreading complex contagions recovers a similar inverted u-shape distribution that resembles the empirical patterns observed by Rajkumar et al. (2022), potentially providing a novel approach to explaining these surprising findings. 

\begin{figure}[H]
    \centering
    \includegraphics[width=1\linewidth]{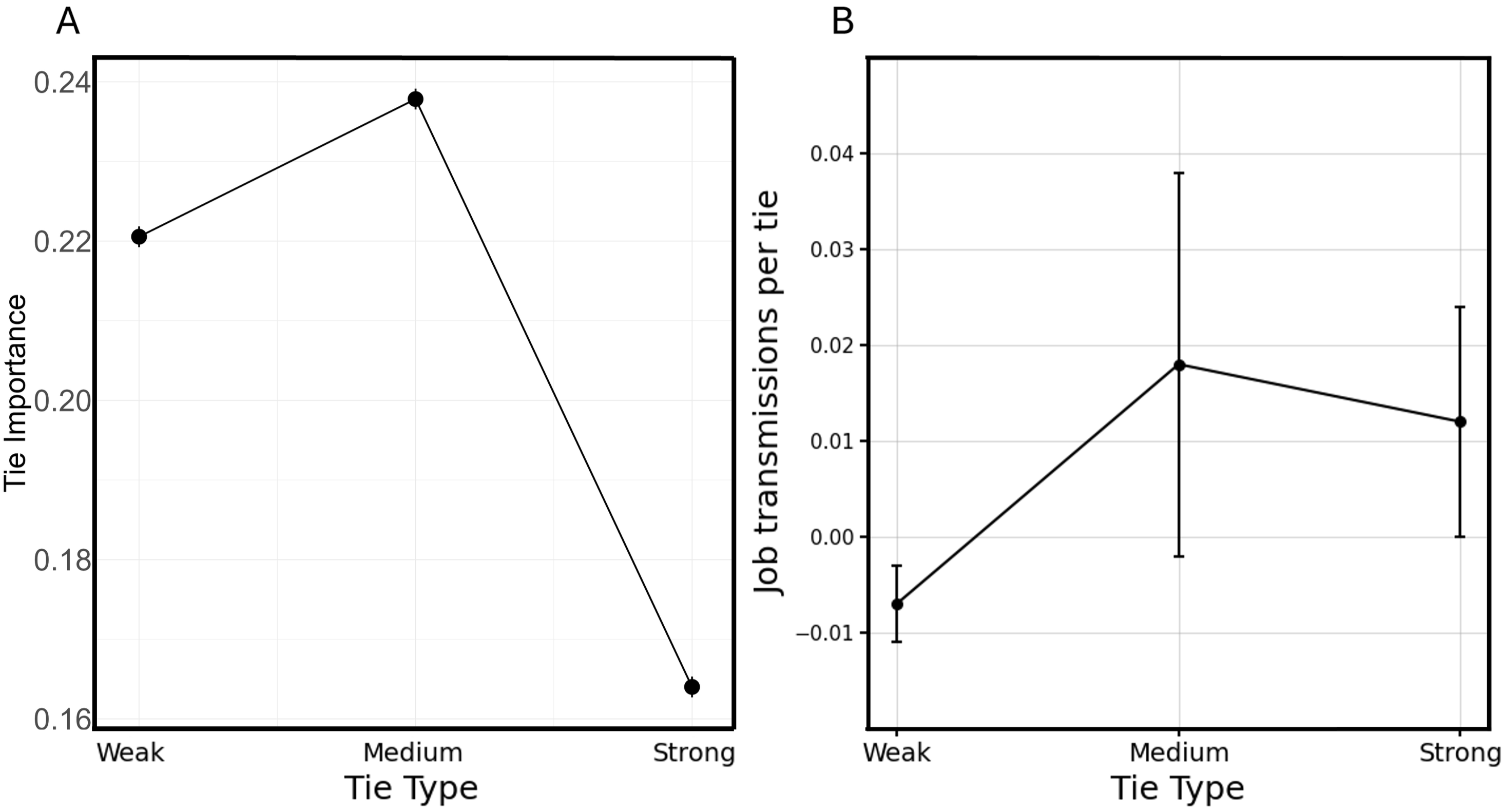}
    \caption{\textbf{Tie importance in complex contagion spreading and job mobility.} (A) Mean tie importances for weak, medium, and strong ties under a complex contagion model across three network datasets: the Addhealth and Banerjee et al. (2012) empirical networks, and scale-free network with tunable clustering \cite{holme2002growing} ($n=1000$) generated for degrees $k \in {2, 3, 4, 5}$ and clustering probabilities $p$ linearly spaced in $[0, 1]$, matching the number of AddHealth graph instances. For each giant component on each network, we define a simulation scenario based on its specific graph structure and a given threshold value $\theta_i \in [0.1, 0.15, 0.2, 0.25, 0.3, 0.35]$. For each scenario, $5N$ independent and randomly chosen clustered seed sets comprised of $2\%$ of all nodes are generated to compute the Tie Importance values of each tie, where $N$ denotes the number of nodes in the graph. Data points show the partial effect of tie type estimated by an OLS regression that includes fixed effects by dataset, network, and threshold. (B) Empirical LinkedIn job mobility data showing the effect of weak, medium, and strong ties (defined by mutual friend terciles) on job transmissions, reproduced from \cite{rajkumar2022causal} based on values extracted from the original figure, for visual comparison. Both results, using the same tie strength definition, highlight the disproportionate impact of medium-strength ties in diffusion. Error bars indicate 95\% confidence intervals.}
    \label{fig:panel_inverse_V}
\end{figure}

We explain this nonlinearity as follows. Our approach rests on the intuition that job diffusion is likely a complex contagion (at least minimally), because whether or not one finds a job opportunity relevant and worth applying to involves a certain amount of trust and credibility in the source of this recommendation, as well as perceived legitimacy of the target organization, and both credibility and legitimacy are key drivers of complex contagions \cite{centola2007complex, guilbeault2018complex}. Following this logic, the intuition holds (bolstered by prior empirical work \cite{guilbeault2021topological}) that very weak ties, characterized by low neighborhood overlap, often fail to meet reinforcement conditions, making them ineffective in facilitating contagion spread. In contrast, moderately strong ties provide just enough local reinforcement to sustain complex contagion while still acting as bridges between otherwise disconnected communities, enabling them to facilitate exposure to novel and relevant information \cite{aral2011diversity}. However, very strong ties, with high neighborhood overlap, are constrained by excessive clustering, which limits exposure to active neighbors outside their local communities, reducing the likelihood of exposure facilitating access to novel information. Additionally, this high clustering creates redundancies, making most very strong ties inefficient and therefore less causally important individually.

This nonlinearity in causal tie importance for the spread of complex contagion spread is visualized in Figure~\ref{fig:panel_inverse_V}, where we show that for synthetic power-law networks, the AddHealth dataset, and the Banerjee Village networks, tie importance peaks at an intermediate level of tie strength, forming an inverse U-shaped curve. The data points in this figure show the partial effect of tie type estimated by an OLS regression that includes fixed effects by dataset, network, and threshold (T in [0.1, 0.15, 0.2, 0.25, 0.3]). The original figure displaying this inverted-u effect in the empirical LinkedIn dataset from Rajkumar et al. (2022) is copied in panel B for visual comparison. According to our measures, we observe the same nonlinear causal impact of weak ties in networks of very different sizes and origin compared to the LinkedIn setting, suggesting that this inverted u-shape in causal tie impact may be a generalizable feature of tie impact in the context of complex contagions. 

These findings illustrate how accounting for emergent directedness in complex contagions can help explain puzzling nonlinearities in the causal impact of ties. An interesting question arises when connecting these findings with our results indicating that increasingly complex contagions (those with higher thresholds) are increasingly directed in nature -- namely, how does varying the complexity of a contagion impact which nodes and ties are most impactful for shaping global influence dynamics? In the following section, we highlight one particularly surprising yet clear pattern that contradicts standard notions of centrality as concerns its role in measuring influence: as the level of local reinforcement ($T$) required for a contagion increases, global influence dynamics realign by giving more power to the periphery over the core. 

\subsection{The Power of the Periphery}
\begin{figure}[H]
    \centering
    \includegraphics[width=1\linewidth]{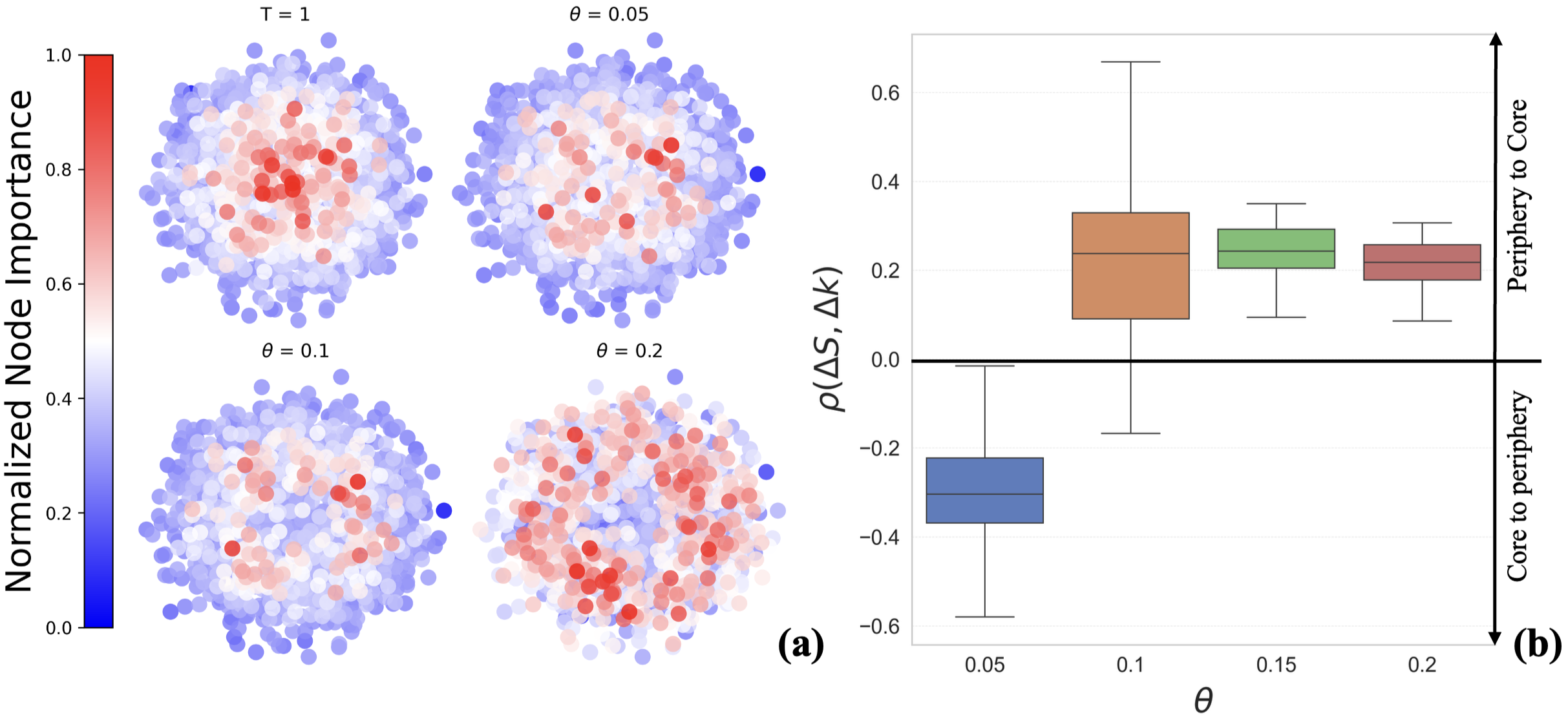}
    \caption{\textbf{Global Realignments of Node Importance and Influence Flow Induced by Complex Contagions.} (a) Visualization of the shift in degree-normalized node importance from the network core to the periphery as the spreading mechanism transitions from simple ($T=1$; T is used in place of $\theta$ for absolute thresholds) to complex contagions with increasing relative thresholds ($\theta = 0.05, 0.1, 0.2$) on a scale-free network with tunable clustering ($n = 1000$, $m = 4$, $p = 0.4$) \cite{holme2002growing}. Node color indicates degree-normalized importance, with red representing high importance and blue representing low importance. As $\theta$ increases, influence moves away from central hubs toward peripheral nodes. (b) Boxplot of the correlation $\rho(\Delta S, \Delta k)$ on the AddHealth network, quantifying how tie importance asymmetry ($\Delta S$) relates to degree differences ($\Delta k$) across varying $\theta$. Negative correlations reflect core-to-periphery spreading, while positive values indicate a reversal toward periphery-to-core flow. For each threshold configuration and graph (including panel (a) and each graph in the AddHealth dataset for (b)), $100N$ independent clustered seed sets covering $5\%$ of nodes were used. Results for panel (b) include only simulation scenarios achieving full network activation.}
    \label{fig:panel_periphery_core_shift}
\end{figure}

Our results demonstrate that asymmetry in contagion flow significantly impacts global influence dynamics, with important implications for standard network-based centrality measures that typically assume symmetric ties and diffusion patterns. Standard network theory often posits that central nodes (high-degree hubs) are the most influential in diffusion processes. However, our findings indicate that for complex contagions governed by relative thresholds, influence does not radiate outward from network cores. Instead, as the need for reinforcement increases, contagions flow inward from the periphery to the core. This inversion of influence is illustrated in Figure~\ref{fig:panel_periphery_core_shift}(a), showing that the degree-normalized Causal Node Importance (NI) moves from core nodes to peripheral ones with increasing relative thresholds ($\theta$) on a scale-free network with tunable clustering ($n=1000$, $m=4$, p=$0.4$) \cite{holme2002growing}. We use this network topology as a demonstration because it is designed to yield the kind degree heterogeneity that classically captures the qualitative distinction between the high-degree core and the low-degree but tight-knit periphery. This core-to-periphery shift does not occur under absolute thresholds ($T$). This qualitative impact of threshold type aligns with the insight that relative thresholds capture the salience of non-adopters, who exert countervailing influences against adoption \cite{centola2018behavior}. Consequently, central individuals in highly connected positions face disproportionately higher ``risk" in adopting behaviors that require reinforcement, thereby diminishing their influence. Our results show how the structural impact of these risk dynamics has immediate consequences for emergent directedness by systematically reducing the causal impact of seemingly central nodes, while empowering peripheral regions in diffusion processes. This reversal in global influence direction is quantified by the correlation $\rho(\Delta S, \Delta k)$, which measures the relationship between tie-importance asymmetry ($\Delta S$) and degree differences ($\Delta k$) between connected nodes. Panel (b) of Figure~\ref{fig:panel_periphery_core_shift} demonstrates this analysis across thresholds for the Addhealth dataset. When $\rho < 0$, influence flows predominantly via core-to-periphery transmissions; when $\rho > 0$, peripheral regions become the primary diffusion drivers. The outsized power of the periphery dissipates as thresholds become high ($\theta \geq 0.2$) largely because spreading is much harder and overall diminished once thresholds require this level of reinforcement in the Addhealth network. Comprehensive results for absolute thresholds and alternative seeding conditions are presented in the appendix \ref{app:periphery_shift}.

The explanation for this reversal extends beyond the effect of relative thresholds alone and includes the structural organization of peripheral regions—specifically, the presence of wide bridges, chains of reinforcing ties crucial for sustaining diffusion. Nodes in peripheral regions typically have fewer connections compared to those at the network core, yet these connections often exhibit significant overlap and local clustering. Such decentralized clustering strongly reinforces diffusion locally, facilitating an inward shift of influence from the periphery toward the core. This structural effect becomes increasingly important as higher relative thresholds require greater neighborhood overlap to maintain contagion momentum. The supportive role of peripheral clustering is illustrated by the high importance of ties with low tie ranges in Figure~\ref{fig:lineplot_mean_cpc_tie_range} in the appendix. By contrast, ties connecting nodes separated by greater social distances (high tie range)—often characteristic of core-to-periphery connections—exhibit notably lower tie importance. Consequently, complex contagions preferentially propagate via the clustered, low-range ties typical of peripheral structures. Together, relative threshold dynamics and peripheral clustering jointly drive this inversion of influence, challenging the conventional assumption that network hubs constitute the most influential nodes in diffusion processes. Under complex contagion dynamics governed by relative thresholds, peripheral nodes gain disproportionate influence because central nodes become more resistant to influence and typically lack the highly reinforcing structures found in peripheral regions—an effect that standard centrality measures fail to capture. These findings carry significant theoretical and practical implications for understanding cultural diffusion processes, especially regarding the adoption of ideas, behaviors, and innovations that depend on peer reinforcement. In particular, our results suggest that peripheral actors may exert an outsized influence on cultural innovation and evolution, consistent with recent observational studies indicating that peripheral individuals disproportionately originate and diffuse innovative ideas \cite{vicinanza2023deep, shi2023surprising, ma2025identifying}. A full exploration of the practical implications of these findings is beyond the scope of this paper; but in the interest of demonstrating an application of broad relevance across a range contexts, we use our measures to provide insight into the directed nature of bridges that endogenously form as networks evolve, along with structural and cultural constraints that can increase the probability of symmetrical, integrative bridges. 

\subsection{Bridge Formation: Asymmetric by Nature, Symmetric by Design}
Networks often evolve as new connections form between previously disconnected actors and network regions, spawning new network bridges when sufficiently concentrated. In the wild, bridges in social networks form through a mixture of random and non-random processes. In terms of random processes, new connections often form through spontaneous encounters between people; it is random processes of this kind that inspired the ``small-world" network terminology, in reference to the common expression ``what a small world" which people use to mark the fortuitous discovery that two people from seemingly distant social worlds are connected \cite{watts1998collective}. Importantly, tie formation is also constrained by various non-random selection processes, such as pressures toward triadic closure (e.g., for ties to form among mutual contacts such as friends of friends) \cite{romero2010directed, mosleh2025tendencies}. One driver of triadic closure is homophily (a preference to form connections with like others) \cite{kossinets2009origins}; however, homophily often leads to reinforced connections within rather than between network communities, given the increased similarity between  members of their own community. Triadic closure can also be shaped by cultural norms, such as the preference in collectivist cultures to form ties within rather than between communities (e.g., via friendships and marriages) \cite{gelfand2024norm}. Another key driver of triadic closure is consolidation (i.e., the increased probability to connect with proximate others you frequently encounter in overlapping social contexts) \cite{centola2015social}. The latter identifies the potential for organizational strategies that curate social contexts that promote interactions between members of disconnected communities \cite{green2025consonance}. Yet, empirical evidence suggests that strategic initiatives within organizations often promote tie formation through quasi-random processes, such as spontaneous encounters at informal social functions \cite{gray2010building, landis2016personality, sevcenko2024office}. This suggests that even in highly structured social environments, both random and non-random processes of tie formation govern network evolution. How bridges form through random and non-random processes -- and whether these bridges promote influence inequalities or mutual integration across communities -- is of direct interest to organizational strategy and cultural evolution more broadly. 

In what follows, we simulate the endogenous formation of functional network bridges (i.e., those that can spread complex contagions) while testing whether systematically varying the probability of triadic closure in tie formation can alter the probability of creating symmetric bridges. As a baseline, we simulate the formation of bridges through purely random sampling. Then, we compare this random process of bridge formation to quasi-random processes with varying levels of simulated constraints promoting triadic closure. These simulated constraints can be imagined as endogenous cultural norms that increase the probability of triadic closure, or outside strategic interventions that engineer consolidation by facilitating repeated interactions between agents in a constrained fashion that promote triadic closure \cite{green2025consonance}. This provides a thought experiment for whether structural and cultural interventions can tip the scales from asymmetric to symmetric bridge formation by increasing triadic closure in the emergence of new ties. All newly added ties in these simulations are undirected at the dyadic level; any directedness in the resulting flow dynamics is emergent at the macro level according to our definition. 

\begin{figure}[H]
    \centering
    \includegraphics[width=1\textwidth]{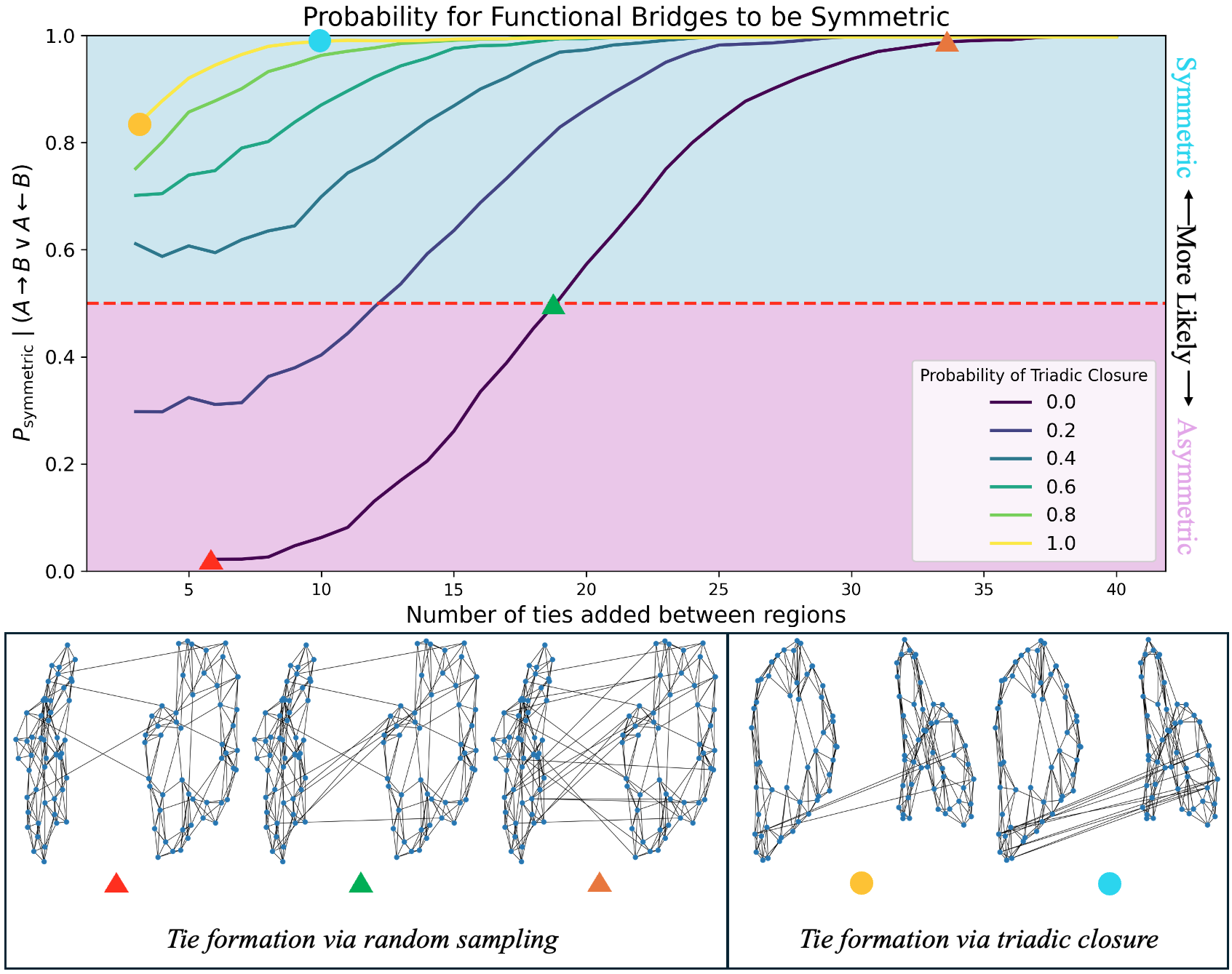}
    \caption{The probability of a \textbf{Functional Bridge}—a connection capable of spreading a complex contagion—being symmetric, as a function of the number of ties added between two previously disconnected networks (Watts–Strogatz graphs with $n=100$, mean degree $k=6$, and rewiring probability $\beta=0.1$). The shaded regions indicate the likelihood of symmetry: pink represents a higher chance of asymmetric bridges, while blue indicates a transition to predominantly symmetric bridges. Contours represent different probabilities \textit{c} of adding ties of forming \textbf{triadic closure} (where otherwise ties are added randomly at probability 1 - \textit{c}), showing how the probability of symmetrical bridges emerging as a result of shifting from random to structural constrained tie formation. Higher levels of triadic closure (yellow/green) promote symmetry earlier, while lower levels (dark blue) require more ties before symmetry becomes predominant. The lines in each case represent the average over 1000 independent simulation runs.}
    \label{fig:symmetry_progression}
\end{figure}

We observe that, when formed randomly, endogenous bridges are much more likely to be directed by nature, promoting the spread of culture from one group to the other, rather than integration. Figure~\ref{fig:symmetry_progression} illustrates this effect by plotting the probability of forming a symmetric bridge as a function of the number of ties added between two previously disconnected Watts-Strogatz graphs. Initially, most bridges remain asymmetric, meaning that influence is significantly more likely to flow from one side to another rather than bidirectionally. The underlying mechanism driving this effect is the reinforcement requirement inherent to complex contagions: for a bridge to enable spreading, the receiving node must be exposed to multiple active neighbors. As a result, bridges formed randomly will often only satisfy this condition in one direction, leading to structural asymmetry in influence flow. Beyond simulations, this fact can be proven mathematically from the foundations of probability theory, given that the set of possible asymmetric bridges is much larger than the set of symmetric bridges (Supplementary appendix \ref{app:math_proof_bridges}). 

Figure~\ref{fig:symmetry_progression} further shows that symmetric, integrative flow patterns only emerge between communities when very large bridges form through quasi-random endogenous processes. Only after crossing a critical density of ties does symmetry become probable. This suggests that endogenous random bridge formation, which tends to be sparse and based on the emergence of only a few ties, is unlikely to yield balanced, symmetrical influence dynamics of the kind that facilitates cultural integration; instead, random endogenous bridge formation is likely to trigger the spread of complex contagions from one group to the other. Our simulations further demonstrate that the threshold conditions of a contagion play a crucial role in determining the likelihood of symmetric bridges. When adoption requires high reinforcement (i.e., multiple exposures), the probability that a randomly formed bridge will support bidirectional spreading declines sharply, suggesting that the challenges for strategically building symmetric bridges are higher. 

We then examine the effects of increasing triadic closure. For this, we include an additional parameter that tunes the formation of new ties by a variable \textit{c}, which defines the probability (from 0 to 1) that a newly added tie satisfies triadic closure by closing a triangle among already connected ties. Figure~\ref{fig:symmetry_progression} shows that when the pressures toward triadic closure are weak (e.g., \textit{c} = 0.2), this is not enough to stem the tide of emergent directedness in endogenous bridge formation. Even when pressures toward triadic closure are moderately high, a considerable number of asymmetric bridges prevail (e.g., when \textit{c} = 0.4, bridges are only slightly more likely than chance to be symmetric). It is only when pressures toward triadic closure are very high (\textit{c} $>$ 0.6) that symmetrical bridges stably form. These findings show that in order to form symmetric bridges, homophily or consolidation-based factors of tie formation have to be very strong to facilitate mutual influence in the flow of complex contagions between communities. 

Given that the bridges we observe in empirical networks are much more likely to be asymmetric, this suggests that triadic closure in real-world network evolution is not strong enough to ensure symmetrical bridges reliably emerge between network communities. Recent experimental findings on tie formation in the wild confirm this interpretation. In examining new connection formation over Twitter, Mosleh et al. (2025) observe that the rate of triadic closure is approximately 12\%, and a targeted experimental manipulation that highlights the salience of an open triad to social media users was only able to raise the rate of triadic closure to just over 20\%. Our measure of the rate of asymmetric bridges in empirical social networks can be viewed as an indicator that random processes likely drive to endogenous bridge formation, consistent with the canonical argument that small-world networks, which are driven by randomly forming weak ties, are strikingly common in biological and social systems. Whether this differs across social contexts -- for example, as a function of the tightness or looseness of norms across cultures -- is a compelling topic for future research \cite{gelfand2024norm, burt2024guanxi}. Implications for policy and strategic network interventions are considered in the discussion section. 

\section{Discussion}
In this study, we show that as soon as contagions require even minimal reinforcement from multiple peers to spread, directed ties and asymmetric flow dynamics emerge within networks composed of symmetric, undirected ties, with no individual-level differences among nodes. This directedness arises through the interaction of network topology and contagion complexity, and is not reducible to either in isolation. Emergent directedness thus appears to be inescapable at a fundamental level in modeling the spread of complex contagions. This has important implications for social science research which aims to examine patterns of cultural diffusion characterized by complex contagions -- such as the spread of norms, categories, attitudes, practices, and technologies -- while using static and often undirected representations of network structure. Our findings suggest that static undirected representations of networks are insufficient for capturing the evolutionary dynamics of cultural diffusion as a complex system. Instead, modeling differential spreadability of complex contagions among nodes requires capturing emergent patterns of directed flow that are shaped by the amount of reinforcement required for the contagion across all nodes, and this information is not directly encoded into the underlying static and undirected network representation. This raises the interesting possibility that more dynamic directed models of network structure, involving the interaction between contagion complexity and underlying social structure, are closer to the essence of how culture spreads and evolves as a complex system. 

These insights shed new light on canonical theories of how weak ties and small-world networks contribute to cultural diffusion. Traditionally, weak ties and small-world networks are viewed as accelerating mutual influence between network regions, facilitating an optimal balance of local segregation and global integration \cite{granovetter1973strength, sporns2006small}. Yet, this theory breaks down once contagions require even minimal local reinforcement from multiple peers. We find that when weak ties are able to spread complex contagions, they preferentially do so in one direction, leading to global influence inequalities with some network regions systematically enacting more influence over others. This does not refute the `strength' of weak ties \textit{per se}, but instead raises the question of what kind of strength they possess. Contrary to facilitating integration across network communities, the strength of weak ties may lie in their ability to expand the global reach of particular local cultures that are (potentially unwittingly) advantageously positioned within the overall population. 

On the one hand, these findings highlight potential challenges for policy makers seeking to facilitate cultural integration across social groups. Our findings suggest that when weak ties form through quasi-random processes -- as is characteristic of small-world networks -- the resulting bridges are most likely to be asymmetric. This provides a valuable cautionary consideration for policy makers and organizations that seek to build network bridges. For example, organizations often seek to build bridges between departments and teams through social functions that randomly pair employees to facilitate new connections \cite{gray2010building, landis2016personality} or through random events such as spontaneous ride-sharing pairings to company events and conferences \cite{sevcenko2024office}. Quasi-random processes of bridge building also unfold in high stakes strategic contexts. For instance, a large body of work based on contact theory attempts to reduce inter-group conflict by facilitating as much interaction as possible between members of conflicting groups in the hopes of promoting shared understanding and cultural integration \cite{pettigrew2006meta, zhou2019extended}. Such strategies frequently yield \textit{ad hoc} network bridges between communities that frequently backfire, exacerbating conflict \cite{tausch2010intergroup, al2013intergroup, bail2018exposure}. Our insights suggest that insofar as quasi-random processes are leveraged to form network bridges, these bridges are likely to give some groups more influence over others in the spread of culture, which may heighten tensions and contribute to exacerbating rather than mitigating conflict. If policy makers seek to maximize cultural integration via network bridges, then strategic interventions such as promoting triadic closure in bridge formation are required. In all cases, our measures for capturing the directed pathways that emerge as a function of bridge formation are helpful for guiding strategic interventions. 

That said, there are many scenarios where asymmetrical flows are not an obstacle, but rather a wave that strategic interventions can benefit from riding. Such scenarios include cases where organizations seek to enable knowledge transfer from one community to another \cite{podolny1996networks, argote2000knowledge}, as in the case of mergers \cite{weber2003cultural} or in foreign policy. Relatedly, contagion-based policy interventions in broader populations are often aimed at particular target populations. For example, anti-smoking initiatives are aimed at smokers \cite{guilbeault2020networked}; many non-smokers who have never smoked in their adult life are at very low risk of taking up smoking, and so are likely inefficient to target when seeding such contagions. Our measures hold promise for informing seeding strategies that are engineered to trigger spreading patterns that target particular groups of interest, while potentially minimizing spill over effects of contagion initiatives on non-target groups. The implications of these measures for understanding competitive ecologies among contagions is rich terrain for future research. 

Beyond social contagions, our finding that simple local constraints on diffusion can induce emergent directed behavior in collective dynamics may have significant implications for the study of collective behavior in complex systems more broadly. Recent work reveals complex contagion dynamics in weighted belief networks \cite{aiyappa2024emergence, goldberg2018beyond}, raising a fascinating topic for future research: does emergent directedness characterize the activation of semantic associations in human cognition and artificial neural networks? Relatedly, biological research observes complex contagion dynamics in animal group behavior -- such as fish schooling and insect swarms \cite{rosenthal2015revealing, poel2022subcritical} -- and even in the collective behavior of cancer cells \cite{deisboeck2009collective}. An open question in this area is how directedness in collective behavior emerges in these biological systems, including the emergence of directed navigation patterns in insect swarms, as well as the critical transition from random interactions among cancer cells to the sudden emergence of metastasis, whereby cancer cells start moving collectively through the body in a directed fashion \cite{deisboeck2009collective}. To the extent these systems consist of networked interactions, our methods may help identify the underlying mechanisms producing the emergence of directed contagion dynamics in collective behavior. One particularly promising direction for future exploration is the finding that animal groups harness peripheral sensing for some tasks (such as fish schools detecting predators) while relying on more centralizing coordination by leaders for others \cite{berdahl2013emergent, hein2015evolution}. Our methods for identifying core-periphery dynamics in emergent directedness are poised to advance future research investigating how collective behavior of various kinds may oscillate between directed regimes (e.g., between core-to-periphery diffusion and periphery-to-core diffusion) as a way of enabling adaptive social learning \cite{almaatouq2020adaptive}. 

More generally, we anticipate that these future inquiries will be enriched by integrating our findings with recent developments in hypergraph network modeling and the study of simplicial contagions \cite{iacopini2019simplicial, de2020social}. While complex paths bear similarity to hyperpaths, fully explicating the connections between these frameworks is a ripe frontier for future exploration. To date, many hypergraph approaches make similar assumptions regarding the undirected nature of pairwise network connections prior to their integration into macro hyperedges for the purposes of modeling contagion dynamics \cite{iacopini2019simplicial, de2020social}. Several studies have investigated the mathematical structure of directed hypergraphs, though how these structures relate to contagion dynamics is a novel and underexplored topic, and the few recent studies examining this connection assume directed ties by design rather than as an emergent property \cite{ausiello2017directed,li2024enhancing}. A benefit of our focus on complex contagions is that it highlights the foundational role of emergent directedness in canonical models of diffusion dynamics that remain in popular use throughout sociology, psychology, and biology. A key implication of our findings is that strikingly simple local constraints (such as a minimal complex contagion threshold in agent behavior, whether deterministic or probabilistic) may be sufficient for generating emergent higher-order network dynamics, suggesting that the presence of complex hypergraph structures may not be inconsistent with simple, formal threshold mechanisms governing agent behavior at the local pairwise level. A promising call for future research is to explore how hypergraphs can enrich the study of emergent directedness in complex contagion, and vice versa. 

\section{Code Availability}
All code for replicating the methods and results of this paper is publicly available at the following github: 
\href{https://github.com/ftschofenig/EmergentDirectedness}{https://github.com/ftschofenig/EmergentDirectedness}.

\section{Data Availability}
All data for replicating the results of this paper is publicly available at the following github: \href{https://github.com/ftschofenig/EmergentDirectedness}{https://github.com/ftschofenig/EmergentDirectedness}.

\section{Acknowledgements}
D.G. gratefully acknowledges support from the Business, Government \& Society Initiative and from the Spencer Faculty Scholar program at the Stanford Graduate School of Business. The authors thank participants in Stanford's macro OB seminar, as well as Damon Centola, Fred Feinberg, Amir Goldberg, Ethan Nadler, Maria Nareklishvili, Patrick Park, Paul Reginato, Hayagreeva Rao, Ruggiero Lo Sardo, Sameer B. Srivastava, and Yian Yin for helpful comments on early drafts of this paper.

\newpage

\bibliographystyle{unsrt}
\bibliography{references} 

\begin{thebibliography}{10}

\bibitem{centola2007complex}
Damon Centola and Michael Macy.
\newblock Complex contagions and the weakness of long ties.
\newblock {\em American journal of Sociology}, 113(3):702--734, 2007.

\bibitem{guilbeault2018complex}
Douglas Guilbeault, Joshua Becker, and Damon Centola.
\newblock Complex contagions: A decade in review.
\newblock {\em Complex spreading phenomena in social systems: Influence and contagion in real-world social networks}, pages 3--25, 2018.

\bibitem{centola2005emperor}
Damon Centola, Robb Willer, and Michael Macy.
\newblock The emperor’s dilemma: A computational model of self-enforcing norms.
\newblock {\em American Journal of Sociology}, 110(4):1009--1040, 2005.

\bibitem{monsted2017evidence}
Bjarke M{\o}nsted, Piotr Sapie{\.z}y{\'n}ski, Emilio Ferrara, and Sune Lehmann.
\newblock Evidence of complex contagion of information in social media: An experiment using twitter bots.
\newblock {\em PloS one}, 12(9):e0184148, 2017.

\bibitem{puglisi2008cultural}
Andrea Puglisi, Andrea Baronchelli, and Vittorio Loreto.
\newblock Cultural route to the emergence of linguistic categories.
\newblock {\em Proceedings of the National Academy of Sciences}, 105(23):7936--7940, 2008.

\bibitem{guilbeault2021experimental}
Douglas Guilbeault, Andrea Baronchelli, and Damon Centola.
\newblock Experimental evidence for scale-induced category convergence across populations.
\newblock {\em Nature communications}, 12(1):327, 2021.

\bibitem{karsai2014complex}
M{\'a}rton Karsai, Gerardo Iniguez, Kimmo Kaski, and J{\'a}nos Kert{\'e}sz.
\newblock Complex contagion process in spreading of online innovation.
\newblock {\em Journal of The Royal Society Interface}, 11(101):20140694, 2014.

\bibitem{tornberg2018echo}
Petter T{\"o}rnberg.
\newblock Echo chambers and viral misinformation: Modeling fake news as complex contagion.
\newblock {\em PLoS one}, 13(9):e0203958, 2018.

\bibitem{centola2010spread}
Damon Centola.
\newblock The spread of behavior in an online social network experiment.
\newblock {\em science}, 329(5996):1194--1197, 2010.

\bibitem{traag2016complex}
Vincent~A Traag.
\newblock Complex contagion of campaign donations.
\newblock {\em PloS one}, 11(4):e0153539, 2016.

\bibitem{pinus2025emotion}
Michael Pinus, Yajun Cao, Eran Halperin, Alin Coman, James~J Gross, and Amit Goldenberg.
\newblock Emotion regulation contagion drives reduction in negative intergroup emotions.
\newblock {\em Nature communications}, 16(1):1387, 2025.

\bibitem{romero2011differences}
Daniel~M Romero, Brendan Meeder, and Jon Kleinberg.
\newblock Differences in the mechanics of information diffusion across topics: idioms, political hashtags, and complex contagion on twitter.
\newblock In {\em Proceedings of the 20th international conference on World wide web}, pages 695--704, 2011.

\bibitem{state2015diffusion}
Bogdan State and Lada Adamic.
\newblock The diffusion of support in an online social movement: Evidence from the adoption of equal-sign profile pictures.
\newblock In {\em Proceedings of the 18th ACM Conference on Computer Supported Cooperative Work \& Social Computing}, pages 1741--1750, 2015.

\bibitem{saetre2025protests}
Juliette Saetre.
\newblock How protests spread: Diasporas, wide bridges, and the transnational diffusion of un violador en tu camino.
\newblock {\em American Journal of Sociology}, 2025.

\bibitem{chaitanya2022hardness}
Meher Chaitanya and Ulrik Brandes.
\newblock Hardness results for seeding complex contagion with neighborhoods.
\newblock In {\em Complex Networks \& Their Applications X: Volume 2, Proceedings of the Tenth International Conference on Complex Networks and Their Applications COMPLEX NETWORKS 2021 10}, pages 207--216. Springer, 2022.

\bibitem{wolfram2003new}
Stephen Wolfram and M~Gad-el Hak.
\newblock A new kind of science.
\newblock {\em Appl. Mech. Rev.}, 56(2):B18--B19, 2003.

\bibitem{epstein2012generative}
Joshua~M Epstein.
\newblock Generative social science: Studies in agent-based computational modeling.
\newblock In {\em Generative Social Science}. Princeton University Press, 2012.

\bibitem{guilbeault2021topological}
Douglas Guilbeault and Damon Centola.
\newblock Topological measures for identifying and predicting the spread of complex contagions.
\newblock {\em Nature communications}, 12(1):4430, 2021.

\bibitem{milgram1967small}
Stanley Milgram et~al.
\newblock The small world problem.
\newblock {\em Psychology today}, 2(1):60--67, 1967.

\bibitem{newman2018networks}
Mark Newman.
\newblock {\em Networks}.
\newblock Oxford university press, 2018.

\bibitem{newill1964generalization}
William GOFFMAN-Vaun~A NEWILL.
\newblock Generalization of epidemic theory: an application to the transmission of ideas.
\newblock {\em Nature}, 204:225--228, 1964.

\bibitem{granovetter1983threshold}
Mark Granovetter and Roland Soong.
\newblock Threshold models of diffusion and collective behavior.
\newblock {\em Journal of Mathematical sociology}, 9(3):165--179, 1983.

\bibitem{centola2013homophily}
Damon~M Centola.
\newblock Homophily, networks, and critical mass: Solving the start-up problem in large group collective action.
\newblock {\em Rationality and society}, 25(1):3--40, 2013.

\bibitem{flache2014small}
Andreas Flache and Michael~W Macy.
\newblock Small worlds and cultural polarization.
\newblock In {\em Micro-Macro Links and Microfoundations in Sociology}, pages 146--176. Routledge, 2014.

\bibitem{dellaposta2015liberals}
Daniel DellaPosta, Yongren Shi, and Michael Macy.
\newblock Why do liberals drink lattes?
\newblock {\em American Journal of Sociology}, 120(5):1473--1511, 2015.

\bibitem{centola2015social}
Damon Centola.
\newblock The social origins of networks and diffusion.
\newblock {\em American journal of sociology}, 120(5):1295--1338, 2015.

\bibitem{centola2018behavior}
Damon Centola.
\newblock {\em How Behavior Spreads: The Science of Complex Contagions}.
\newblock Princeton University Press, 2018.

\bibitem{wasserman1994social}
Stanley Wasserman and Katherine Faust.
\newblock {\em Social network analysis: Methods and applications}.
\newblock Cambridge university press, 1994.

\bibitem{easley2010networks}
David Easley, Jon Kleinberg, et~al.
\newblock {\em Networks, crowds, and markets: Reasoning about a highly connected world}, volume~1.
\newblock Cambridge university press Cambridge, 2010.

\bibitem{burt2013social}
Ronald~S Burt, Martin Kilduff, and Stefano Tasselli.
\newblock Social network analysis: Foundations and frontiers on advantage.
\newblock {\em Annual review of psychology}, 64(1):527--547, 2013.

\bibitem{burt1994measuring}
Ronald~S Burt and Don Ronchi.
\newblock Measuring a large network quickly.
\newblock {\em Social networks}, 16(2):91--135, 1994.

\bibitem{banerjee2013diffusion}
Abhijit Banerjee, Arun~G Chandrasekhar, Esther Duflo, and Matthew~O Jackson.
\newblock The diffusion of microfinance.
\newblock {\em Science}, 341(6144):1236498, 2013.

\bibitem{burt2024guanxi}
Ronald~S Burt and Sonja Opper.
\newblock Guanxi and structural holes: Strong bridges from relational embedding.
\newblock {\em American Journal of Sociology}, 130(1):1--43, 2024.

\bibitem{paluck2016changing}
Elizabeth~Levy Paluck, Hana Shepherd, and Peter~M Aronow.
\newblock Changing climates of conflict: A social network experiment in 56 schools.
\newblock {\em Proceedings of the National Academy of Sciences}, 113(3):566--571, 2016.

\bibitem{airoldi2024induction}
Edoardo~M Airoldi and Nicholas~A Christakis.
\newblock Induction of social contagion for diverse outcomes in structured experiments in isolated villages.
\newblock {\em Science}, 384(6695):eadi5147, 2024.

\bibitem{kempe2003maximizing}
David Kempe, Jon Kleinberg, and {\'E}va Tardos.
\newblock Maximizing the spread of influence through a social network.
\newblock In {\em Proceedings of the ninth ACM SIGKDD international conference on Knowledge discovery and data mining}, pages 137--146, 2003.

\bibitem{shakarian2015independent}
Paulo Shakarian, Abhivav Bhatnagar, Ashkan Aleali, Elham Shaabani, Ruocheng Guo, Paulo Shakarian, Abhinav Bhatnagar, Ashkan Aleali, Elham Shaabani, and Ruocheng Guo.
\newblock The independent cascade and linear threshold models.
\newblock {\em Diffusion in social networks}, pages 35--48, 2015.

\bibitem{eckles2024long}
Dean Eckles, Elchanan Mossel, M~Amin Rahimian, and Subhabrata Sen.
\newblock Long ties accelerate noisy threshold-based contagions.
\newblock {\em Nature Human Behaviour}, pages 1--8, 2024.

\bibitem{granovetter1973strength}
Mark~S Granovetter.
\newblock The strength of weak ties.
\newblock {\em American journal of sociology}, 78(6):1360--1380, 1973.

\bibitem{watts1998collective}
Duncan~J Watts and Steven~H Strogatz.
\newblock Collective dynamics of ‘small-world’networks.
\newblock {\em nature}, 393(6684):440--442, 1998.

\bibitem{park2018strength}
Patrick~S Park, Joshua~E Blumenstock, and Michael~W Macy.
\newblock The strength of long-range ties in population-scale social networks.
\newblock {\em Science}, 362(6421):1410--1413, 2018.

\bibitem{sporns2006small}
Olaf Sporns and Christopher~J Honey.
\newblock Small worlds inside big brains.
\newblock {\em Proceedings of the National Academy of Sciences}, 103(51):19219--19220, 2006.

\bibitem{rajkumar2022causal}
Karthik Rajkumar, Guillaume Saint-Jacques, Iavor Bojinov, Erik Brynjolfsson, and Sinan Aral.
\newblock A causal test of the strength of weak ties.
\newblock {\em Science}, 377(6612):1304--1310, 2022.

\bibitem{steinert2017spontaneous}
Zachary~C Steinert-Threlkeld.
\newblock Spontaneous collective action: Peripheral mobilization during the arab spring.
\newblock {\em American Political Science Review}, 111(2):379--403, 2017.

\bibitem{hassanpour2016leading}
Navid Hassanpour.
\newblock {\em Leading from the periphery and network collective action}, volume~42.
\newblock Cambridge University Press, 2016.

\bibitem{vicinanza2023deep}
Paul Vicinanza, Amir Goldberg, and Sameer~B Srivastava.
\newblock A deep-learning model of prescient ideas demonstrates that they emerge from the periphery.
\newblock {\em PNAS nexus}, 2(1):pgac275, 2023.

\bibitem{clement2018searching}
Julien Clement and Phanish Puranam.
\newblock Searching for structure: Formal organization design as a guide to network evolution.
\newblock {\em Management Science}, 64(8):3879--3895, 2018.

\bibitem{centola2021change}
Damon Centola.
\newblock {\em Change: How to make big things happen}.
\newblock Hachette UK, 2021.

\bibitem{mosleh2025tendencies}
Mohsen Mosleh, Dean Eckles, and David~G Rand.
\newblock Tendencies toward triadic closure: Field experimental evidence.
\newblock {\em Proceedings of the National Academy of Sciences}, 122(27):e2404590122, 2025.

\bibitem{harris2011national}
Kathleen~Mullan Harris.
\newblock The national longitudinal study of adolescent health: Research design.
\newblock {\em http://www. cpc. unc. edu/projects/addhealth/design}, 2011.

\bibitem{lyu2022investigating}
Ding Lyu, Yuan Yuan, Lin Wang, Xiaofan Wang, and Alex Pentland.
\newblock Investigating and modeling the dynamics of long ties.
\newblock {\em Communications physics}, 5(1):87, 2022.

\bibitem{holme2002growing}
Petter Holme and Beom~Jun Kim.
\newblock Growing scale-free networks with tunable clustering.
\newblock {\em Physical review E}, 65(2):026107, 2002.

\bibitem{aral2011diversity}
Sinan Aral and Marshall Van~Alstyne.
\newblock The diversity-bandwidth trade-off.
\newblock {\em American journal of sociology}, 117(1):90--171, 2011.

\bibitem{shi2023surprising}
Feng Shi and James Evans.
\newblock Surprising combinations of research contents and contexts are related to impact and emerge with scientific outsiders from distant disciplines.
\newblock {\em Nature Communications}, 14(1):1641, 2023.

\bibitem{ma2025identifying}
Sibo Ma and Julian Nyarko.
\newblock Identifying emerging concepts in large corpora.
\newblock {\em arXiv preprint arXiv:2502.21315}, 2025.

\bibitem{romero2010directed}
Daniel Romero and Jon Kleinberg.
\newblock The directed closure process in hybrid social-information networks, with an analysis of link formation on twitter.
\newblock In {\em Proceedings of the International AAAI Conference on Web and Social Media}, volume~4, pages 138--145, 2010.

\bibitem{kossinets2009origins}
Gueorgi Kossinets and Duncan~J Watts.
\newblock Origins of homophily in an evolving social network.
\newblock {\em American journal of sociology}, 115(2):405--450, 2009.

\bibitem{gelfand2024norm}
Michele~J Gelfand, Sergey Gavrilets, and Nathan Nunn.
\newblock Norm dynamics: Interdisciplinary perspectives on social norm emergence, persistence, and change.
\newblock {\em Annual Review of Psychology}, 75(1):341--378, 2024.

\bibitem{green2025consonance}
Paul Green, Matthew Yeaton, Grace Cormier, Lara Yang, and Sameer Srivastava.
\newblock Consonance versus dissonance: How exposure to unfamiliar colleagues within and across network communities affects social belonging and network change.
\newblock 2025.

\bibitem{gray2010building}
Daniel Gray~Wilson.
\newblock Building bridges for change: how leaders enable collective change in organizations.
\newblock {\em Development and Learning in Organizations: An International Journal}, 24(1):21--23, 2010.

\bibitem{landis2016personality}
Blaine Landis.
\newblock Personality and social networks in organizations: A review and future directions.
\newblock {\em Journal of Organizational Behavior}, 37:S107--S121, 2016.

\bibitem{sevcenko2024office}
Victoria Sevcenko, Charles Ayoubi, Prithwiraj Choudhury, and Sujin Jang.
\newblock Office at offsite: How temporary colocation shapes communication in an all-remote organization.
\newblock {\em Available at SSRN 4825368}, 2024.

\bibitem{pettigrew2006meta}
Thomas~F Pettigrew and Linda~R Tropp.
\newblock A meta-analytic test of intergroup contact theory.
\newblock {\em Journal of personality and social psychology}, 90(5):751, 2006.

\bibitem{zhou2019extended}
Shelly Zhou, Elizabeth Page-Gould, Arthur Aron, Anne Moyer, and Miles Hewstone.
\newblock The extended contact hypothesis: A meta-analysis on 20 years of research.
\newblock {\em Personality and Social Psychology Review}, 23(2):132--160, 2019.

\bibitem{tausch2010intergroup}
Nicole Tausch and Miles Hewstone.
\newblock Intergroup contact.
\newblock {\em The SAGE handbook of prejudice, stereotyping and discrimination}, pages 544--560, 2010.

\bibitem{al2013intergroup}
Ananthi Al~Ramiah and Miles Hewstone.
\newblock Intergroup contact as a tool for reducing, resolving, and preventing intergroup conflict: evidence, limitations, and potential.
\newblock {\em American Psychologist}, 68(7):527, 2013.

\bibitem{bail2018exposure}
Christopher~A Bail, Lisa~P Argyle, Taylor~W Brown, John~P Bumpus, Haohan Chen, MB~Fallin Hunzaker, Jaemin Lee, Marcus Mann, Friedolin Merhout, and Alexander Volfovsky.
\newblock Exposure to opposing views on social media can increase political polarization.
\newblock {\em Proceedings of the National Academy of Sciences}, 115(37):9216--9221, 2018.

\bibitem{podolny1996networks}
Joel~M Podolny, Toby~E Stuart, and Michael~T Hannan.
\newblock Networks, knowledge, and niches: A sociological examination of worldwide competition in the semiconductor industry.
\newblock {\em American Journal of Sociology}, 102:659--689, 1996.

\bibitem{argote2000knowledge}
Linda Argote and Paul Ingram.
\newblock Knowledge transfer: A basis for competitive advantage in firms.
\newblock {\em Organizational behavior and human decision processes}, 82(1):150--169, 2000.

\bibitem{weber2003cultural}
Roberto~A Weber and Colin~F Camerer.
\newblock Cultural conflict and merger failure: An experimental approach.
\newblock {\em Management science}, 49(4):400--415, 2003.

\bibitem{guilbeault2020networked}
Douglas Guilbeault and Damon Centola.
\newblock Networked collective intelligence improves dissemination of scientific information regarding smoking risks.
\newblock {\em PLoS One}, 15(2):e0227813, 2020.

\bibitem{aiyappa2024emergence}
Rachith Aiyappa, Alessandro Flammini, and Yong-Yeol Ahn.
\newblock Emergence of simple and complex contagion dynamics from weighted belief networks.
\newblock {\em Science Advances}, 10(15):eadh4439, 2024.

\bibitem{goldberg2018beyond}
Amir Goldberg and Sarah~K Stein.
\newblock Beyond social contagion: Associative diffusion and the emergence of cultural variation.
\newblock {\em American Sociological Review}, 83(5):897--932, 2018.

\bibitem{rosenthal2015revealing}
Sara~Brin Rosenthal, Colin~R Twomey, Andrew~T Hartnett, Hai~Shan Wu, and Iain~D Couzin.
\newblock Revealing the hidden networks of interaction in mobile animal groups allows prediction of complex behavioral contagion.
\newblock {\em Proceedings of the National Academy of Sciences}, 112(15):4690--4695, 2015.

\bibitem{poel2022subcritical}
Winnie Poel, Bryan~C Daniels, Matthew~MG Sosna, Colin~R Twomey, Simon~P Leblanc, Iain~D Couzin, and Pawel Romanczuk.
\newblock Subcritical escape waves in schooling fish.
\newblock {\em Science Advances}, 8(25):eabm6385, 2022.

\bibitem{deisboeck2009collective}
Thomas~S Deisboeck and Iain~D Couzin.
\newblock Collective behavior in cancer cell populations.
\newblock {\em Bioessays}, 31(2):190--197, 2009.

\bibitem{berdahl2013emergent}
Andrew Berdahl, Colin~J Torney, Christos~C Ioannou, Jolyon~J Faria, and Iain~D Couzin.
\newblock Emergent sensing of complex environments by mobile animal groups.
\newblock {\em Science}, 339(6119):574--576, 2013.

\bibitem{hein2015evolution}
Andrew~M Hein, Sara~Brin Rosenthal, George~I Hagstrom, Andrew Berdahl, Colin~J Torney, and Iain~D Couzin.
\newblock The evolution of distributed sensing and collective computation in animal populations.
\newblock {\em Elife}, 4:e10955, 2015.

\bibitem{almaatouq2020adaptive}
Abdullah Almaatouq, Alejandro Noriega-Campero, Abdulrahman Alotaibi, PM~Krafft, Mehdi Moussaid, and Alex Pentland.
\newblock Adaptive social networks promote the wisdom of crowds.
\newblock {\em Proceedings of the National Academy of Sciences}, 117(21):11379--11386, 2020.

\bibitem{iacopini2019simplicial}
Iacopo Iacopini, Giovanni Petri, Alain Barrat, and Vito Latora.
\newblock Simplicial models of social contagion.
\newblock {\em Nature communications}, 10(1):2485, 2019.

\bibitem{de2020social}
Guilherme~Ferraz de~Arruda, Giovanni Petri, and Yamir Moreno.
\newblock Social contagion models on hypergraphs.
\newblock {\em Physical Review Research}, 2(2):023032, 2020.

\bibitem{ausiello2017directed}
Giorgio Ausiello and Luigi Laura.
\newblock Directed hypergraphs: introduction and fundamental algorithms—a survey.
\newblock {\em Theoretical Computer Science}, 658:293--306, 2017.

\bibitem{li2024enhancing}
Juyi Li, Xiaoqun Wu, Jinhu L{\"u}, and Ling Lei.
\newblock Enhancing predictive accuracy in social contagion dynamics via directed hypergraph structures.
\newblock {\em Communications Physics}, 7(1):129, 2024.

\end{thebibliography}


\begin{thebibliography}{1}

\bibitem{guilbeault2021topological}
Douglas Guilbeault and Damon Centola.
\newblock Topological measures for identifying and predicting the spread of complex contagions.
\newblock {\em Nature communications}, 12(1):4430, 2021.

\bibitem{harris2013add}
Kathleen~Mullan Harris.
\newblock The add health study: Design and accomplishments.
\newblock 2013.

\bibitem{banerjee2013diffusion}
Abhijit Banerjee, Arun~G Chandrasekhar, Esther Duflo, and Matthew~O Jackson.
\newblock The diffusion of microfinance.
\newblock {\em Science}, 341(6144):1236498, 2013.

\bibitem{rajkumar2022causal}
Karthik Rajkumar, Guillaume Saint-Jacques, Iavor Bojinov, Erik Brynjolfsson, and Sinan Aral.
\newblock A causal test of the strength of weak ties.
\newblock {\em Science}, 377(6612):1304--1310, 2022.

\end{thebibliography}

\AddToHook{enddocument/afteraux}{\immediate\write18{cp output.aux main_arxiv_paper.aux}}
\end{document}


\maketitle
\vspace*{-1cm}
\tableofcontents 
\appendix
\renewcommand{\thefigure}{S\arabic{figure}}  
\setcounter{figure}{0}  
\section{Appendix}
\subsection{Mathematical Formalisms (Extended)}
\subsubsection*{Notation and Setup}
Let $G = (V, E)$ be a network with node set $V$ and edge set $E$. We denote by $R$ the set of all possible seed configurations, where each $R_k \subseteq V$ specifies which nodes are initially active (at $t=0$). 

We focus on two particular families of seed sets:
$$
R^{RS} \subseteq R 
\quad\text{and}\quad 
R^{RCS} \subseteq R.
$$
In $R^{RS}$, each seed set $R_k$ is chosen uniformly at random from $V$ so that $\frac{|R_k|}{|V|} = p$, where $p \in [0,1]$ is the fraction of active nodes. In $R^{RCS}$, each $R_k$ satisfies the same size constraint but with the additional condition that every node in $R_k$ has at least one neighbor in $R_k$, ensuring the seeds form a clustered subset.

We denote the state of node $i \in V$ at time $t$ by $\sigma^k_i(t) \in \{0,1\}$, where $\sigma^k_i(t)=1$ indicates that $i$ is active at time $t$ in the case of seed set $R_k$. Each node $i$ has an activation threshold $T_i$, which is the minimum number of active neighbors required for $i$ to become active.

We formalize the update rule as follows:
\begin{equation}
\sigma^k_i(t+1) = 
\begin{cases}
    1, & \text{if } \sum_{j \in N[i]} \sigma^k_j(t) \geq T_i, \\
    0, & \text{otherwise}.
\end{cases}
\end{equation}

For a relative threshold $\theta_i \in [0,1]$, the threshold $T_i$ can be defined as $T_i = \left\lceil \theta_i |N[i]| \right\rceil$, where $\lceil \cdot \rceil$ denotes the ceiling function, ensuring that the threshold is an integer by rounding up.

Each set $R_k \in R$ represents a unique diffusion scenario. The initial conditions are set at $t = t_0$ by defining for all $i \in V$:

\begin{equation}
\sigma^k_i(t_0) =
\begin{cases}
    1, & \text{if } i \in R_k, \\
    0, & \text{if } i \notin R_k.
\end{cases}
\end{equation}

The system then evolves at each timestep according to the update rule above.

Define $\tau_i^k$ as the first timestep at which node $i \in V$ becomes active, given a seed set $R_k$:

\begin{equation}
\tau_i^k = 
\begin{cases} 
\min \{ t \mid \sigma^k_i(t) = 1 \}, & \text{if } \exists t \text{ such that } \sigma^k_i(t) = 1, \\ 
-1, & \text{otherwise}.
\end{cases}
\end{equation}

This transition function and discrete timesteps are executed until convergence, therefore we define convergence as a state where no further changes in node activity occur across successive timesteps:

\begin{equation}
\sum_{i \in V} \sigma^k_i(t) = \sum_{i \in V} \sigma^k_i(t - 1) = \sum_{i \in V} \sigma^k_i(t_{max}).
\end{equation}

Upon convergence, let $I_k \subseteq G$ be the induced subgraph of $G$, consisting of all nodes that got activated after starting from the seed set $R_k$. Thus, the vertexes $V(I_k)$ and edges $E(I_k)$ of the induced subgraph are defined by:

\begin{equation}
V(I_k) = \{ i \in V \mid \sigma^k_i(t_{max}) = 1 \},
\end{equation}

\begin{equation}
E(I_k) = \{ (i, j) \in E \mid \sigma^k_i(t_{max}) = 1, \sigma^k_j(t_{max}) = 1 \}.
\end{equation}

\subsubsection*{Causal Subgraph, Causal Tie Importance, Causal Node Importance}
Now we want to derive the algorithm for our proposed Causal Tie Importance (\textit{TI}) and Causal Node Importance (\textit{NI}). For each seed set $R_k$, the spreading yields an induced subgraph $I_k$. Recursive back-calculation generates causal subgraphs $C_{k,m}$, and $\text{NI}(i)$ and $\text{TI}(i,j)$ are obtained by summing over all $C_{k,m}$ across all $R_k$, followed by normalization.

\begin{figure}[H]
    \centering
    \includegraphics[width=0.6\linewidth]{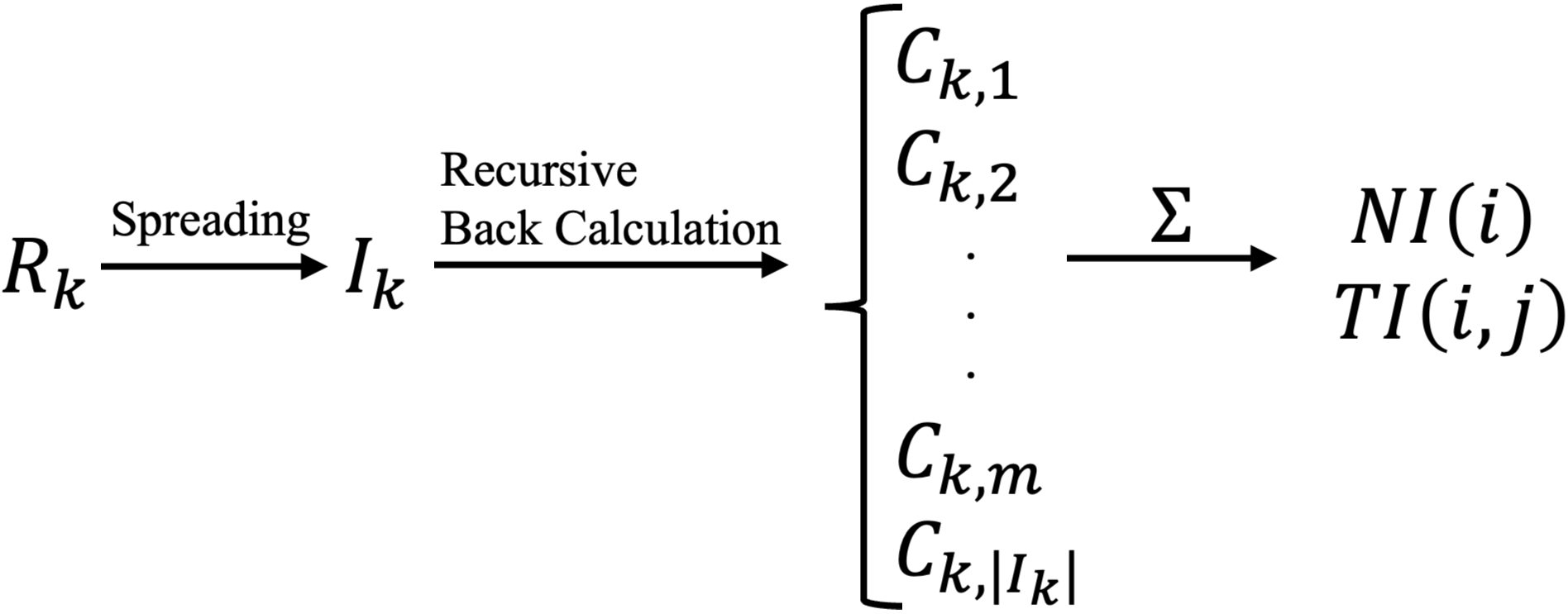}
\end{figure}
For each $m \in V(I_k)$, we define the causal subgraph $C_{k,m}$ as the set of nodes in $I_k$ that causally contribute to the activation of $m$. Specifically, $C_{k,m}$ includes all neighbors of $m$ and recursively all predecessors for each of these neighbors until reaching the seed node:

\begin{equation}
C_{k,m} =  \{ m \} \cup \{ C_{k,j} \mid j \in N[m], \tau^k_j < \tau^k_m \}
\end{equation}

We define the Causal Node Importance of a node $i$ as the number of occurrences of $i$ in all causal subgraphs $C_{k,m}$ in $I_k$ for all seed sets $R_k \in R$:

\begin{equation}
\text{NI}(i) = \sum_{C_{k,m}} \mathbf{1}_{\{ i \in C_{k,m} \}},
\end{equation}

where $\mathbf{1}_{\{ i \in C_{k,m} \}}$ is an indicator function that equals 1 if $i \in C_{k,m}$ and 0 otherwise.

Similarly, the Causal Tie Importance of an edge $(i, j)$ is defined as follows:

\begin{equation}
\text{TI}(i, j) = \sum_{C_{k,m}} \mathbf{1}_{\{ i \in C_{k,m}, j \in C_{k,m}, \tau^k_i < \tau^k_j \}}.
\end{equation}

Finally, we normalize the Causal Tie and Causal Node Importance by the largest value in each respective category. Specifically:

In the case of Causal Node Importance:

\begin{equation}
\text{NI}(i) = \frac{\sum_{C_{k,m}} \mathbf{1}_{\{ i \in C_{k,m} \}}}{\max\limits_{j} \sum_{C_{k,m}} \mathbf{1}_{\{ j \in C_{k,m} \}}}
\end{equation}

In the case of Causal Tie Importance:

\begin{equation}
\text{TI}(i, j) =\frac{\sum_{C_{k,m}} \mathbf{1}_{\{ i \in C_{k,m},\ j \in C_{k,m},\ \tau^k_i < \tau^k_j \}}}{\max\limits_{u,v} \sum_{C_{k,m}} \mathbf{1}_{\{ u \in C_{k,m},\ v \in C_{k,m},\ \tau^k_u < \tau^k_v \}}}
\end{equation}

\subsubsection*{Causal Flow Symmetry Measure,}
To evaluate the symmetry of spreading dynamics on a network $G=(V,E)$, we introduce the \emph{Causal Flow Symmetry Measure}, denoted by $\Xi_s$. This measure is computed as the correlation coefficient between the Causal Tie Importance (TI) values of each directed edge $(i,j)$ and its reverse $(j,i)$. In essence, $\Xi_s$ quantifies how evenly causal influence flows in both directions across the edges of $G$. Higher values of $\Xi_s$ indicate more balanced (symmetric) flow, whereas lower values reflect stronger asymmetries.

Let $\text{TI}(i, j)$ denote the Causal Tie Importance of the directed edge $(i, j)$, and similarly $\text{TI}(j, i)$ for the reverse edge. Considering each undirected edge $\{(i, j), (j, i)\}$ in $E$, we define
$$
\Xi_s 
= \text{Corr}\Bigl(\{\text{TI}(i, j)\}_{(i, j) \in E}, \{\text{TI}(j, i)\}_{(i, j) \in E}\Bigr),
$$
where $\text{Corr}(\cdot,\cdot)$ denotes any standard correlation coefficient (e.g., Pearson's). 

This measure complements the Causal Tie and Node Importance framework, adding the ability to measure macroscopic symmetries and asymmetries in spreading dynamics across the network.

\subsubsection*{Uncovering Flow Dynamics between Core and Periphery}
To further quantify the relationship between asymmetry in tie importance and structural network properties, we introduce the measure $\rho(\Delta S, \Delta k)$. This metric captures the correlation between the local asymmetry in tie strength ($\Delta S$) and the degree difference ($\Delta k$) between connected nodes. Specifically, $\rho(\Delta S, \Delta k)$ provides insight into how diffusion pathways categorize flow between highly connected core nodes and less connected peripheral nodes.

We compute $\Delta S$ as the absolute difference in Causal Tie Importance between the two directions of a given edge $(i, j)$:
\begin{equation}
\Delta S_{ij} = TI(i,j) - TI(j,i).
\end{equation}

Similarly, the degree difference $\Delta k$ between nodes $i$ and $j$ is defined as:
\begin{equation}
\Delta k_{ij} = k(j) - k(i),
\end{equation}
where $k_i$ and $k_j$ denote the degrees of nodes $i$ and $j$, respectively.

The Pearson correlation coefficient $\rho(\Delta S, \Delta k)$ across all edges in a given network serves as an indicator of the predominant directionality of complex contagion flow. A negative value of $\rho(\Delta S, \Delta k)$ suggests that contagions preferentially spread from high-degree core nodes to low-degree peripheral nodes, amplifying hierarchical influence structures. Conversely, a positive value indicates stronger flow from the periphery to the core, highlighting the potential of the periphery to influence the core stronger than the core the periphery and therefore inverting typical influence dynamics identified by standard network centralities.

\subsubsection*{Correlation Between Node Importance and Degree}

Another fundamental aspect of network influence dynamics is the relationship between Causal Node Importance (NI) and the degree of nodes. To quantify this relationship, we introduce the measure $\rho(NI(i), k(i))$ which is the correlation between the Causal Node Importance values of nodes $NI(i)$ and their degree $k(i)$ which captures the correlation between a node's importance in contagion spreading and its degree in the network. Positive values can be interpreted that higher degree nodes are also more important, values close to zero as no significant impact of degree on the importance of nodes and negative values as high degrees being detrimental.

\subsubsection*{Degree-Normalized Tie Importance Correlation}
To further explore the role of node degree in determining tie importance, we introduce the degree-normalized correlation $\rho(\frac{NI(i)}{k(i)}, k(i))$. This measure captures the relationship between the degree-normalized Causal Node Importance (NI divided by node degree) and the node degree itself, providing insights into if the degree of a node has linearly proportional impact on its Node Importance or not. This measure allows us to assess whether high-degree nodes exhibit disproportionately high or low tie importance relative to their connectivity and giving us the ability to investigate if the impact of node degrees is more than linearly proportionally increasing the Causal Node Importance.
\subsection{Experimental Design and Simulations}

We analyze synthetic and empirical networks, including:
\begin{itemize}
    \item \textbf{Watts-Strogatz Networks:} Small-world networks with varying rewiring probability $\beta$.
    \item \textbf{Clustered Power Law Networks: } Scale-free networks with tunable clustering.
    \item \textbf{AddHealth Dataset:} High school friendship networks.
    \item \textbf{Banerjee Village Networks:} Rural economic and social networks.
\end{itemize}
For each network, we conduct extensive simulations with:
\begin{itemize}
    \item Various seeding strategies (random and randomly clustered).
    \item Absolute Thresholds ranging from $T = 1$ (simple contagion) to $T > 1$ (complex contagion).
    \item Relative Thresholds ranging from $\theta = 0.05$ to $\theta = 0.35$.
    \item Noisy threshold-based contagions with varying probabilities for subthreshold activation $5\%$ up to $30\%$.
    \item Independent Cascade Model with transmission probabilities $\beta = 0.05$ to $\beta = 0.2$.
    \item Linear Threshold model with homogeneous edge weights and also Gaussian sampled edge weights.
    \item Large-scale Monte Carlo runs to ensure robustness. These simulations were typically carried out with 2 to 4 sweeps, where a sweep is defined as the number of nodes in the graph.
\end{itemize}

We systematically measure the impact of weak ties, bridge formation, and periphery-to-core influence shifts. Convergence is assessed via correlation analysis across multiple runs.

The full source code and datasets will be made available upon publication. 

\subsection{Algorithms}
Based on this derivation, we now introduce pseudo code for how the Causal Node Importance for nodes and Causal Tie Importance for edges can be implemented. 
\begin{algorithm}
\caption{Causal Tie and Node Importance Calculation}
\begin{algorithmic}[1]
\REQUIRE Graph $G = (V, E)$, seed sets $R$
\ENSURE NI and TI values

\STATE Initialize $\text{NI}[i] = 0$ for all $i \in V$
\STATE Initialize $\text{TI}[i, j] = 0$ for all $(i, j) \in E$

\FOR{each seed set $R_i \in R$}
    \STATE Set $\sigma_i = 0$ for all $i \in V$
    \STATE Set $\sigma_i = 1$ for all $i \in R_i$ at $t = t_0$
    \STATE Initialize $\text{activation\_times}[i] = -1$ for all $i \in V$
    \STATE Set $\text{activation\_times}[i] = t_0$ for all $i \in R_i$
    \STATE $t = t_0$

    \STATE changed = \text{True}
    \WHILE{changed}
        \STATE changed = \text{False}

        \FOR{each node $i \in V$}
            \STATE $\sigma_i(t+1) = \sigma_i(t)$ \COMMENT{carry over active state}
            \IF{ $\sigma_i(t) = 0$ and $\sum_{j \in N[i]} \sigma_j(t) \ge T_i$ }
                \STATE $\sigma_i(t+1) = 1$
                \STATE $\text{activation\_times}[i] = t + 1$
                \STATE changed = \text{True}
            \ENDIF
        \ENDFOR

        \STATE $t = t + 1$
    \ENDWHILE

    \STATE $t_{\max} = t - 1$ \COMMENT{the final time step after convergence}

    \STATE Define $I = (V(I), E(I))$ as the subgraph of nodes $i$ with $\sigma_i(t_{\max}) = 1$

    \FOR{each node $k \in V(I)$}
        \STATE Initialize $\text{causal\_subgraph} = \emptyset$
        \STATE \textbf{ComputeCausalContributers}($k$, $\text{causal\_subgraph}$, $\text{activation\_times}$, $t_{\max}$)

        \FOR{each node $i \in \text{causal\_subgraph}$}
            \STATE $\text{NI}[i] \mathrel{+}= 1$
        \ENDFOR

        \FOR{each edge $(i, j)$ in $\text{causal\_subgraph}$ where $\text{activation\_times}[i] < \text{activation\_times}[j]$}
            \STATE $\text{TI}[i, j] \mathrel{+}= 1$
        \ENDFOR
    \ENDFOR
\ENDFOR

\RETURN $\text{NI}$, $\text{TI}$

\STATE
\STATE \textbf{Subroutine: ComputeCausalContributers}($\text{target\_node}$, $\text{causal\_subgraph}$, $\text{activation\_times}$, $t_{\max}$)
\STATE \textbf{Input:} $\text{target\_node}$, $\text{causal\_subgraph}$, $\text{activation\_times}$, $t_{\max}$
\STATE \textbf{Output:} Updated $\text{causal\_subgraph}$ for $\text{target\_node}$

\STATE Add $\text{target\_node}$ to $\text{causal\_subgraph}$

\FOR{each $\text{neighbor\_node} \in N[\text{target\_node}]$}
    \IF{$\text{activation\_times}[\text{neighbor\_node}] < \text{activation\_times}[\text{target\_node}]$
        \textbf{ and } $\text{neighbor\_node} \notin \text{causal\_subgraph}$
        \textbf{ and } $\sigma_{\text{neighbor\_node}}(t_{\max}) = 1$}
        \STATE Add $\text{neighbor\_node}$ to $\text{causal\_subgraph}$
        \STATE \textbf{ComputeCausalContributers}($\text{neighbor\_node}$, $\text{causal\_subgraph}$, $\text{activation\_times}$, $t_{\max}$)
        \COMMENT{Recursively trace causal predecessors}
    \ENDIF
\ENDFOR
\end{algorithmic}
\end{algorithm}

\newpage

\subsection{Extensive Symmetry figure results}\label{sec:extensive_symmetry_figs}
In this section we show the stability of emerging asymmetries across the General Influence Model, the Linear Threshold Model, the Independent Cascade Model and the Noisy Threshold-based contagions.

\subsubsection{Linear Threshold Model}\label{app:LTM}
\begin{figure}[H]
    \centering
    \includegraphics[width=0.6\linewidth]{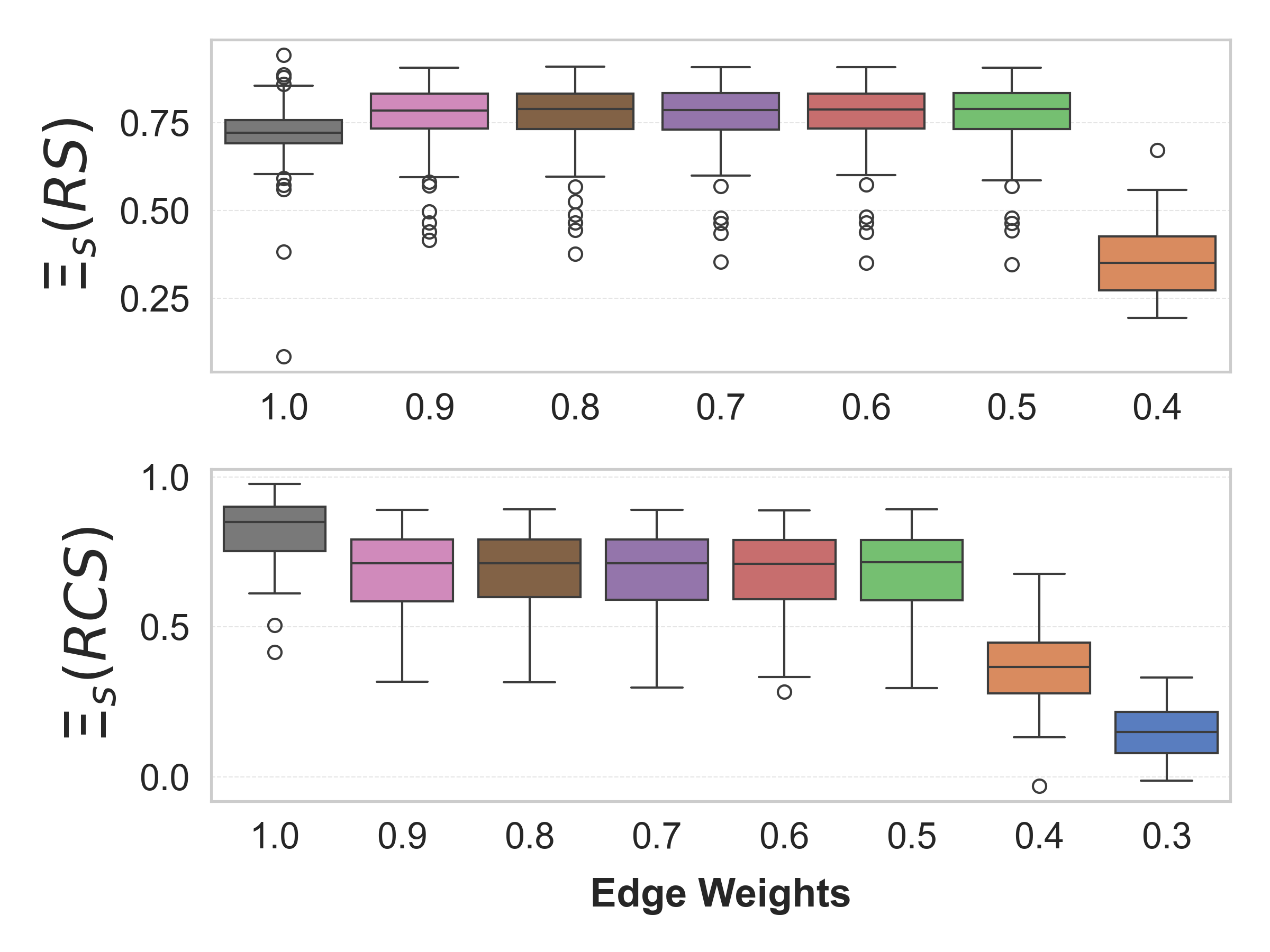}
    \caption{Box plots illustrating the relationship between symmetry $\Xi_s$ and edge weights for the Linear Threshold Model in simulations on the Add Health Dataset. (a) uses random seeding of $2\%$ of nodes, while (b) uses random clustered seeding of $2\%$ of nodes. For each giant component on each network, we define a simulation scenario based on its specific graph structure and a given threshold value. For each scenario, $10N$ independent and randomly chosen clustered seed sets comprised of $5\%$ of all nodes are generated to compute the symmetry measure, where $N$ denotes the number of nodes in the graph. Symmetry declines as the systems need for reinforcement/complexity increases, reflecting more asymmetrical spreading dynamics.}
    \label{fig:boxplot_symmetry_LTM}
\end{figure}
\begin{figure}[H]
    \centering
    \includegraphics[width=0.6\linewidth]{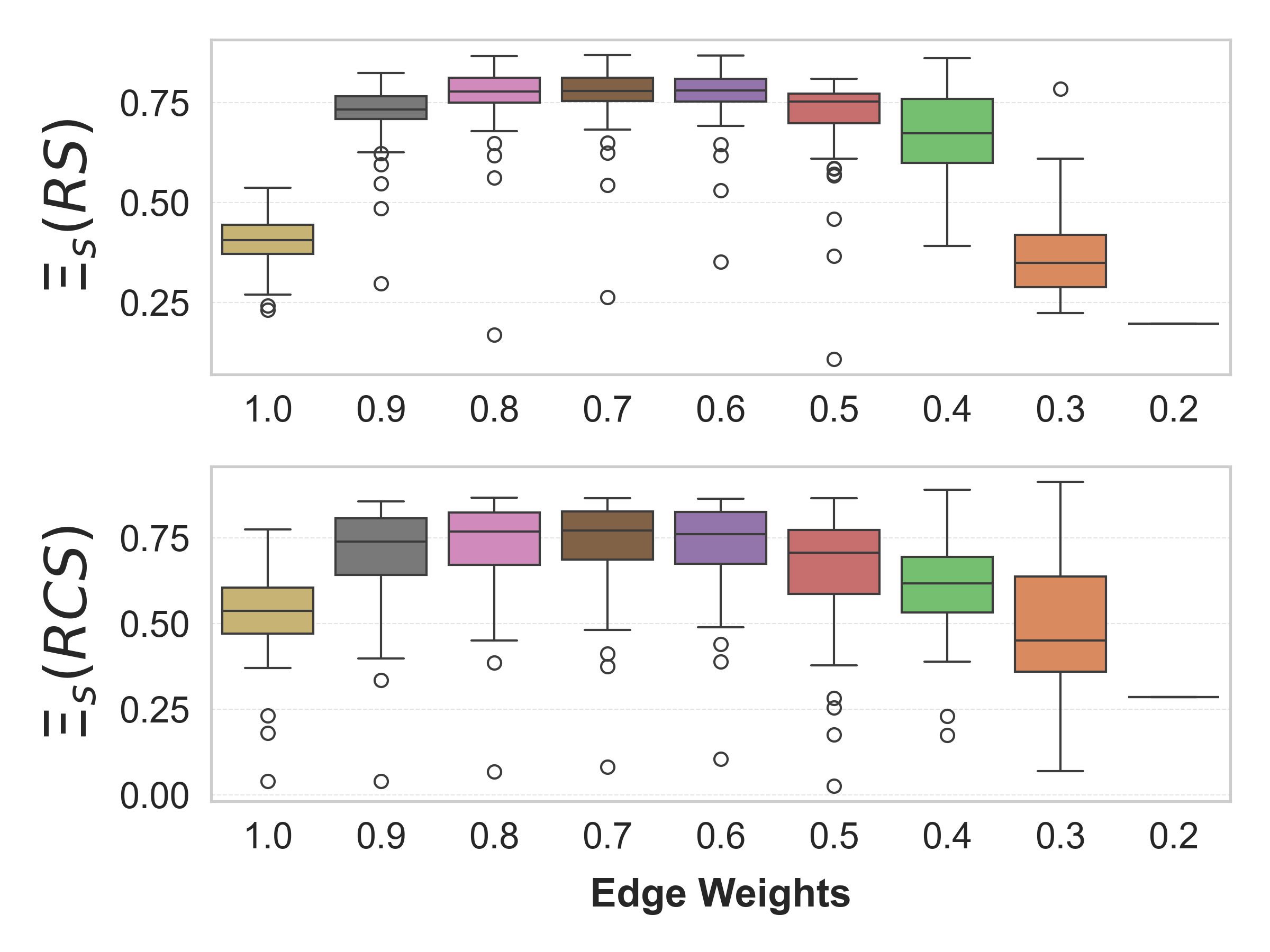}
    \caption{Box plots illustrating the relationship between symmetry $\Xi_s$ and edge weights sampled from a Gaussian distribution with $\sigma = 0.05$ for the Linear Threshold Model in simulations on the Add Health Dataset. (a) uses random seeding of $5\%$ of nodes, while (b) uses random clustered seeding of $5\%$ of nodes. For each giant component on each network, we define a simulation scenario based on its specific graph structure and a given threshold value. For each scenario, $10N$ independent and randomly chosen clustered seed sets comprised of $5\%$ of all nodes are generated to compute the symmetry measure, where $N$ denotes the number of nodes in the graph. Symmetry declines as the systems need for reinforcement/complexity increases, reflecting more asymmetrical spreading dynamics.}
    \label{fig:boxplot_symmetry_LTM_Gauss}
\end{figure}
\subsubsection{Independent Cascade Model}\label{app:ICM}
\begin{figure}[H]
    \centering
    \includegraphics[width=0.6\linewidth]{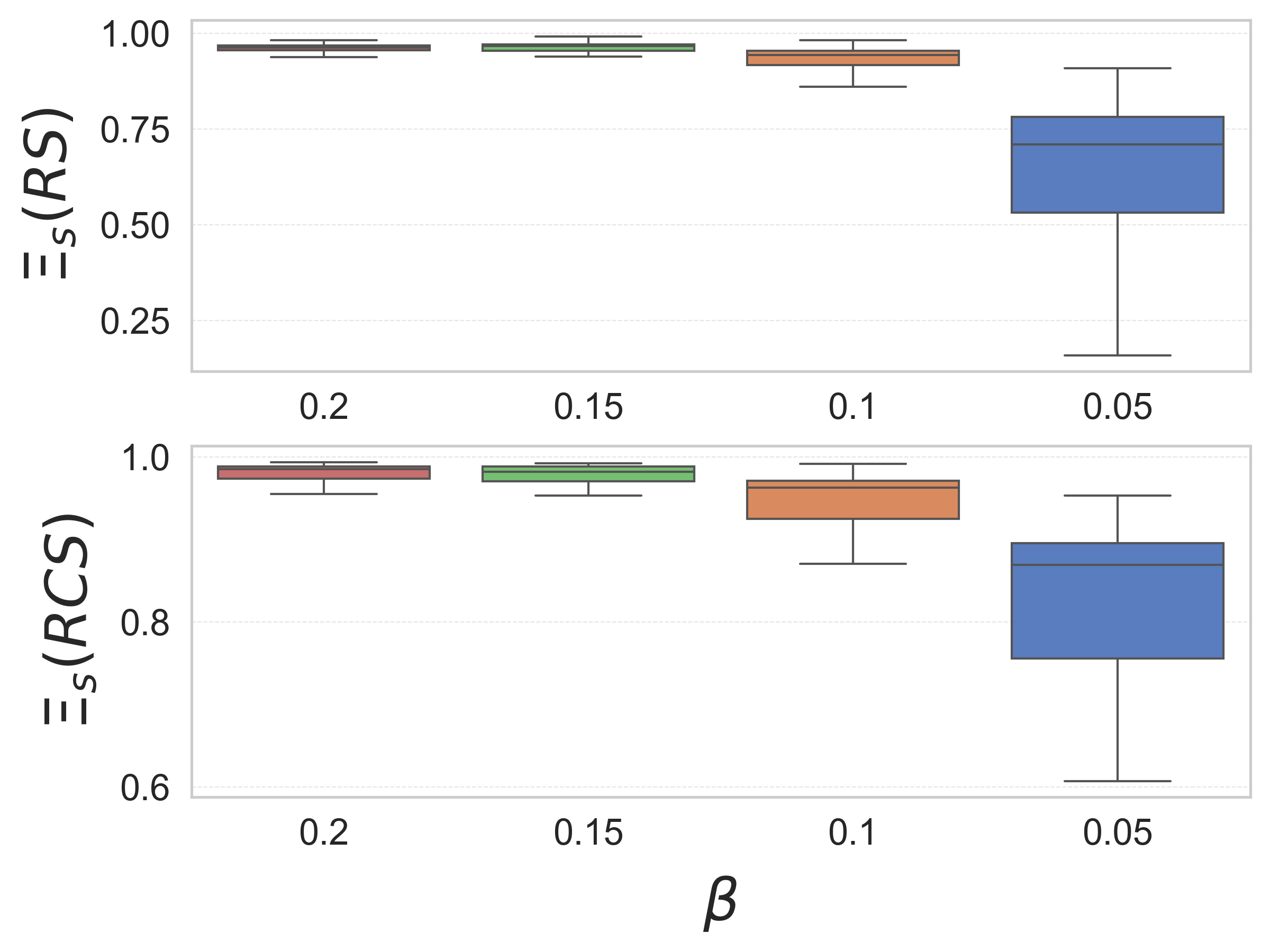}
    \caption{Box plots illustrating the relationship between symmetry $\Xi_s$ and edge transmission probability for the Independent Cascade Model in the range of Complex Contagion behavior for simulations on the Add Health Dataset \cite{guilbeault2021topological}. (a) uses random seeding of $2\%$ of nodes, while (b) uses random clustered seeding of $2\%$ of nodes. Each color represents a distinct edge transmission probability value, with box plots indicating distribution, interquartile range. For each giant component on each network, we define a simulation scenario based on its specific graph structure and a given edge transmission probability. For each scenario, $10N$ independent and randomly chosen clustered seed sets comprised of $5\%$ of all nodes are generated to compute the symmetry measure, where $N$ denotes the number of nodes in the graph. Symmetry declines as the systems need for reinforcement/complexity increases, reflecting more asymmetrical spreading dynamics.}
    \label{fig:boxplot_symmetry_ICM}
\end{figure}
\subsubsection{Noisy-threshold based contagions}\label{app:noisy}

In the noisy threshold-based contagion model, each node adopts a behavior based on the number of adopting neighbors, similar to the standard threshold model, but with added stochasticity. Specifically, if a node has fewer than $\theta$ active neighbors—i.e., below the deterministic threshold—it can still adopt with a small, non-zero probability $q > 0$. We also consider a \emph{modified} version of the Noisy threshold model where a node can adopt through subthreshold transmission \emph{only once}, mirroring the single-attempt dynamics of the Independent Cascade Model (ICM).

\begin{figure}[H]
    \centering
    \includegraphics[width=1\linewidth]{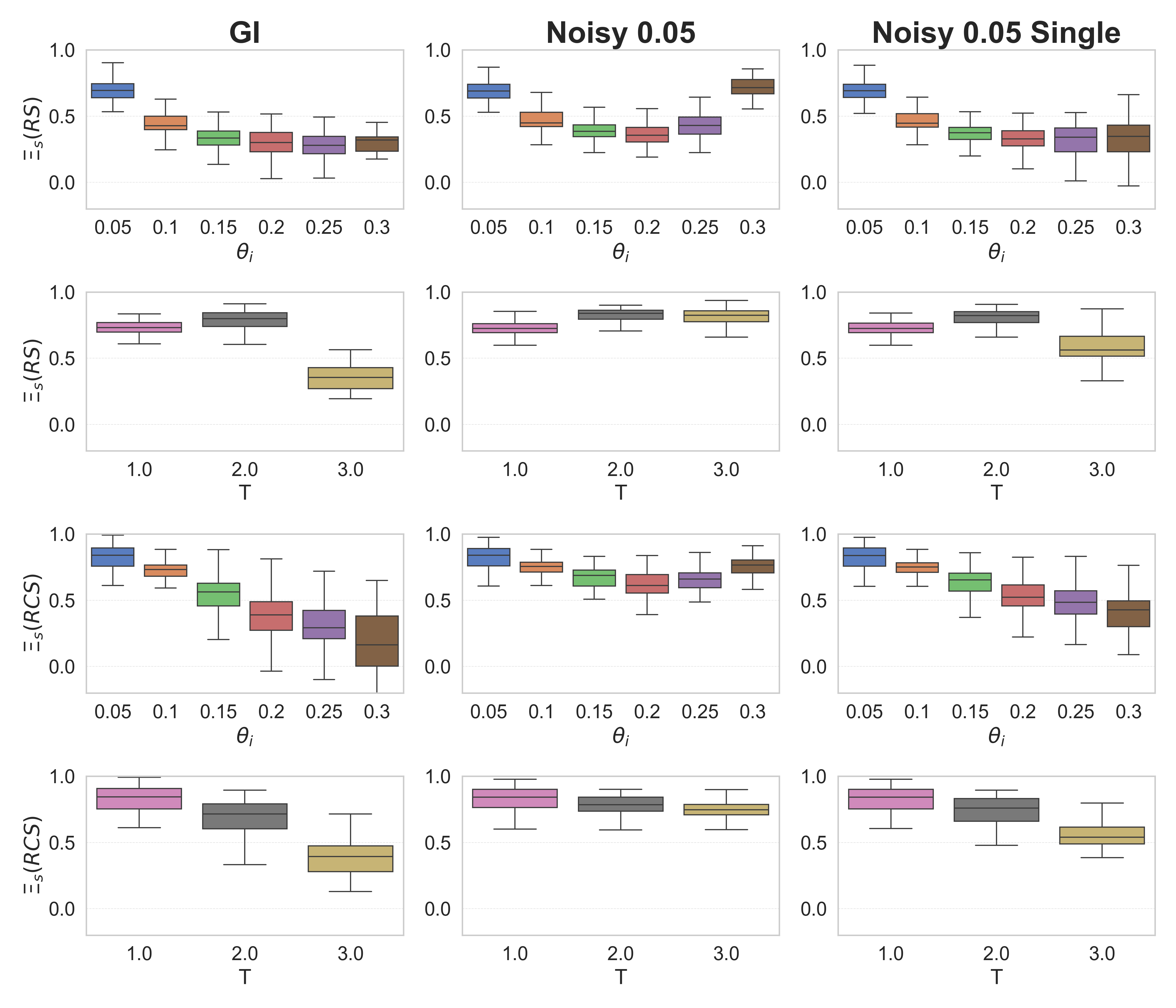}
    \caption{ 
    \textbf{Symmetry ($\Xi$) across three contagion models with random and clustered seeding strategies.} 
    Each panel compares the symmetry outcomes for one of the contagion models (GI, Noisy, Noisy Single). The rows correspond to different seeding strategies: random (top two rows) and random clustered (bottom two rows). Thresholds are either absolute ($T$) or relative ($\theta_i$). For each giant component on each network, we define a simulation scenario based on its specific graph structure and a given threshold value. For each scenario, $10N$ independent and randomly chosen clustered seed sets comprised of $5\%$ of all nodes are generated to compute the symmetry measure, where $N$ denotes the number of nodes in the graph. The figure highlights how restricting the Noisy threshold model to a single transmission changes the resulting symmetry and makes it resemble that of the deterministic GI model more closely.
    }
    \label{fig:combined_boxplots_noisy}
\end{figure}

Figure~\ref{fig:combined_boxplots_noisy} presents the symmetry ($\Xi$) of the final adoption patterns for three models: 
1) the deterministic General Influence (GI) model, 
2) the standard Noisy threshold model (Noisy), and 
3) the single-transmission variant of the Noisy model (Noisy Single). 
The x-axis indicates different threshold formulations: $T$ (absolute threshold) and $\theta_i$ (relative threshold). The first and second rows use \emph{random} seeding of 2\% of the nodes, whereas the third and fourth rows use \emph{random clustered} seeding of 2\%. All three models exhibit significant emerging symmetries (non-zero values of $\Xi$), yet the patterns differ. In particular, restricting Noisy threshold contagion to a single transmission attempt causes the asymmetries to more closely align with those of the deterministic GI model. This underscores the strong influence of deterministic thresholds on the propagation process when nodes only have one chance to transmit (subthreshold) contagion.

\begin{figure}[H]
    \centering
    \includegraphics[width=0.6\linewidth]{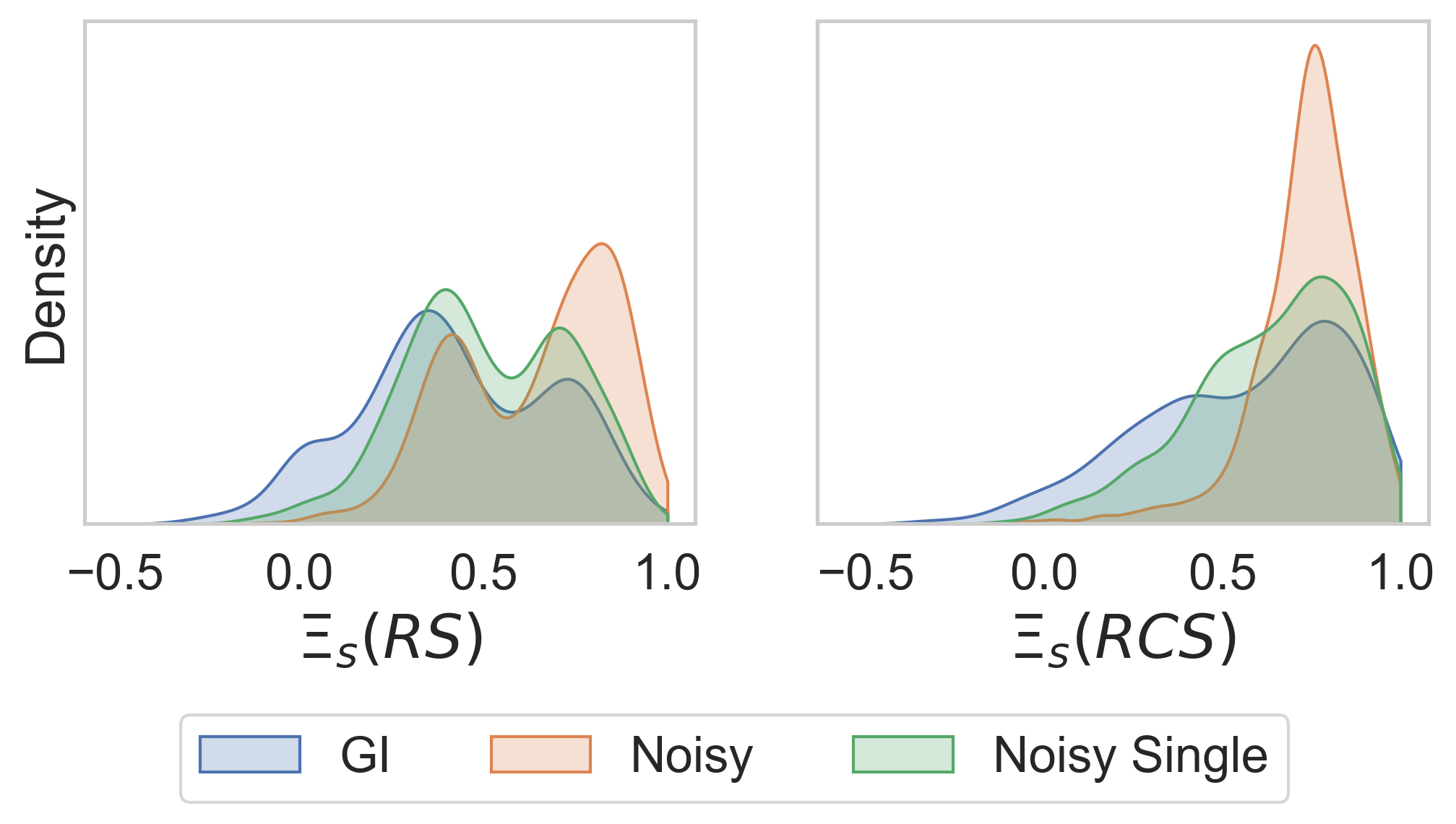}
    \caption{
    \textbf{Density plots of symmetry values ($\Xi_s$) under different models and seeding strategies.}
    The left plot shows the distribution of symmetries for random seeding of 2\% of the nodes, and the right plot shows the distribution for random clustered seeding of 2\%. Colors differentiate the three contagion models (GI, Noisy, and Noisy Single), and each distribution is aggregated over all tested threshold values. For each giant component on each network, we define a simulation scenario based on its specific graph structure and a given threshold value. For each scenario, $10N$ independent and randomly chosen clustered seed sets comprised of $5\%$ of all nodes are generated to compute the symmetry measure, where $N$ denotes the number of nodes in the graph. The consistently high concentration near 0.5 indicates that asymmetries in final adoption are a common outcome across diverse contagion dynamics and seeding patterns.
    }
    \label{fig:kdeplot_symmetry_RC_RCS_GI_NOISY_SINGLE}
\end{figure}

Figure~\ref{fig:kdeplot_symmetry_RC_RCS_GI_NOISY_SINGLE} shows density plots of the symmetry values ($\Xi_s$), aggregated across all threshold types, under random seeding (left) and random clustered seeding (right). Each plot compares the distribution of symmetries for the three models: GI, Noisy, and Noisy Single. Despite some variation, the distribution of symmetries consistently centers around 0.5, indicating that moderate degrees of asymmetry frequently emerge regardless of the contagion model or seeding strategy.

\subsubsection{General Influence Model}
\begin{figure}[H]
    \centering
    \includegraphics[width=1\linewidth]{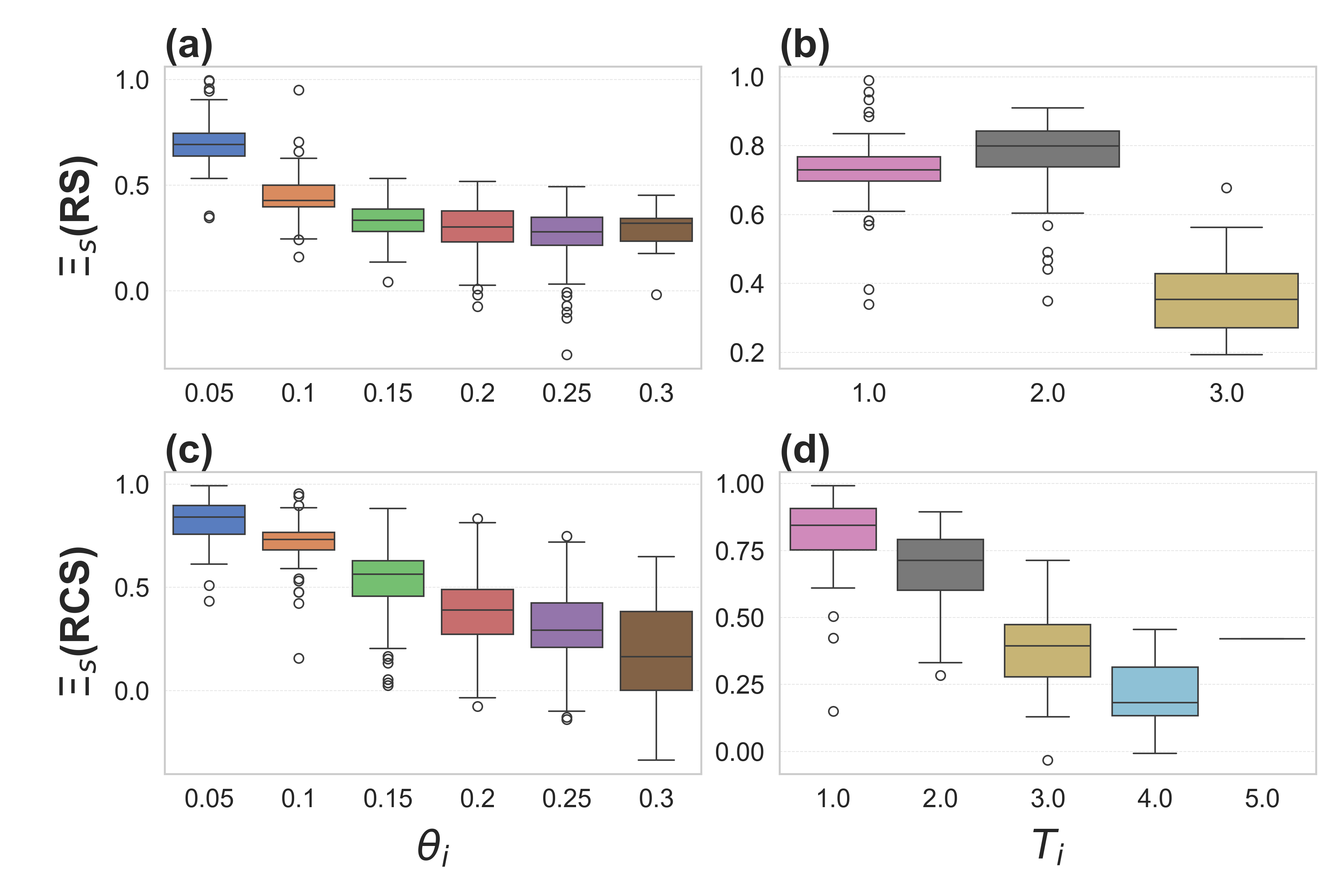}
    \caption{Box plots illustrating the relationship between symmetry $\Xi_s$ and threshold parameters in simulations on the Add Health Dataset. Only graphs with meaningful spreading (spreading density $>10\%$ of nodes) are shown. (a) and (b) use random seeding of $2\%$ of nodes, while (c) and (d) use random clustered seeding of $2\%$ of nodes. (a) and (c) show symmetry at varying levels of the relative threshold $\theta_i$, while (b) and (d) display symmetry for the absolute threshold $T_i$. Each color represents a distinct threshold value, with box plots indicating distribution, interquartile range, and outliers. For each giant component on each network, we define a simulation scenario based on its specific graph structure and a given threshold value. For each scenario, $100N$ independent and randomly chosen clustered seed sets comprised of $5\%$ of all nodes are generated to compute the symmetry measure, where $N$ denotes the number of nodes in the graph. Symmetry consistently declines as thresholds increase, reflecting more asymmetrical spreading dynamics.}
    \label{fig:boxplot_symmetry}
\end{figure}

\subsection{Transition Through the Small-World Regime and Its Impact on Spreading and Asymmetry}\label{app:rewiring_symmetry}

To investigate how the transition from regular to random topologies affects complex contagion dynamics, we simulate spreading processes on Watts--Strogatz (WS) networks while increasing the rewiring probability \(\beta\). This transition passes through the small-world regime, where the introduction of shortcuts dramatically lowers path lengths while maintaining high clustering, for this we track both the spreading density and the symmetry.

\begin{figure}[H]
\centering
\includegraphics[width=0.9\textwidth]{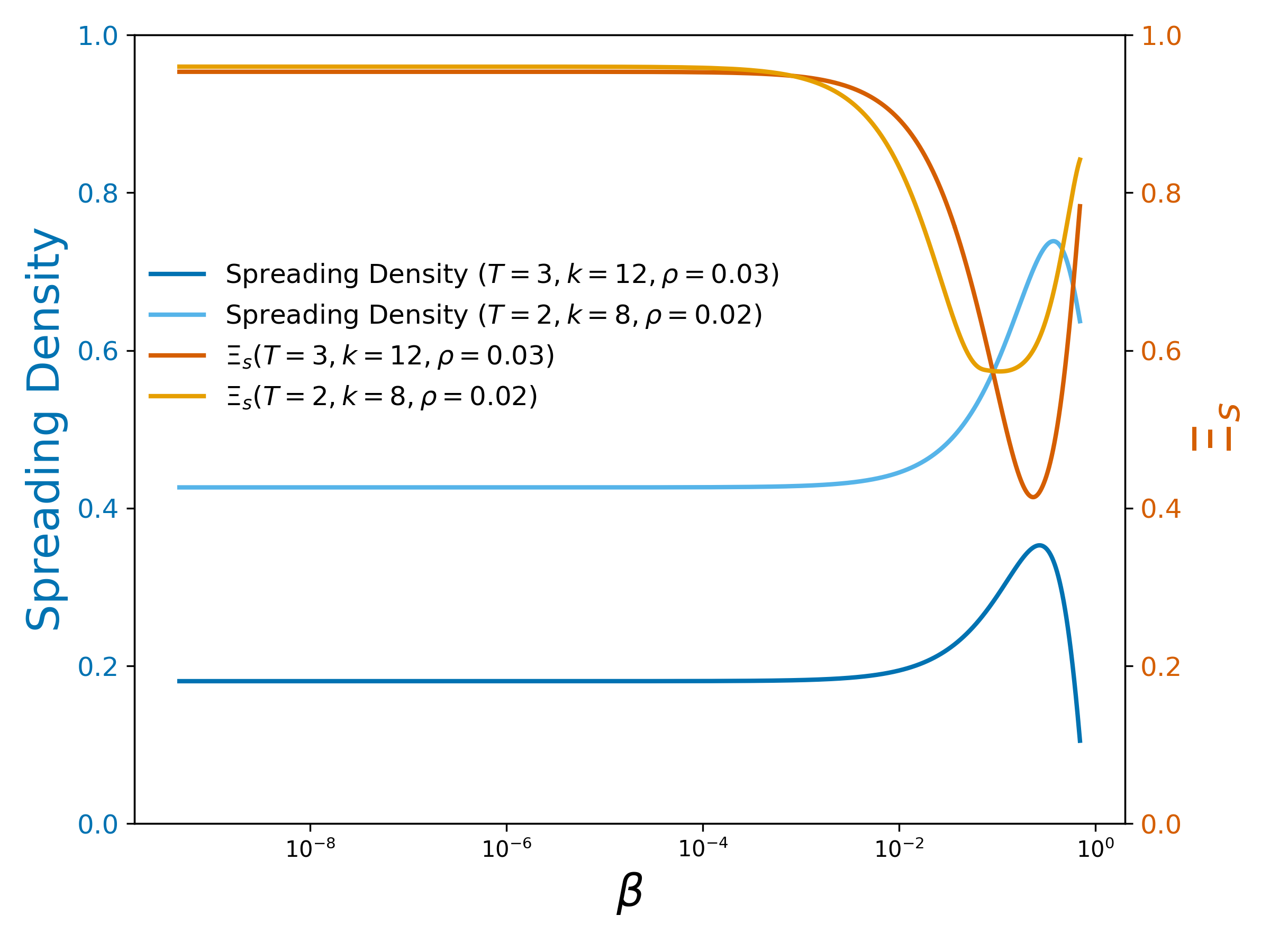}
\caption{\textbf{Impact of the Small-World Transition on Spreading Efficiency and Directional Symmetry.} Spreading Density (blue curves, left y-axis) and symmetry (orange curves, right y-axis) as a function of the rewiring probability \(\beta\) in Watts--Strogatz graphs. Two parameter settings are shown: ($T = 3, k = 12, \rho = 0.03$) (dark curves) and $(T = 2, k = 8, \rho = 0.02)$ (light curves). As \(\beta\) increases, the network transitions through a small-world regime that initially facilitates spreading due to the formation of shortcuts. However, further randomization reduces spreading density by disrupting the clustered structures needed for reinforcement. At the same time, the symmetry of spreading directions declines and shows the lowest values where the ability to spread is highest.}
\label{fig:rewiring_dip_plot}
\end{figure}

\subsection{Convergence Analysis of Causal Tie Importance, Causal Node Importance, and Symmetry}\label{sec:convergence}
To assess the convergence properties of our Causal Tie Importance, Causal Node Importance and Causal Flow Symmetry measures, we analyze how the computed values stabilize as a function of the number of sweeps performed. In our methodology, each sweep consists of $N$ independent realizations of the spreading process with random starting conditions over the network, where $N$ represents the number of nodes in the graph. For each graph we initialize the General Influence model with absolute thresholds $T \in [1,2,3]$ and relative thresholds $\theta \in [0.05, 0.1, 0.15, 0.2, 0.25, 0.3]$.

\subsubsection{Convergence Criteria}
We define convergence as the point where additional sweeps introduce negligible changes in the computed values. Specifically, we establish three independent analyses to assess convergence:

\begin{enumerate}
    \item \textbf{Pearson correlation of importance values} across consecutive runs to evaluate the stability of the Tie and Node Importance measures.
    \item \textbf{Causal Flow Symmetry measure convergence} as an additional structural validation of stability.
    \item \textbf{Graph size effects on convergence} to determine computational efficiency and scalability.
\end{enumerate}
We define convergence quantitatively as the point where:
\begin{itemize}
    \item The Pearson correlation between consecutive sweeps exceeds $0.95$.
    \item The relative difference in symmetry values between consecutive runs falls below $2\%$. 
\end{itemize}
\subsubsection{Pearson Correlation of Tie and Node Importance}
To assess the stability of importance values across simulation runs, we perform two independent runs—each with an identical number of sweeps—on all Banerjee giant component graphs. We then compute the Pearson correlation coefficient between Causal Node Importance and Causal Tie Importance values obtained from the two runs.

Pearson correlation is appropriate in this context because it quantifies the degree of linear association between two sets of values. This allows us to measure whether fluctuations in importance values due to stochastic effects preserve the overall structure of variation across nodes and ties. In other words, a high Pearson correlation indicates that nodes (or ties) which are estimated to be more important in one run tend to also receive proportionally high importance in the other run, even if the absolute values differ. This makes Pearson correlation a suitable metric for assessing the reproducibility of the importance measures under repeated simulations.

\begin{figure}[h]
    \centering
    \includegraphics[width=0.85\textwidth]{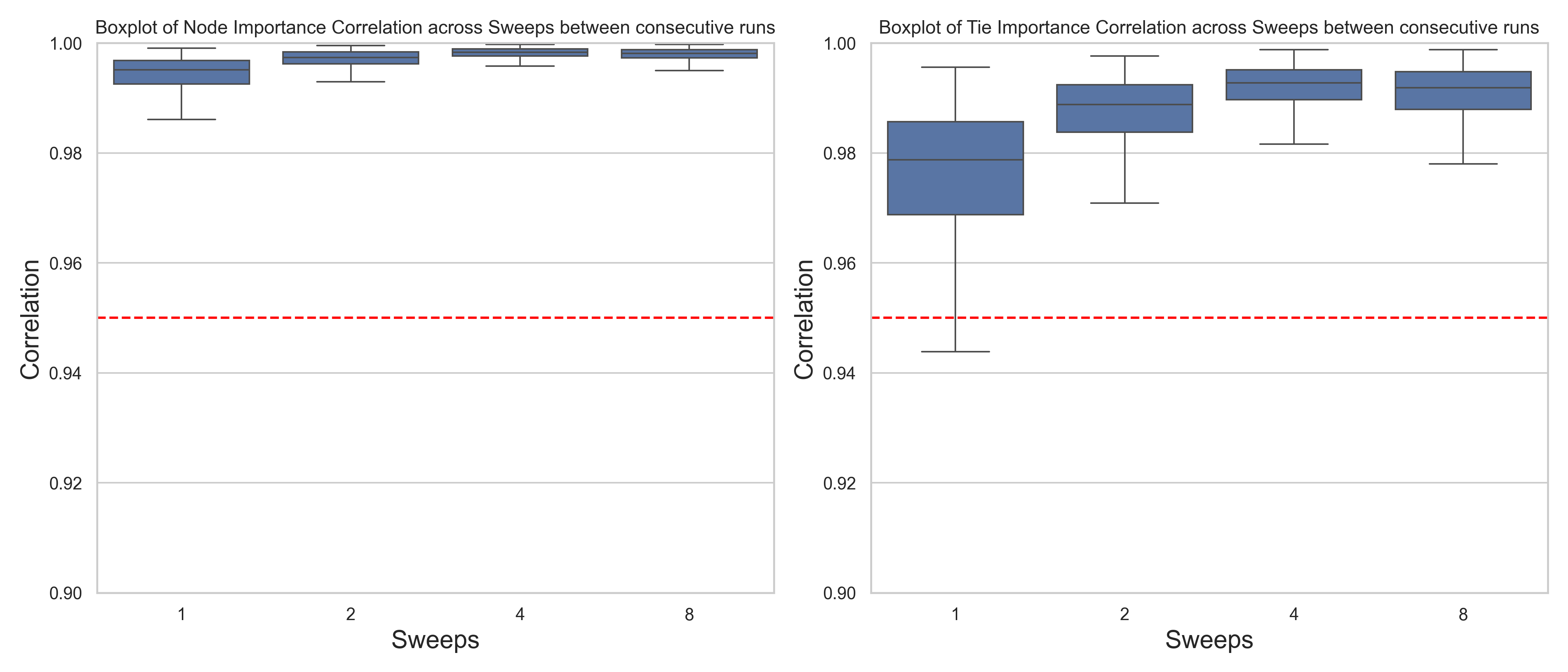}
    \caption{Boxplots showing the Pearson correlation coefficients for Node (left) and Tie (right) importance measures across sweeps between two consecutive runs. The x-axis represents the number of sweeps, and the y-axis represents the correlation values. As the number of sweeps increases, the correlation stabilizes close to $1.0$ (indicating almost perfect alignment), demonstrating the convergence of the node and edge-level causal importance measures. The red dashed line indicates a correlation of $0.95$ as a reference.}
    \label{fig:boxplot_node_edge_correlation}
\end{figure}

Figure~\ref{fig:boxplot_node_edge_correlation} displays the distribution of Pearson correlation coefficients across sweeps for the two simulation runs.

\subsubsection{Statistical Convergence of our Symmetry Measure}
\begin{figure}[h]
    \centering
    \includegraphics[width=0.85\textwidth]{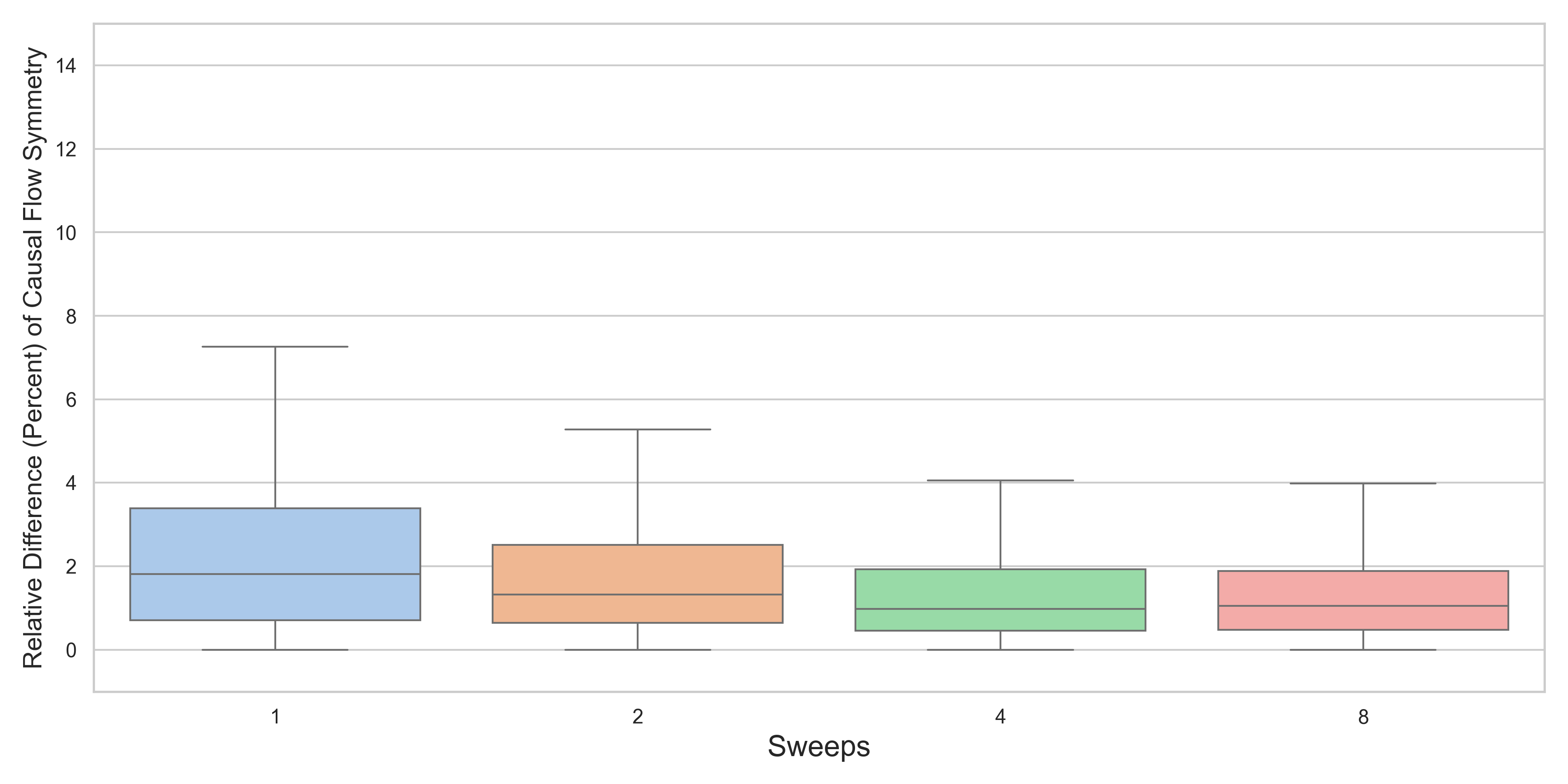}
    \caption{Boxplot showing the relative differences in the symmetry measure across sweeps. The x-axis represents the number of sweeps, and the y-axis shows the relative difference (in percentage) of the symmetry measure between consecutive runs. As the number of sweeps increases, both the relative differences and the variance decrease, highlighting improved stability and convergence of the symmetry measure. One of the crucial findings is that the results for the symmetry measure is already very stable for 1 sweep.}
    \label{fig:boxplot_relative_diff_symmetry}
\end{figure}

To evaluate the convergence of symmetry, we calculate the symmetry measure for each graph and determine the difference between the symmetry values of consecutive runs. If convergence is achieved, the symmetry measure should stabilize alongside the importance values, as it reflects a sufficiently large number of sweeps in the random sampling process for the values to stabilize. Figure~\ref{fig:boxplot_relative_diff_symmetry} presents the relative differences in symmetry values across increasing sweeps and shows the very small deviations between consecutive runs even for a small number of sweeps.

\subsubsection{Graph Size and Convergence Scaling}
To assess the effect of graph size on convergence, we group the graphs in the dataset into five categories: very small, small, medium, and large and very large, each representing one-fifth of the graphs. Our results show that, for the same number of sweeps, larger graphs exhibit better convergence.

A likely reason for this faster stabilization is that on larger graphs, the spreading process is more extensive, allowing our method to extract more information from the same number of different starting conditions. With each sweep, larger networks provide a richer set of activation patterns, leading to a more comprehensive and stable estimation of Causal Tie Importance and Causal Node Importance. Connecting to the mathematical description, more $C_k,m$ and be extracted for each $I_k$.

\begin{figure}[H]
    \centering
    \includegraphics[width=1\textwidth]{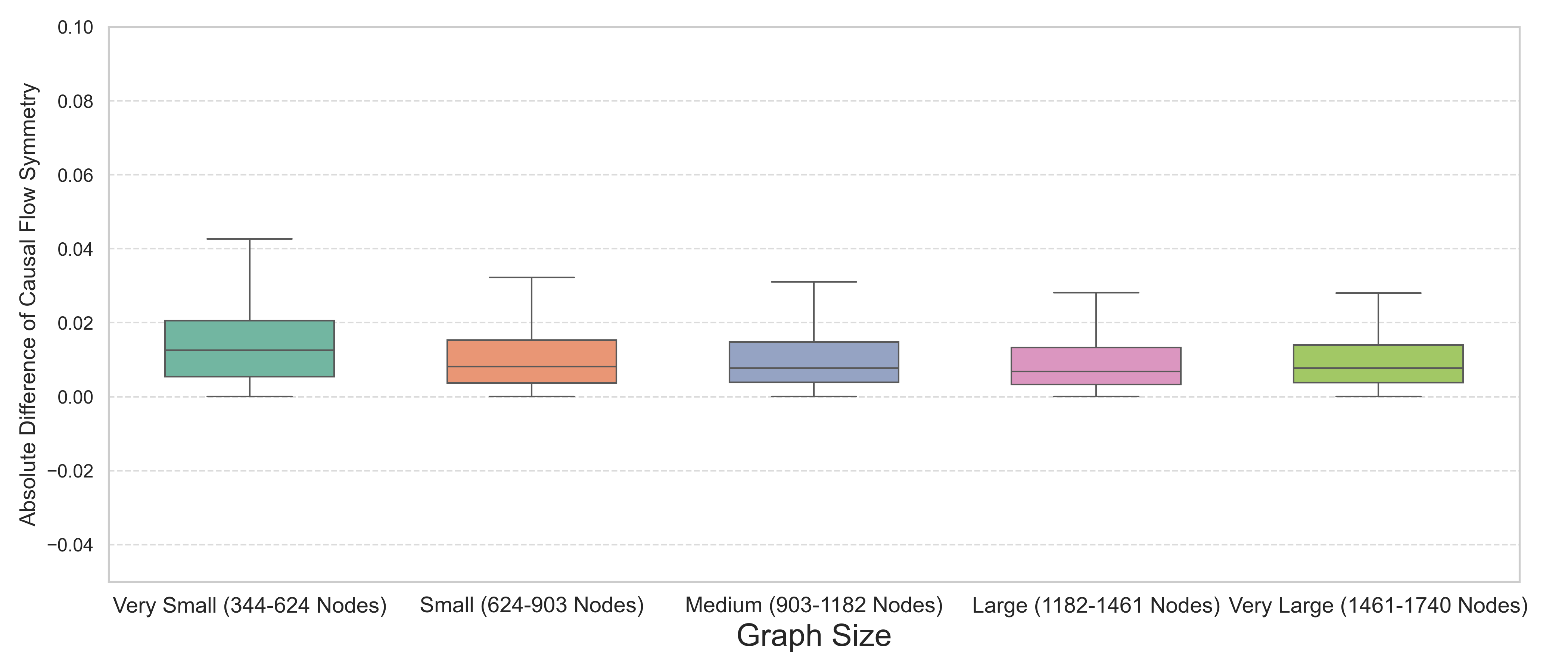}
    \caption{Boxplot showing the absolute differences in the symmetry measure across very small, small, medium, and large and very large graph size groups over $1$ to $8$ sweeps. The x-axis represents the graph size categories, and the y-axis shows the absolute differences of the symmetry measure between consecutive runs. Larger graphs tend to exhibit smaller differences and reduced variance for the same number of sweeps, demonstrating that convergence improves with increasing graph size. This indicates that fewer sweeps are required for stable results in larger networks.}
    \label{fig:boxplot_abs_diff_symmetry_grouped}
\end{figure}

Figure~\ref{fig:boxplot_abs_diff_symmetry_grouped} visualizes this effect, showing the tendency that the symmetry measure stabilizes more quickly for larger graphs.

\subsubsection{Computational Efficiency}
To quantify the computational efficiency of our method, we analyze the scaling behavior of the number of sweeps required for convergence across different graph sizes. We find that the number of sweeps required scales sub-linearly with network size, meaning that our method remains computationally feasible even for large-scale networks. Our results confirm that the method achieves stable results without requiring an excessive number of sweeps, ensuring both accuracy and practical scalability.

\subsubsection{Conclusion}
Our analysis demonstrates that the Causal Tie Importance and Causal Node Importance measures reliably converge after a low number of sweeps. The use of multiple independent convergence criteria (importance correlation, symmetry stability, and graph-size scaling) strengthens the validity of our results.

Furthermore, we find that larger graphs require fewer sweeps to achieve stability, indicating an inherent efficiency gain in our approach. The computational burden seems to scale sub-linearly with graph size, making our method well-suited for analyzing large-scale networks.

\subsection{Important Weak Ties are more Asymmetrical}
One of the central findings in our study is that the importance of weak ties in complex contagion processes is closely linked to their asymmetry. That is, the most consequential weak ties for spreading complex contagions tend to exhibit pronounced directional biases. This observation challenges traditional assumptions that weak ties are uniformly bi-directional facilitators of diffusion and suggests that their real influence is more nuanced.

\begin{figure}[H]
    \centering
    \includegraphics[width=0.8\linewidth]{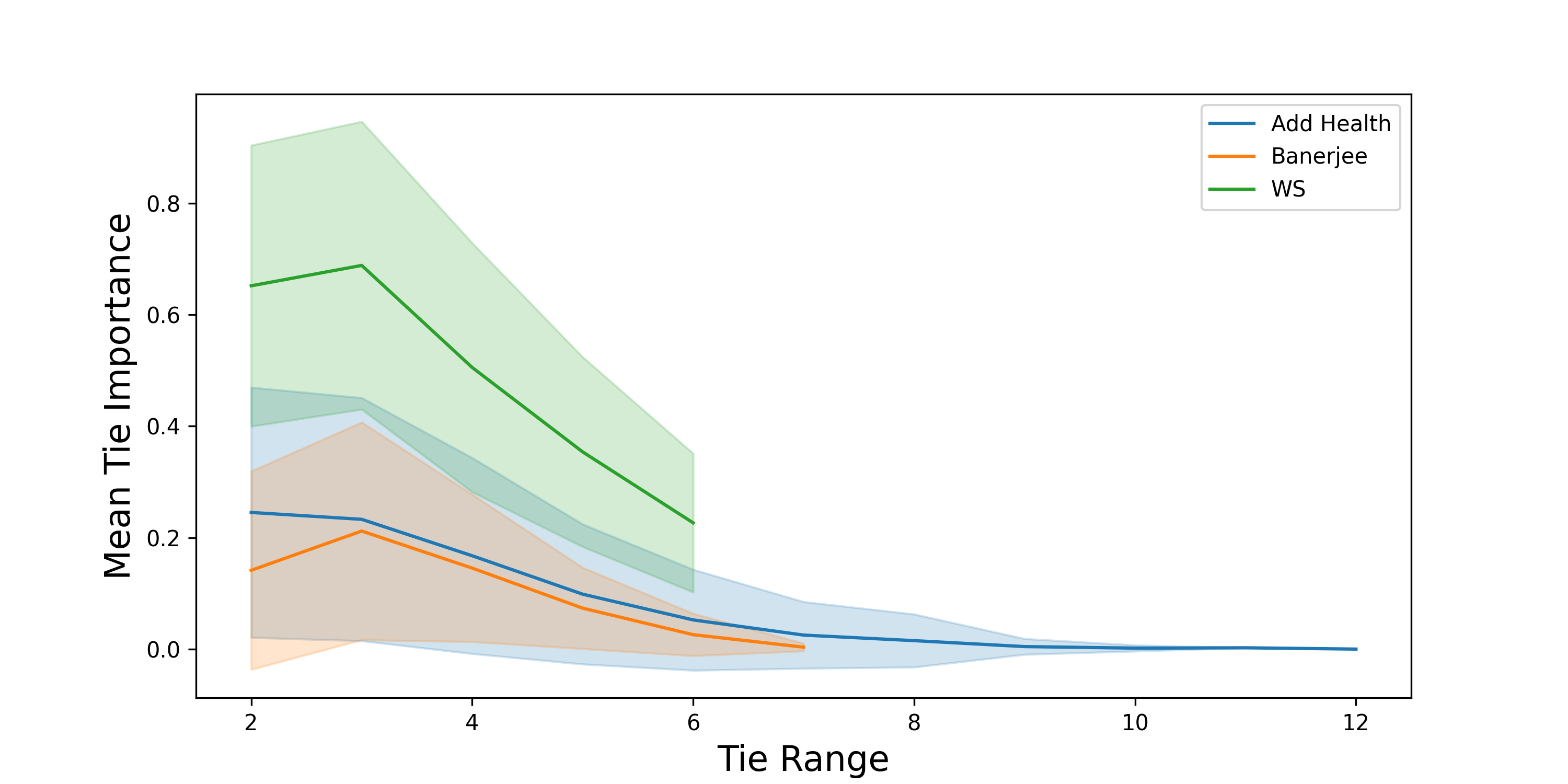}
    \caption{Mean tie importance as a function of tie range across three network structures: Add Health (blue), Banerjee (orange), and Watts-Strogatz $N=200$, $k=9$, $\beta=0.15$, $T=2$ (WS, green). The shaded regions represent the standard error. On average, Tie Importance decreases with increasing tie range.}
    \label{fig:lineplot_mean_cpc_tie_range}
\end{figure}

To systematically investigate the relationship between Tie Range and Causal Flow Symmetry, we analyze two empirical network datasets: the AddHealth high school friendship network and the Banerjee rural village social network as described in \ref{sec:network_datasets}. We calculate the Causal Tie Importance (TI) for each edge and examine the asymmetry in importance by comparing the directional TI values for each pair of nodes. We further classify ties based on their Tie Range, defined as the second shortest path length between two adjacent nodes (see Figure~\ref{fig:tie_range_heatmaps}), and assess how asymmetry scales with increasing tie range.

Figure~\ref{fig:tie_range_heatmaps}(b)-(d) presents heatmaps of tie importance asymmetry ($\Delta$) as a function of maximum TI of the two directions of a tie and Tie Range in both datasets, along with results from synthetic Watts-Strogatz (WS) graphs.

\begin{table}[ht]
\centering
\caption{Regression results without interaction for the AddHealth, Banerjee, and Watts–Strogatz datasets displayed in Figure~\ref{fig:tie_range_heatmaps}. TR = Tie Range, MI = Max of the Tie Importances for each Tie. Standard errors are in parentheses. Significance levels: * $p<0.05$, ** $p<0.01$, *** $p<0.001$.}
\label{app:regression_no_interaction}
\begin{tabular}{lccc}
\toprule
\textbf{Term} & \textbf{AddHealth} & \textbf{Banerjee} & \textbf{Watts–Strogatz} \\
\midrule
\textbf{Intercept} & -0.0118*** (0.0001) & -0.0127*** (0.0001) & 0.0036*** (0.0001) \\
\textbf{TR} & 0.0039*** (0.0000) & 0.0064*** (0.0000) & -0.0018*** (0.0000) \\
\textbf{MI} & 0.4320*** (0.0002) & 0.4622*** (0.0001) & 0.2158*** (0.0001) \\
\midrule
\textbf{R-squared} & 0.600 & 0.688 & 0.252 \\
\textbf{Adj. R-squared} & 0.600 & 0.688 & 0.252 \\
\bottomrule
\end{tabular}
\end{table}

\begin{table}[ht]
\centering
\caption{Regression results with interaction for the AddHealth, Banerjee, and Watts-Strogatz datasets displayed in Figure~\ref{fig:tie_range_heatmaps}. TI = Tie Importance Delta, TR = Tie Range, MI = Max of the Tie Importances for each Tie. Standard errors are in parentheses. Significance levels: * $p<0.05$, ** $p<0.01$, *** $p<0.001$.}
\label{app:regression_with_interaction}
\begin{tabular}{lccc}
\toprule
\textbf{Term} & \textbf{AddHealth} & \textbf{Banerjee} & \textbf{Watts-Strogatz} \\
\midrule
\textbf{Intercept} & -0.0025*** (0.0001) & 0.0082*** (0.0001) & 0.0089*** (0.0001) \\
\textbf{TR} & 0.0003*** (0.0000) & -0.0037*** (0.0001) & -0.0037*** (0.0000) \\
\textbf{MI} & 0.3244*** (0.0005) & 0.2640*** (0.0007) & 0.1997*** (0.0002) \\
\textbf{TR × MI} & 0.0447*** (0.0002) & 0.0937*** (0.0003) & 0.0062*** (0.0001) \\
\midrule
\textbf{R-squared} & 0.605 & 0.692 & 0.252 \\
\textbf{Adj. R-squared} & 0.605 & 0.692 & 0.252 \\
\bottomrule
\end{tabular}
\end{table}

Our findings reveal several key insights:

\begin{itemize}
\item Weak ties (as defined by having high Tie Ranges) exhibit significantly greater asymmetry in their importance compared to short-range ties.
\item High-importance ties are more likely to display strong directional biases, indicating that the most critical ties for contagion spread do not equally support flow in both directions.
\item The observed patterns are consistent across both the AddHealth and Banerjee networks, and are replicated in WS graphs, suggesting a generalizable structural property of social networks.
\item The WS graph results demonstrate that such asymmetry patterns emerge even in synthetic networks, reinforcing that the observed phenomenon is not merely an artifact of specific empirical datasets.
\end{itemize}

This finding has profound implications for theories of network-mediated diffusion. It suggests that weak ties, often assumed to facilitate mutual exchange and integration between distant communities, in fact predominantly channel influence in a single direction. This observation aligns with the broader pattern we uncover: complex contagions tend to propagate along emergent directed paths rather than symmetrically diffusing through undirected ties.

A further implication of our findings is that network-based interventions relying on the strategic use of weak ties must account for their directional biases. If policymakers or organizations aim to optimize the spread of beneficial behaviors or information, simply increasing the number of weak ties may not suffice. Instead, understanding and leveraging the asymmetries in tie importance is crucial for effectively guiding diffusion processes.

Our results also shed light on previously unexplained discrepancies in the effectiveness of weak ties in empirical studies. For instance, the observed nonlinear effects of weak ties in large-scale job diffusion on LinkedIn may be partially attributable to the asymmetry of weak ties: averaging across their directional influence underestimates their true impact. Future work should consider asymmetry as a key variable when evaluating the role of weak ties in social and economic outcomes.

In sum, our findings reveal that important weak ties seem to be inherently more asymmetrical, fundamentally altering our understanding of their role in complex contagion dynamics. The replication of this pattern in synthetic WS graphs suggests that it is a fundamental network property rather than a dataset-specific artifact. By integrating this insight into network theory, we move closer to a more accurate and predictive model of social diffusion processes.

\subsection{Simulation-Based Evidence for the Rarity of Symmetric Bridges}\label{app:computational_bridges}
One of the key findings in our study is that emergent directedness naturally arises in the spread of complex contagions, even in undirected graphs. A crucial implication of this phenomenon is that bridges between different network communities tend to be asymmetric in their ability to facilitate diffusion. While prior work has explored the role of weak ties and network bridges in diffusion processes, the likelihood of forming symmetric vs. asymmetric bridges in complex contagions has not been systematically examined likely due to the lack of knowledge about emerging asymmetries. In this section, we present an computational analysis demonstrating that, under random bridge formation, asymmetric connections between two groups are significantly more probable than symmetric ones and that the introduction of a higher portion of triadic closure is a probable strategies to build more symmetric bridges.

\subsubsection{Experimental Setup}
To systematically assess the probability of forming symmetric and asymmetric bridges between two initially disconnected groups, we conduct the following simulation:

\begin{itemize}
    \item Generate Two Disconnected Watts-Strogatz (WS) Graphs:
    \begin{itemize}
        \item Each WS graph consists of $N = 100$ nodes with mean degree $k = 6$ and rewiring probability $\beta = 0.1$.
        \item These graphs represent two separate communities with high clustering and low shortest path lengths with no initial connections between them.
    \end{itemize}

    \item Adding Ties between the two Groups A and B: Randomly or to form Triadic Closures. The portion of added ties forming triadic closure is adjusted.
    
    \item The process continues iteratively, allowing bridges to gradually form between the groups.
    
    \item Assess the Directionality of Complex Contagion Flow:
    \begin{itemize}
        \item After each new tie is added, we determine whether a complex contagion can spread:
        \begin{itemize}
            \item From group A to group B.
            \item From group B to group A.
            \item In both directions.
        \end{itemize}
        \item If a contagion can only spread in one direction, the bridge is classified as \textit{asymmetric}.
        \item If it can spread in both directions, the bridge is classified as \textit{symmetric}.
    \end{itemize}
    
    \item Empirical Probability Calculation:
    \begin{itemize}
        \item We record the fraction of cases where a symmetric bridge forms out of the total cases where spreading was possible in at least one direction.
        \item The final probability is computed as:
        \[
        P_{symmetric} = \frac{\text{Number of symmetric bridges}}{\text{Total number of bridges where spreading was possible}}
        \]
    \end{itemize}
    
    \item Simulation Repetitions:
    \begin{itemize}
        \item The entire simulation was run for 10,000 realizations of these networks to obtain statistically robust results.
        \item A complex contagion threshold of $T = 3$ was used, meaning that a node required activation from at least three neighbors to adopt the contagion.
    \end{itemize}
\end{itemize}

\subsubsection{Results}
Our simulations consistently show that the probability of forming an \textit{asymmetric bridge} is significantly higher than that of forming a symmetric one, we observe:

\begin{itemize}
    \item Asymmetry is the norm: On average, only a small fraction of bridges facilitate bidirectional contagion flow and the ones that do need a high number of ties when the ties are predominantly added randomly.
    \item Higher reinforcement thresholds amplify asymmetry: As the threshold for adoption increases, symmetric bridges become even rarer.
    \item Tie length matters: Short-range ties within groups increase the chance of forming symmetric wide bridges, but weak ties across groups almost always favor one direction over the other.
    \item Progression of Symmetry Formation: As shown in Figure~\ref{fig:symmetry_progression}, the probability of forming a symmetric connection gradually increases as more ties are added between regions. However, during the early stages of bridge formation, as soon as spreading from one region to the other becomes possible, it is much more likely to be asymmetric than symmetric. With more added ties, symmetry becomes increasingly probable, and eventually, when a very high number of ties is added, the bridge becomes so wide and reinforced that the two initially separate regions essentially merge into one, causing all bridges at this stage to be symmetric.
    \item The higher the portion of Ties added by Triadic Closure, the higher the likelihood that they form symmetric bridges.
\end{itemize}

\subsubsection{Implications}
These findings have profound implications for network-based interventions, policy design, and our understanding of cultural diffusion:

\begin{itemize}
    \item Bridges between groups rarely enable mutual influence: Instead of promoting balanced exchange, newly formed ties often direct influence in one direction, reinforcing inequalities rather than mitigating them.
    \item Network interventions should consider asymmetry: Attempts to build integrative network structures should explicitly measure and optimize for symmetric connections if mutual influence is the goal.
    \item Wide bridges require strategic formation: Simply adding ties between groups does not ensure bidirectional spread; ensuring symmetry requires careful planning of bridge structure and reinforcement dynamics. As shown in our simulations, introducing triadic closures seems to be a simple yet effective strategy to achieve this.
\end{itemize}

\subsection{Proofs for Higher Likelihood of Asymmetric Bridges in Complex Contagion Spreading}\label{app:math_proof_bridges}
Our goal in this section is to show \emph{combinatorially} that asymmetric bridges are much more likely than symmetric bridges. We begin by introducing the graph setup of two disconnected network communities $A$ and $B$ and then clarify when a set of newly added ties between these communities should be counted as a single bridge or as distinct bridges. We then establish that every minimal bridge uses exactly $T$ edges converging on one target node (Lemma 1); prove that two opposite-direction bridges can overlap in at most one edge, with such an overlap yielding a symmetric bridge (Lemma 2); and finally demonstrate that the number of asymmetric configurations strictly exceeds the symmetric ones whenever $|A|\,|B|>2T^{2}$ (Propositions 1–2). These results together entail that, under random tie formation, the cross-group diffusion of complex contagions is overwhelmingly more likely to proceed via \emph{asymmetric} bridges.

\subsection*{Setup}
Let $G=(V,E)$ be an undirected graph consisting of two disconnected subgraphs \textit{A} and \textit{B}.  Every node follows a homogeneous threshold rule: it activates once at least $T\in\mathbb N$ of its neighbours are active.  Initially all nodes in $A$ are active and all nodes in $B$ are inactive or vice versa.

\paragraph{Bridge.}  A set $E_{A\to B}\subseteq A\times B$ of inter--group edges is a \textbf{bridge from $A$ to $B$} if it activates at least one node of $B$ under the above initial condition.  It is \textbf{minimal} if no proper subset is a bridge.

\paragraph{Distinct bridges.}  We regard two inter--group tie sets as \emph{separate bridges} whenever they fail to overlap in even a single edge, i.e.\ when their intersection is empty.  Only when two edge sets coincide edge-for-edge are they treated as the same bridge.

\paragraph{Symmetric vs.\ asymmetric.} A pair of minimal bridges $E_{A\to B}$ and $E_{B\to A}$ is called 	\textbf{symmetric} if the \emph{same} set of inter--group edges triggers activation whichever group starts active—i.e.\ when the roles of $A$ and $B$ are swapped, $E_{A\to B}\cup E_{B\to A}$ still activates a node in the opposite side.  Otherwise the pair is 	\textbf{asymmetric}.  Lemma 1 implies that symmetry forces the two minimal bridges to overlap in exactly one edge, whereas asymmetric pairs are completely disjoint.  We exploit this necessary overlap later when counting how many symmetric versus asymmetric configurations exist.

\subsection*{Lemma 1: Minimal Bridges}\label{lem:min-bridge}
Any minimal bridge from $A$ to $B$ consists of exactly $T$ $A\!\leftrightarrow\!B$ edges, all incident to a single target node $b\in B$ and to $T$ distinct sources in $A$.

\textit{Proof.}  With only $A$ initially active, a node $b\in B$ can activate at $t=1$ iff it has $T$ \emph{distinct} neighbours in $A$.  Hence at least $T$ edges incident to one $b$ are necessary.  Minimality forbids superfluous edges, giving exactly $T$ ties of the form $\{a_i,b\}$ with $a_i\in A$, $i=1,\dots,T$.\hfill$\square$

\subsection*{Lemma 2: Maximum Overlap of Two Minimal Bridges \& Creation of a Symmetric Bridge}
For minimal bridges $E_{A\to B}$ and $E_{B\to A}$ we have $|E_{A\to B}\cap E_{B\to A}|\le1$.

\textit{Proof.}  By Lemma 1 we may write $E_{A\to B}=\{\{a_1,b\},\dots,\{a_T,b\}\}$ for some $b\in B$ and distinct $a_i\in A$, and similarly $E_{B\to A}=\{\{a,b_1\},\dots,\{a,b_T\}\}$ for some $a\in A$.  The only possible common edge is $\{a,b\}$, so the intersection has size at most one. This overlap also then creates a symmetric bridge from the formerly non overlapping asymmetric bridges in the counteracting directions.\hfill$\square$

\subsection*{Proposition 1: Tie Budget for Symmetric vs.\ Asymmetric Bridges}
\begin{itemize}
  \item \textbf{Asymmetric case.}  If $E_{A\to B}\cap E_{B\to A}=\varnothing$, each bridge contributes $T$ ties, so $|E_{A\to B}\cup E_{B\to A}|=2T$.
  \item \textbf{Symmetric case.}  Sharing the single allowed edge (Lemma~2) yields $|E_{A\to B}\cup E_{B\to A}|=T+(T-1)=2T-1$.  Thus symmetry saves exactly one inter--group tie.
\end{itemize}\hfill$\square$

\subsection*{Proposition 2: Asymmetric Bridges Are More Numerous}
Call a pair of bridges $E_{A\to B},E_{B\to A}$ \textbf{symmetric} if the two edge sets overlap on exactly one edge, and \textbf{asymmetric} if they are disjoint. With $n_A=|A|\ge T$ and $n_B=|B|\ge T$ define

$$
  \mathcal P_{\mathrm{sym}} = \bigl\{(E_{A\to B},E_{B\to A}):|E_{A\to B}\cap E_{B\to A}|=1\bigr\}, \quad
  \mathcal P_{\mathrm{asym}} = \bigl\{(E_{A\to B},E_{B\to A}):|E_{A\to B}\cap E_{B\to A}|=0\bigr\}.
$$

\paragraph{Counting $\mathcal P_{\mathrm{sym}}$.}
Fix the common edge $\{a^\ast,b^\ast\}\in A\times B$ ($n_A n_B$ choices) and choose $T-1$ additional sources on either side:
$$
  |\mathcal P_{\mathrm{sym}}| = n_A n_B 
    \binom{n_A-1}{T-1} \binom{n_B-1}{T-1}.
$$

\paragraph{Counting $\mathcal P_{\mathrm{asym}}$.}
A directional bridge is specified by its target in the opposite group and $T$ distinct sources at home (Lemma~\ref{lem:min-bridge}), so
$$
  |\mathcal P_{\mathrm{asym}}| = n_A n_B 
    \binom{n_A}{T} \binom{n_B}{T} - |\mathcal P_{\mathrm{sym}}|.
$$

\paragraph{Ratio.}
Using $\displaystyle \binom{n_A}{T}=\frac{n_A}{T}\binom{n_A-1}{T-1}$ (and the analogous identity for $B$):
$$
  \frac{|\mathcal P_{\mathrm{sym}}|}{|\mathcal P_{\mathrm{asym}}|} = \frac{T^{2}}{\,n_A n_B - T^{2}}.
$$

\paragraph{Conclusion.}
Whenever $n_A n_B > 2T^{2}$ this ratio is below one, i.e.
$$
  |\mathcal P_{\mathrm{sym}}| < |\mathcal P_{\mathrm{asym}}|,
$$
so there are \emph{more} ways to realise an asymmetric pair of minimal bridges than a symmetric one.
\hfill$\square$

\subsection*{Take‑away}
Even though a symmetric configuration uses one tie less, the combinatorial space of asymmetric bridges is much larger, making asymmetric cross--group influence via bridges the more likely outcome under random tie formation.

\subsection*{Lemma 3: Increasing thresholds reduces likelihood for symmetric bridge formation}

Let the two communities satisfy \(|A|=|B|=N\) with $N \gg T$.
Fix an integer \(T\) with \(2\le T\le 9\).
Draw \(2T\) inter-community ties between $A$ and $B$.
For each \(T\) write
\[
P_{N}(T)\;:=\;
\Pr\!\bigl[\text{some vertex in }A\cup B\text{ is incident to at least }T
\text{ of these ties}\bigr].
\]

Then
\[
\boxed{\;
  P_{N}(T)\;\le\;2N\,(T+1)\Bigl(\tfrac{4}{N}\Bigr)^T
  \;}\qquad\text{and}\qquad
P_{N}(T+1)\;<\;P_{N}(T)\quad(\,2\le T\le 8\,).
\]

\textit{Proof: }
\textbf{Incidence distribution.}
For a fixed vertex \(v\in A\cup B\) let \(X_v\) be its incidence count
among the \(2T\) ties.
Because each tie chooses its \(A\)-endpoint (and likewise its
\(B\)-endpoint) uniformly at random,
\[
X_v\sim\mathrm{Binomial}\bigl(2T,\tfrac1N\bigr).
\]

\textbf{Union bound.}
There are \(2N\) vertices, so for any fixed vertex \(v_0\)
\[
P_{N}(T)=\Pr\!\bigl[\exists v:\,X_v\ge T\bigr]
\;\le\;
2N\,\Pr[X_{v_0}\ge T].
\]

\textbf{Bounding a binomial tail.}
With \(p:=1/N\),
\[
\Pr[X_{v_0}\ge T]
   =\sum_{j=T}^{2T}\binom{2T}{j}p^{j}(1-p)^{2T-j}
   \;\le\;(T+1)\binom{2T}{T}\,p^{T},
\]
because each of the \(T+1\) terms in the sum is at most
\(\binom{2T}{T}p^{T}\) (we drop the factor \((1-p)^{2T-j}\le1\)).
The central binomial coefficient satisfies
\(\binom{2T}{T}\le 4^{T}\), so
\[
\Pr[X_{v_0}\ge T]\;\le\;(T+1)4^{T}N^{-T}.
\]

\textbf{Combine.}
Substituting in the union bound gives
\[
P_{N}(T)\;\le\;2N\,(T+1)\Bigl(\tfrac{4}{N}\Bigr)^{\!T}.
\]

\textbf{Strict monotonicity.}
The event
\(\{\exists v:X_v\ge T+1\}\) is a subset of
\(\{\exists v:X_v\ge T\}\); hence
\(P_{N}(T+1)<P_{N}(T)\) for every \(T\).

\subsection*{Proposition 3: Increasing thresholds decrease likelihood for symmetric bridges}
Under random tie formation between $A$ and $B$ via $2T$ ties, let $P_N(T)$ be the probability of a minimal symmetric bridge in one direction.  Assuming independence between directions, define:

1. The probability of a symmetric bridge (minimal bridges both ways) is
$$
P_{\mathrm{sym}}(T) = [P_N(T)]^2.
$$

2. The probability of an asymmetric bridge (which can spread in exactly one direction) is
$$
P_{\mathrm{asym}}(T) = 2\,P_N(T)\bigl(1 - P_N(T)\bigr).
$$

3. The ratio of symmetric to asymmetric probabilities:
$$
\frac{P_{\mathrm{sym}}(T)}{P_{\mathrm{asym}}(T)}
= \frac{[P_N(T)]^2}{2\,P_N(T)\bigl(1 - P_N(T)\bigr)}
= \frac{P_N(T)}{2\,(1 - P_N(T))}.
$$
Since $P_N(T)$ is strictly decreasing in $T$ and lies in $(0,1)$, the ratio also decreases with $T$, therefore making symmetric bridges increasingly unlikely for increasing thresholds. Figure~\ref{fig:panel_declining_symmetry_increasing_threshold} presents strong statistical evidence in support of this proposition by examining the distribution of bridges across a wide range of synthetic and real-world social networks. 

\subsection{The Network Datasets}\label{sec:network_datasets}

In our study, we analyze two distinct network datasets. For both datasets, we focus exclusively on the giant component of each network to eliminate isolated communities and ensure that the dynamics captured are representative of the dominant social structure. 

\paragraph{AddHealth.}
The AddHealth dataset originates from the National Longitudinal Study of Adolescent Health, a comprehensive survey designed to investigate the health-related behaviors of adolescents in the United States \cite{harris2013add}. The dataset includes rich information on social connections, making it an invaluable resource for studying friendship networks within high schools. In our analysis, we extract the friendship networks and focus solely on the giant component of each network. 

\paragraph{Banerjee Village.}
The Banerjee Village dataset stems from research conducted on the social and economic interactions within a rural village \cite{banerjee2013diffusion}. This dataset captures detailed information on the interpersonal relationships among residents, providing insights into the fabric of rural social networks. Consistent with our approach to the AddHealth dataset, we restrict our analysis to the giant component of the network. 

\subsection{Seeding Strategies}
\label{subsec:seeding-strategies}
\begin{figure}[H]
    \centering
    \includegraphics[width=0.6\linewidth]{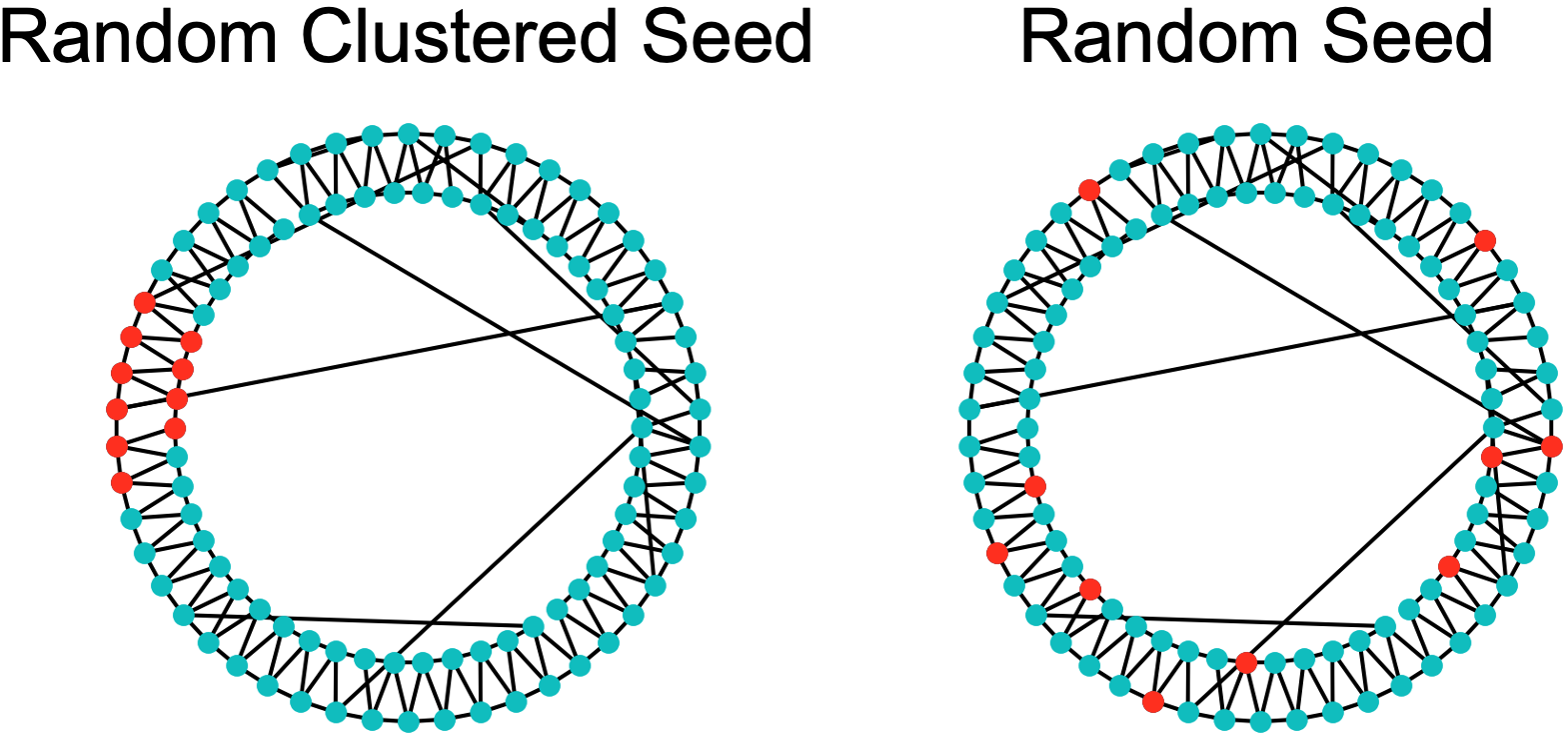}
    \caption{Two seeding strategies illustrated on a Watts-Strogatz network. 
    In the Random Clustered Seed approach (left), seed nodes (red) are confined to a dense subregion, each adjacent to at least one other seed node, accelerating a strong local cascade. By comparison, the Random Seed approach (right) spreads seed nodes more widely across a graph, potentially activating multiple regions at once. }
    \label{fig:Seed_topology_example}
\end{figure}

Figure~\ref{fig:Seed_topology_example} illustrates two broad seeding strategies—Random Clustered Seed (RCS) and Random Seed (RS)—that we use throughout this paper. Although displayed here on a Watts--Strogatz (WS) network for demonstration, these approaches can be applied to various other graphs. In both strategies, the seed size is always set to a certain fraction of the total number of nodes in the network (e.g., 0.05 for 5\% of the nodes). RCS and RS differ only in how these seeds are positioned topologically.

In the Random Clustered Seed configuration (left panel), the seed nodes (red) are placed in a single dense cluster, promoting strong local reinforcement where each seed node is adjacent to at least one other seed node. Because many seeds lie near one another, they can quickly activate their neighbors under a threshold model, often forming a rapid local cascade.

By contrast, the Random Seed configuration (right panel) disperses seed nodes randomly across the entire network, potentially allowing for a broader initial reach. This can spark multiple, simultaneous diffusion fronts, though each localized cascade may be weaker than in the heavily clustered case.

\subsection{Causal Path Visual Example} \label{sec:causal-path-visual}

\begin{figure}[H]
    \centering
    \includegraphics[width=1.0\linewidth]{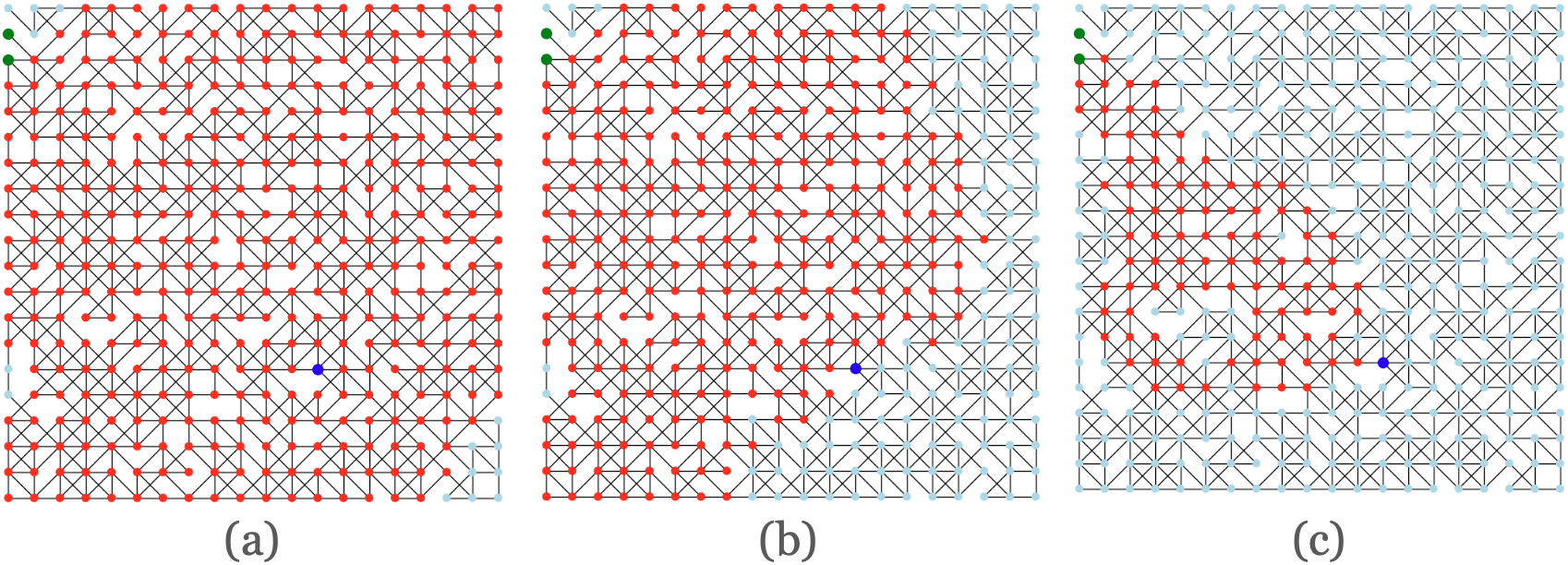}
    \caption{Visualization of the causal subgraph extraction process for a specific target node (blue). The green nodes are initial seeds from which activation spreads under a threshold model, and the red nodes are all activated nodes at convergence. (a) Shows the full spread after diffusion. (b) Retains only nodes that activated earlier than the target node. (c) Depicts the final distilled causal subgraph, which captures only those predecessors that causally contributed to the target node’s activation therefore displaying the outcome of our Causal Identification algorithm.}
    \label{fig:causal_tree_visualization}
\end{figure}

Figure~\ref{fig:causal_tree_visualization} demonstrates how our algorithm identifies the causal predecessors of a particular target node in a threshold-based diffusion process. The green nodes at the top are the initial seeds from which activation spreads. The blue node near the bottom is our “goal node,” for which we want to determine all nodes (and edges) that causally contributed to its activation.

\paragraph{(a) Induced subgraph of active nodes.}
Shows the complex contagion spreading outcome from the green seed nodes, with every active node marked in red. By the end, the blue node also becomes active, but at this stage we have not distinguished which red nodes actually contributed to its activation versus those that happened to activate on separate paths or too late.

\paragraph{(b) Earlier‐Activated Nodes Only.}
Displays only those nodes whose activation time is earlier than the blue node’s activation. This rules out any nodes that activated after the blue node. However, some remaining earlier‐active nodes may still be irrelevant if they never influenced the blue node’s activation chain.

\paragraph{(c) Causal Subgraph.}
Depicts the causal subgraph deduced by the algorithm. We obtain this by recursively “back‐tracing” from the blue node, adding only those predecessors that causally influenced its activation. This yields a directed acyclic graph (DAG) pointing from the seed nodes to the blue node, because the flow of influence can branch and merge.
\newline
A directed acyclic graph (DAG) is a structure in which all edges point in a single direction and no cycles exist. In the context of threshold diffusion, this direction naturally goes from earlier‐activated nodes to later‐activated nodes, and cycles are impossible because a node cannot re‐activate once it is already active. Thus, the flow of influence is strictly “cause to effect,” yielding a time‐ordered DAG of edges. This directional, acyclic nature of causal paths also explains why Tie Importance ($\text{TI}$) can differ between the two directions of the same edge. Even in an undirected network, once we track who activated first, the edge $(i, j)$ is effectively used only in the direction earlier $\to$ later. If, across many seed sets, node $i$ consistently activates before node $j$ (and thus contributes to activating $j$), then $\text{TI}[i, j]$ may be high, whereas $\text{TI}[j, i]$ remains low or zero. In other words, the causal timing imposes a one‐way flow of influence along each edge, causing asymmetry in how frequently each “direction” is counted in the final importance scores.

\subsection{Replication of the Inverse U-Shape Across Tie-Strength Groups}
\label{subsec:inverse-u-shape}

In their work on ``A Causal Test of the Strength of Weak Ties,'' Rajkumar et al.~\cite{rajkumar2022causal} introduce a notion of structural tie strength for Ties $(i,j)$ based on their mutual connections. Formally, they define:
$$
  \text{StructuralTieStrength}_{i,j} \;=\; \frac{M_{ij}}{D_i + D_j \;-\; M_{ij} \;-\; 2},
$$
where $M_{ij}$ is the number of mutual neighbors shared by $i$ and $j$, and $D_i$ and $D_j$ are the degrees of $i$ and $j$, respectively. Intuitively, the more common neighbors two nodes share relative to their degrees, the stronger their structural tie strength.

Following \cite{rajkumar2022causal}, we partition ties into three categories-\textit{weak}, \textit{medium}, and \textit{strong}—by sorting all ties in the graph by their structural tie strength and then grouping them into the bottom, middle, and top thirds of this distribution, respectively.

\begin{figure}[H]
    \centering
    \includegraphics[width=0.5\linewidth]{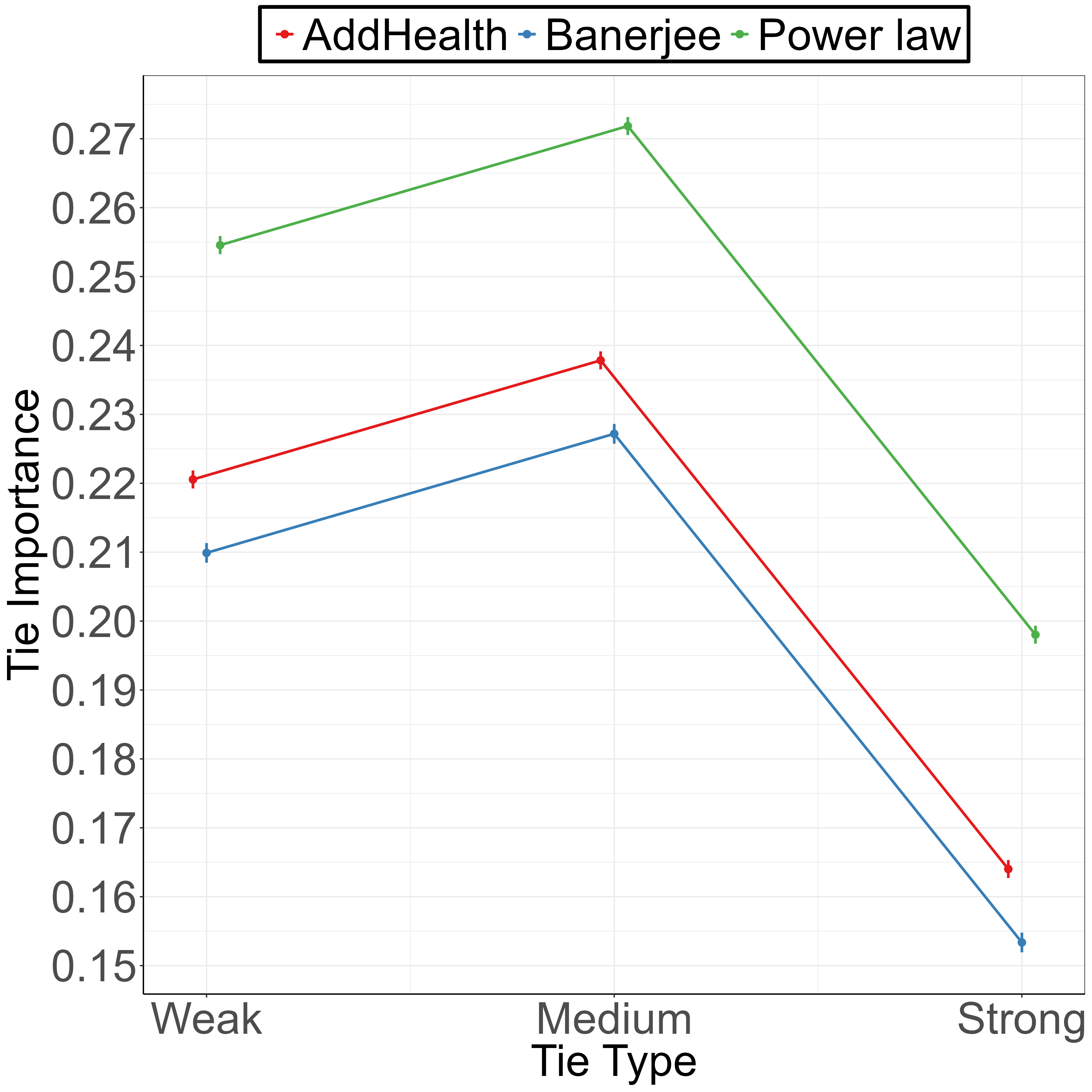}
    \caption{Mean tie importances for three different network datasets under a complex contagion model with different relative thresholds ($\theta$ in [0.1, 0.15, 0.2, 0.25, 0.3, 0.35]) and a seed size of $2\%$ (random clustered seeding). Tie strengths are categorized using the structural tie strength formula in \cite{rajkumar2022causal}, and then grouped into the bottom (weak), middle (medium), and top (strong) thirds of the distribution. (Left) Generated clustered power law networks using the with $n = 1000$, $m \in {2,3,4,5}$, and probability of adding a triangle after adding a random edge $p \in [0, 1]$ (sampled linearly over 85 values). (Middle) The giant components of the Add Health dataset. (Right) The giant components of the Banerjee Village Networks. Across all three datasets, the results exhibit an inverse U-shape, with medium-strength ties showing the highest mean tie importance, aligning with empirical findings on the disproportionately large role of mid-range tie strengths. Error bars represent standard deviations across multiple runs.}
    \label{fig:Average_Tie_Importance_by_Tie_Strength_Group}
\end{figure}

Figure~\ref{fig:Average_Tie_Importance_by_Tie_Strength_Group} shows our average Causal Tie Importance measure grouped by these tie-strength categories. We apply a complex contagion model with a relative threshold ($\theta$ in [0.1, 0.15, 0.2, 0.25, 0.3]) and a random clustered seeding of $\approx 2\%$ of the nodes on all graphs. The figure compares three different network datasets: (1) a synthetic clustered power law network, (2) the Add Health dataset, and (3) the Banerjee Village Networks. In each case, mean tie importance exhibits a distinct inverse U-shape across tie-strength categories, with medium-strength ties showing the greatest overall contribution to diffusion. This pattern reflects the empirical finding in \cite{rajkumar2022causal} that mid-range tie strengths can strike a crucial balance: they are sufficiently ``close'' to facilitate influence, yet not so strong as to be confined to tightly knit sub-communities. This is a replication of the main analysis shown in Figure~\ref{fig:panel_inverse_V}, broken down by dataset. The inverted-u pattern replicates equally strongly across all datasets examined. 

A theoretical replication using our Causal Tie and Node Importance measures is highly relevant because it validates that our framework captures core diffusion phenomena observed in real networks—namely, the disproportionately large role of medium-strength ties. By reproducing the empirical ``inverse U'' pattern previously demonstrated by others, we gain confidence that our causal approach is both methodologically robust and theoretically aligned with established findings on how social influence and information spread. This positions our measures not as an isolated theoretical construct, but as a practical tool that mirrors—and potentially extends—the well-studied properties of tie strength in network contagion.

\subsection{Core to Periphery shift}\label{app:periphery_shift}
\begin{figure}[H]
    \centering
    \includegraphics[width=0.7\linewidth]{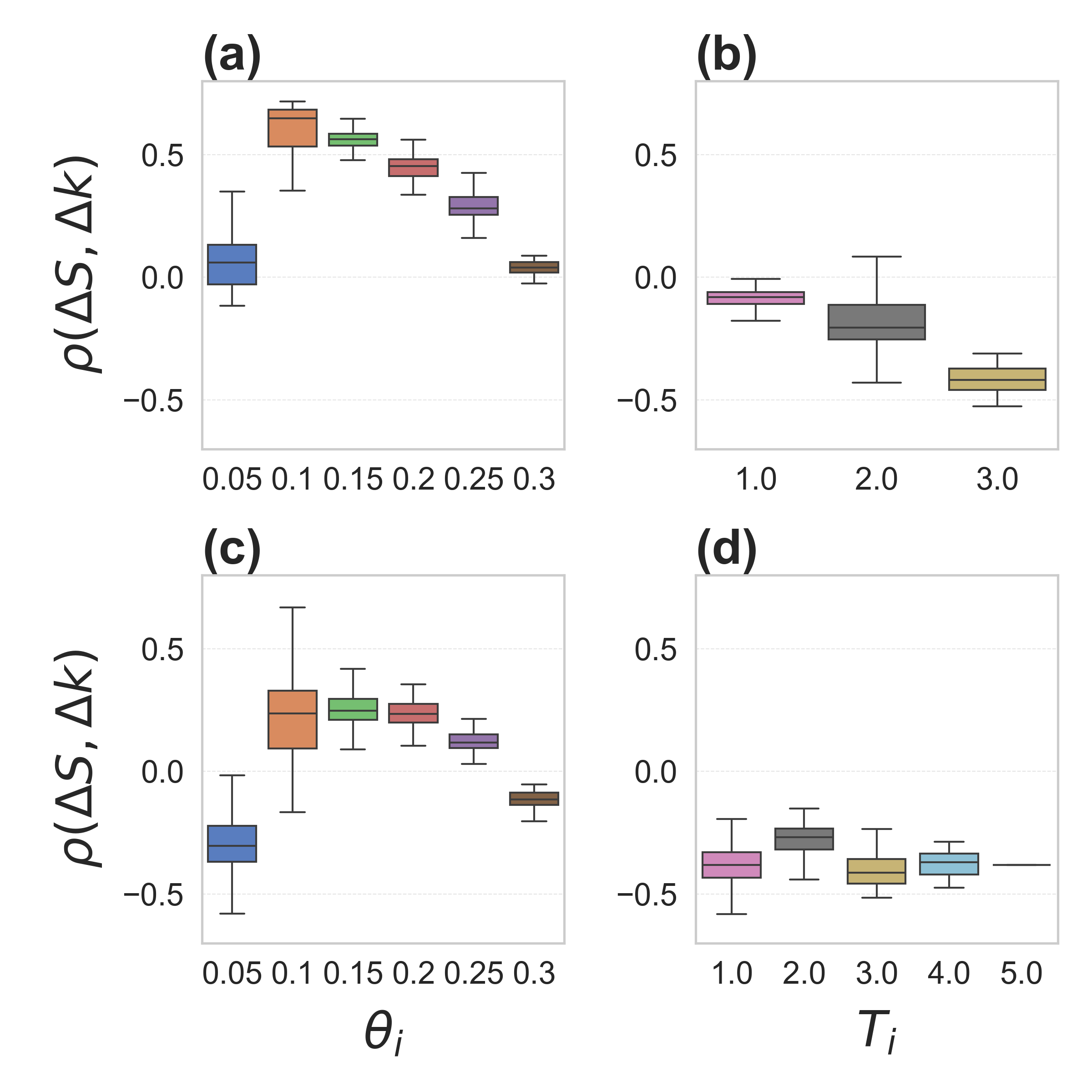}
    \caption{Boxplots of the correlation $\rho(\Delta S, \Delta k)$, capturing the relationship between tie strength asymmetry ($\Delta S$) and node degree difference ($\Delta k$) across varying relative thresholds $\theta$ and absolute thresholds $T_i$. The annotated axis delineates two influence regimes: a core-to-periphery flow (negative correlation) and a periphery-to-core reversal (positive correlation), reflecting structural shifts in the pathways of contagion propagation. under varying threshold values. Only graphs with meaningful spreading (spreading density $>$ 10\% of nodes) are included. Panels (a) and (b) use random seeding of 2\% of nodes, while (c) and (d) apply random clustered seeding of 2\% of nodes. Relative thresholds $\theta_i$ are used in (a) and (c), and absolute thresholds $T_i$ are used in (b) and (d). Positive correlations reflect a periphery-to-core influence pattern, where peripheral (low-degree) nodes increasingly drive spreading, while negative correlations indicate a core-to-periphery flow dominated by central hubs. For each threshold configuration and graph in the AddHealth dataset, $100N$ independent clustered seed sets covering $5\%$ of nodes were used.
    Relative thresholds show a clear transition toward periphery-driven influence at intermediate values, whereas absolute thresholds suppress this reversal and maintain core dominance across all conditions.}
    \label{fig:enter-label}
\end{figure}
\newpage
\bibliographystyle{unsrt}
\bibliography{references} 
\AddToHook{enddocument/afteraux}{\immediate\write18{cp output.aux main_arxiv_appendix.aux}}